\documentclass[prd,10pt,floatfix,amsmath,amssymb,twocolumn,longbibliography,superscriptaddress,nofootinbib]{revtex4-1}

\usepackage{verbatim} 
\usepackage{mathtools}
\usepackage{graphicx}
\usepackage[T1]{fontenc} 
\usepackage[dvipsnames]{xcolor}
\usepackage[toc,page]{appendix}
\usepackage{array} 
\usepackage[pdftex,breaklinks,colorlinks,
linkcolor=Blue,
citecolor=teal,
anchorcolor=red,
urlcolor=cyan]{hyperref}
\usepackage{orcidlink}
\usepackage[normalem]{ulem}

\usepackage{tensor}
\usepackage{url}
\usepackage{mathrsfs} 
\usepackage{braket} 
\graphicspath{{Plots/}}

\usepackage{booktabs} 

\newcommand{\bracket}[1]{\left( #1 \right)}
\newcommand{\deltaT}{\delta T}
\newcommand{\deltapT}{\Delta T}

\newcommand{\rmin}{{r_{\rm min}}}
\newcommand{\rmax}{{r_{\rm max}}}

\newcommand{\pfrac}[2]{\frac{\partial #1}{\partial #2}}
\newcolumntype{M}[1]{>{\centering\arraybackslash}m{#1}} 

\newcommand{\pth}{\text{\normalfont\sffamily\itshape\th\/}}

\font\ec=ecrm0800 at 11pt
\def\th{\hbox{\ec\char'336}}
\def\edth{\hbox{\ec\char'360}}

\begin{document}


\title{Implementation of a Green-Hollands-Zimmerman-Teukolsky puncture scheme for gravitational self-force calculations}

\author{Patrick Bourg\,\orcidlink{0000-0003-0015-0861}}
\affiliation{School of Mathematical Sciences and STAG Research Centre, University of Southampton,
	Southampton SO17 1BJ, United Kingdom}
\affiliation{Institute for Mathematics, Astrophysics and Particle Physics, Radboud University, Heyendaalseweg 135, 6525 AJ Nijmegen, The Netherlands}
\author{Benjamin Leather\,\orcidlink{0000-0001-6186-7271}}
\affiliation{Max Planck Institute for Gravitational Physics (Albert Einstein Institute), Am M{\"u}hlenberg 1, Potsdam 14476, Germany}
\author{Marc Casals\,\orcidlink{0000-0002-8914-4072}}
\affiliation{Institut f\"ur Theoretische Physik, Universit\"at Leipzig,\\ Br\"uderstra{\ss}e 16, 04103 Leipzig, Germany.}
\affiliation{Centro Brasileiro de Pesquisas F\'isicas (CBPF), Rio de Janeiro, 
CEP 22290-180, 
Brazil.}
\affiliation{School of Mathematics and Statistics, University College Dublin, Belfield, Dublin 4, Ireland.}
\author{Adam Pound\,\orcidlink{0000-0001-9446-0638}}
\affiliation{School of Mathematical Sciences and STAG Research Centre, University of Southampton,
	Southampton SO17 1BJ, United Kingdom}
\author{Barry Wardell\,\orcidlink{0000-0001-6176-9006}}
\affiliation{School of Mathematics and Statistics, University College Dublin, Belfield, Dublin 4, Ireland.}

\date{\today}

\begin{abstract}
    Post-adiabatic models of extreme- and intermediate-mass-ratio inspirals will require calculations of second-order gravitational self-force effects in the spacetime of a spinning, Kerr black hole. We take a step toward such calculations by implementing the recently formulated Teukolsky puncture scheme with Green-Hollands-Zimmerman metric reconstruction [CQG 39, 015019 (2022)]. This scheme eliminates the critical obstacle of gauge singularities that arise in the standard ``no-string'' metric reconstruction. Our first proof-of-principle implementation is limited to the simple case of circular orbits in Schwarzschild spacetime, but the method also applies to generic orbits on a Kerr background. We conclude with a discussion of various approaches to the second-order self-force problem in Kerr.
\end{abstract}

\maketitle

\tableofcontents


\section{Introduction}

We are currently standing at the beginning of a golden era in astronomy.  Since their ground-breaking discovery~\cite{LIGOScientific2016}, gravitational waves have opened an entirely new way to probe a vast range of astrophysical settings. Because these waves interact very weakly with their surroundings, we will be able to probe systems at far greater distances than ever before. Furthermore, we will be able to study systems and phenomena which emit very little, if any, electromagnetic waves, such as black holes, dark matter and dark energy.
The upcoming next-generation space-based gravitational-wave detectors, such as LISA~\cite{LISA, LISA-LaunchDate}, promise to open the way to study uncharted territories. Specifically, one of the main targets of the LISA mission is the study of extreme-mass-ratio inspirals (EMRIs)~\cite{Babak2017,Barausse2020}. These systems, consisting of the inspiral of a stellar-mass, compact body into a massive black hole in a galactic core, will serve as unique probes of black-hole physics and enable tests of general relativity with unparalleled precision. 

Currently, the most viable method of modelling these systems is with gravitational self-force theory~\cite{Barack2018,Pound2021}, an asymptotic approximation in the limit $m \ll M$, where $m$ and $M$ are the companion’s and black hole’s respective masses. Moreover, while self-force theory was originally conceived to model EMRIs, even low-order self-force calculations and waveforms have been found to be highly accurate for all mass ratios $m/M \lesssim 10^{-1}$~\cite{Wardell:2021fyy,Albertini:2022rfe}.
At zeroth order in the mass ratio, the compact body, modelled as a test particle, follows a geodesic around the massive black hole. At first order, radiation-reaction effects come into play, inducing a metric perturbation which exerts a so-called \textit{self-force} on the particle, accelerating it away from its background geodesic trajectory. Current state-of-the-art calculations at linear order allow for generic orbits (i.e., inclined and eccentric) around Kerr black holes~\cite{vanDeMeent2016,vanDeMeent2017}.
A detailed scaling argument~\cite{Hinderer2008,Pound2021} and parameter-estimation studies~\cite{Burke:2023lno} show that to achieve the necessary phase accuracy for LISA data analysis, self-force models need to include the \textit{second-order} dissipative effects as well. On the long timescale  of the inspiral, $t\propto M^2/m$, these second-order dissipative effects accumulate to have an impact on the phase evolution comparable to first-order conservative effects: both contribute $O((m/M)^0)$ to the waveform phase~\cite{Hinderer2008,Miller2020,Pound2021}. An overarching goal of the EMRI modeling community is to compute all such necessary first- and second-order effects to produce complete waveform models with so-called first-post-adiabatic (1PA) accuracy~\cite{LISA:2022kgy,LISA:2022yao,LISAConsortiumWaveformWorkingGroup:2023arg,Colpi:2024xhw}, in which phase errors are $O(m/M)$.\footnote{The ``$n$PA'' counting stems from multiscale (or post-adiabatic) expansions of the field equations~\cite{Pound2021,Hinderer2008,Miller2020,Miller:2023ers}, which are now the almost-universal basis for self-force waveform models because they inherently maintain phase accuracy on the long time scale of an inspiral while enabling rapid waveform generation~\cite{Katz:2021yft}. However, the full multiscale framework will not be needed for the calculations in the present paper.}

Intuitively, one can expect the modelling of second-order effects to be substantially more difficult than first-order ones, as the former now include nonlinear information.
This intuition turns out to be correct: while the foundations of second-order self-force theory are well understood~\cite{Pound:2012nt,Pound2012,Pound:2017psq}, and concrete calculations have been performed for the relatively simple case of quasicircular orbits in Schwarzschild spacetime~\cite{Pound2019,Warburton2021,Wardell:2021fyy}, there have not yet been any second-order self-force calculations in the astrophysically realistic case of orbits around a Kerr black hole. This is now one of the central challenges in EMRI modelling.

What are the main differences in second-order calculations in Kerr spacetime as opposed to linear-order ones? To answer this, it is important to remind the reader of the traditional first-order framework.

At linear order in perturbation theory, one solves the linearised Einstein equation for the (first-order) metric perturbation $h_{ab}$, sourced by a point-particle stress-energy tensor $T_{ab}$ that models the inspiraling object:
\begin{equation} \label{eqn:FirstOrderPert}
    \mathcal{E}_{ab}(h) = T_{ab}.
\end{equation}
This is a system of ten coupled linear partial differential equations. In a Schwarzschild background, the equations can be fully separated using a basis of tensor spherical harmonics. A major obstacle in Kerr spacetime is that the equations are not separable using any known basis of functions.
While progress has been made to solve the system directly (see~\cite{Osburn2022} and references therein), historically the focus has instead been on solving a single, fully separable scalar equation, \textit{the Teukolsky equation}, for either of the two gauge-invariant perturbed Weyl scalars $\psi_0$ or $\psi_4$. 

In vacuum regions, each of the Weyl scalars contains almost all the information about the linear metric perturbation $h_{ab}$~\cite{Wald1973}. In fact, there is a well-developed procedure that reconstructs $h_{ab}$ in a radiation gauge from $\psi_0$ (or $\psi_4$). This metric reconstruction procedure, first developed by Chrzanowski, Cohen and Kegeles, is dubbed the \textit{CCK procedure}~\cite{Chrzanowski1975,Kegeles1979}. One of its drawbacks is that it is only applicable in vacuum regions~\cite{Ori2002,Price2006}. Nonetheless, it can still be applied to self-force calculations, by carrying it out separately in the two vacuum regions inside and outside the particle's orbit. The metric reconstructed in this fashion is then in a \textit{no-string radiation gauge}~\cite{Keidl2010,Shah2010,Pound2013,Merlin2016}.
Perhaps surprisingly, this method is still useful in the case of eccentric orbits, where the particle now evolves inside a libration region $\rmin < r < \rmax$, meaning that (in the frequency domain) the entire libration region is nonvacuum. In this case, the CCK procedure is still applicable in the vacuum regions $r< \rmin$ and $r>\rmax$ and the method of extended homogeneous solutions~\cite{Barack2008,Hopper2010} allows one to extend the solutions obtained in the vaccum regions into the libration region.
This is the only method that has been used to compute the first-order self-force on fully generic, inclined and eccentric bound orbits in Kerr spacetime~\cite{vanDeMeent2016, vanDeMeent2017}.

One is, however, faced with two major roadblocks when attempting to apply the above methods at second order.
First, the no-string metric reconstruction at first order is highly singular on the time-dependent sphere that intersects the particle at each instant, containing both jump discontinuities and Dirac-delta singularities there~\cite{Pound2013,Shah2015}. While these do not pose a problem at linear order, the second-order source term is constructed from quadratic combinations of the first-order metric perturbation (and its first and second derivatives). In the no-string radiation gauge it would therefore contain ill-defined products of distributions. Secondly, this source term at second order is not confined to a compact spatial domain, meaning neither the CCK reconstruction nor the method of extended homogeneous solutions is applicable.

Recently, two avenues have emerged to get around these problems. Both methods are based on new reconstruction procedures. Building on earlier work in Refs.~\cite{Berndtson:2007gsc,Aksteiner:2016pjt}, it was shown that the metric perturbation in Lorenz gauge can be reconstructed from solutions to a set of separable Teukolsky equations~\cite{Dolan:2021ijg,Dolan:2023enf}. The Lorenz-gauge metric perturbation has a well-behaved singularity structure confined to the particle's location. This reconstruction procedure has been applied to homogeneous perturbations~\cite{Dolan:2021ijg} and to the inhomogeneous perturbation of a point mass on a circular orbit in Kerr spacetime~\cite{Dolan:2023enf}. Recent work has extended the approach to work with generic, extended sources~\cite{Wardell:2023-talk,Wardell:2023-inprep}. The method requires the solution to up to six\footnote{Strictly speaking, the Teukolsky-Starobinsky identities mean we only need one spin-2 and one spin-1 equation, but they are complex so we still have 6 degrees of freedom} Teukolsky equations: two spin-2, two spin-1 and two spin-0 --- as opposed to just one spin-2 Teukolsky equation in the standard CCK approach --- and the solution of one of the spin-0 equations acts as a noncompact source for the other spin-0 equation (equivalently, the two spin-0 equations can be treated as a coupled system with a compact source).

Another reconstruction procedure, which we adopt here, was introduced by Green, Hollands, and Zimmerman (GHZ)~\cite{Green2019}, who showed that sourced metric perturbations can be obtained by supplementing the CCK procedure with the addition of a \textit{corrector tensor}, $x_{ab}$, which is obtained by solving a sequence of three ordinary differential equations (ODEs), one of which is complex. In Ref.~\cite{Toomani2021} (hereafter Paper~I), based on this new procedure, one of the authors of this paper formulated a \textit{puncture scheme} that avoids the pathological singularities of the no-string radiation gauge. The basic idea is to split the retarded field into two parts: a \textit{puncture} field, $h_{ab}^P$, which encodes the point-particle singularity; and a \textit{residual} field, $h_{ab}^R := h_{ab}-h_{ab}^P$, which is more regular. As in any puncture scheme~\cite{Wardell2015}, $h_{ab}^P$ is calculated analytically, and $h_{ab}^R$ becomes the numerical variable. In the GHZ puncture scheme developed in Paper~I, the puncture is put in the Lorenz gauge (or an even more regular gauge~\cite{Upton2021,Upton:2023tcv}), thereby keeping the singularity in a desirable, non-pathological form. The residual field is then calculated using GHZ reconstruction. The ultimate difference between this approach and the one in Ref.~\cite{Dolan:2023enf} is simply a gauge choice: the procedure in Ref.~\cite{Dolan:2023enf} puts the entirety of the metric perturbation in the Lorenz gauge, whereas the procedure in Paper~I only puts the singular piece of the metric perturbation in the Lorenz gauge, while putting the regular piece in a radiation gauge. 
We discuss the relative merits of the two methods in the conclusion, Sec.~\ref{sec:conclusion}.


Regardless of whether Lorenz-gauge or GHZ reconstruction is used, there are several advantages to using a puncture scheme. Most prominently, since it makes $h^R_{ab}$ the numerical variable, it allows one to work with smoother fields, which translates into more rapid convergence of numerical approximations (including mode sums, discretisations onto a grid, and inverse Fourier transforms). In addition, the self-force exerted on the particle can be calculated directly from $h_{ab}^R$. 
At linear order, puncture schemes 
bring increased computational cost as the residual field has an extended \textit{effective source}, even in the case of circular orbits. However, at second order this is immaterial because the physical source extends over the entire spacetime anyway. Moreover, puncture schemes are currently the only viable approach to second-order self-force theory~\cite{Barack2018,Pound2021}. We discuss the utility of the GHZ puncture scheme at second order in further detail in Sec.~\ref{sec:conclusion}.

In this work, we implement the GHZ puncture scheme for the first time in a realistic scenario: a point mass in circular orbit around a Schwarzschild black hole. (Paper~I had previously demonstrated the scheme in the simpler case of a static particle in flat spacetime.)
In Sec.~\ref{sec:Summary}, we give an overview of the scheme and of our main results. 
In Secs.~\ref{sec:Weyl}--\ref{sec:Gauge}, we compute, step by step, the different ingredients necessary to obtain the complete metric perturbation.
In Sec.~\ref{sec:Results}, as a consistency check, we show we obtain the correct value of the Detweiler redshift using the new GHZ puncture scheme. In that section we also discuss the regularity of the reconstructed GHZ metric. Some review and technical material is relegated to appendices.

We adopt geometric units with $G=c=1$. All plots of numerical results are in units with $M=1$. Unlike Paper I, we use a mostly positive, $(-+++)$ signature. 
\section{Overview}
\label{sec:Summary}

In this section, we review the GHZ puncture scheme, summarize the main features of our implementation, and preview our results. Readers who are uninterested in the more technical details of our calculations can skip directly to Sec.~\ref{sec:Results} after reading this section.

Figures~\ref{fig:CCK_scheme}--\ref{fig:GHZ_Fields} provide a visualization of the puncture scheme and how it contrasts with a standard no-string reconstruction and completion procedure. These figures, along with Table~\ref{table:regularity}, can be used as aids to the text throughout this section. We note that the text in this section refers to the generic scenario of a bound orbit with time-dependent radius $r_p$, but the figures specialize to the specific case of a circular orbit with $r_p=r_0=\text{constant}$, which we specialize to in subsequent sections. Similarly, the text does not necessarily specialize to mode-decomposed fields, but Table~\ref{table:regularity} refers to the (spin-weighted) spherical-harmonic modes that we use in later sections.

\subsection{Punctures and residual fields}

We consider an asymptotic expansion of the metric of the form  ${\sf g}_{ab}=g_{ab}+\epsilon h_{ab}+O(\epsilon^2)$, where $g_{ab}$ is the background Schwarzschild metric of mass $M$, and $\epsilon = m/M$ is the mass ratio of the binary. The linear perturbation $h_{ab}$ satisfies the linearised Einstein equation, Eq.~\eqref{eqn:FirstOrderPert}.
The stress-energy tensor ${\sf T}_{ab}=\epsilon T_{ab}+O(\epsilon^2)$ describes a point mass and is given by 
\begin{equation}\label{eqn:covariant Tab}
    T_{ab} = 8 \pi M\int u_a u_b \delta^4(x,x_p(\tau)) d\tau,
\end{equation}
where the factor of $M$ appears because we have factored out $\epsilon$; $\tau$ is the particle's proper time as defined in the background metric $g_{ab}$; $x_p(\tau)$ is its worldline, which will be approximated as a geodesic\footnote{In a complete treatment, $x_p$ here would instead be the leading-order (non-geodesic) term in a multiscale expansion of the orbital motion~\cite{Miller2020,Pound2021,Miller:2023ers}. However, as discussed in those references, the distinction does not materially affect the calculations in this paper.} in $g_{ab}$; $u_a=g_{ab}\,dx^b_p/d\tau$ is the particle's 4-velocity; and $\delta^4$ is the covariant delta function in $g_{ab}$. For convenience, we incorporate the Einstein equation's usual factor of $8\pi$ into $T_{ab}$.

A core feature of self-force theory is the split of the physical, retarded field $h_{ab}^{ret}$ into two distinct pieces, called the Detweiler-Whiting singular ($h_{ab}^s$) and regular ($h_{ab}^r$) fields, $h_{ab}^{ret} = h_{ab}^s + h_{ab}^r$. The singular field is a particular solution of Eq.~\eqref{eqn:FirstOrderPert}.
It contains only local information about the field created by the particle’s mass and is singular at the particle's position.  The regular field is instead a (smooth) homogeneous solution of~\eqref{eqn:FirstOrderPert}, which contains information about global boundary conditions.
In general, this split is not unique since one can always add a homogeneous solution to the definition of $h_{ab}^s$. However, a judicious split makes it possible to express the self-force only in terms of the regular field~\cite{Detweiler2000,Detweiler:2002mi,Poisson:2011nh}.
Specifically, we can choose the split such that the motion of the particle is a geodesic in the effective metric $\tilde{g}_{ab} := g_{ab} + \epsilon h_{ab}^r+O(\epsilon^2)$.

In most situations, it is not possible to calculate the exact singular field $h_{ab}^s$ and corresponding exact regular field $h_{ab}^r$. Instead, one considers a local expansion for $h_{ab}^s$, written as a series in powers of distance to the particle.
For example, in Fermi normal coordinates centered on the particle, $h_{ab}^s$ behaves at leading order like a Coulomb field, 
\begin{equation}\label{eq:hs}
h_{ab}^\mathfrak{s}= \frac{2M}{\mathfrak{s}} \delta_{ab} + O(\mathfrak{s}),
\end{equation}
where $\mathfrak{s}$ denotes the proper (orthogonal) distance from the particle's worldline and $\delta_{ab}$ is the Kronecker delta. (Like in the leading-order stress-energy tensor, a factor of $M$ appears because we have factored out $\epsilon$.)
This local expansion, or an analogous one in any convenient coordinates, is truncated at some finite order, resulting in an approximate solution to Eq.~\eqref{eqn:FirstOrderPert}, valid only in the vicinity of the particle, called the puncture field, $h_{ab}^\mathcal{P}$. We refer the reader to Refs.~\cite{Heffernan2012,Pound2014} for how to construct the puncture in practice.

The difference between the retarded and puncture field, $h_{ab}^\mathcal{R} := h_{ab}^{ret}-h_{ab}^\mathcal{P}$, is called the residual field. Since $h_{ab}^\mathcal{P}$ is only an approximate solution to~\eqref{eqn:FirstOrderPert}, the residual field is only approximately a homogeneous solution, satisfying the equation
\begin{equation}
\label{eqn:hREqnLocal}
	\mathcal{E}_{ab}(h^\mathcal{R}) = T_{ab}-\mathcal{E}_{ab}(h^\mathcal{P}).
\end{equation}
If the puncture were precisely equal to $h^s_{ab}$, then the right-hand side would vanish identically, and the residual field would coincide with the regular field. If $h_{ab}^\mathcal{P}$ is an $n$th-order puncture, in the sense that it includes terms up to order $\mathfrak{s}^n$ (inclusive), then $\mathcal{E}_{ab}(h^\mathcal{P}) \sim \partial^2 h_{ab}^\mathcal{P}$ is a $C^{n-2}$ field at $\mathfrak{s}=0$ and smooth everywhere else.\footnote{What we call an $n$th-order puncture would often be referred to as an $(n+2)$nd-order puncture. The ``$(n+2)$nd'' label corresponds to the total number of orders from $1/\mathfrak{s}$ to $\mathfrak{s}^n$.} $h^{\cal R}_{ab}$ is then $C^n$ at $\mathfrak{s}=0$, and $h^{\cal R}_{ab}=h^r_{ab}+O(\mathfrak{s}^{n+1})$. At the level of individual $\ell m$ modes in a spherical-harmonic expansion, which will be our focus here, the degree of regularity is generally increased by two due to the integration over the sphere, leading to residual field modes $h^{{\cal R},\ell m}_{ab}$ that are $C^{n+2}$ functions of $r$ at the particle's orbital radius~$r_p$.


The puncture can be extended away from the particle in any convenient way. For example, if we strictly define $h_{ab}^\mathcal{P}$ as a series expansion in coordinate distance from the particle, we can attenuate it away from the particle using a window function, $\mathcal{W}$. 
This window function must be chosen to ensure that
\begin{equation}
h_{ab}^P := h_{ab}^\mathcal{P} \mathcal{W}
\end{equation}
has the same local expansion as $h_{ab}^\mathcal{P}$ at the particle when truncated at the same order. In this paper, we take $\mathcal{W}$ to be a radial box function, such that
\begin{equation}\label{eqn:hP box window}
h_{ab}^P := [\Theta(r-\rmin)-\Theta(r-\rmax)]h_{ab}^\mathcal{P},
\end{equation}
where $\Theta(x)$ is the usual Heaviside function. The radii $r_{\rm min/max}$ are chosen such that the particle's orbital radius $r_p$ always lies in the range 
$\rmin \leq r_p \leq \rmax$.  
Following tradition, we refer to the support of the window function $\mathcal{W}$ as \textit{the worldtube} (though for our choice of box function it is in fact a shell surrounding the black hole).
The amended residual field 
\begin{equation}
h_{ab}^R := h_{ab}^{ret} - h_{ab}^P
\end{equation}
then obeys
\begin{equation}
\label{eqn:hREqnGlobal}
	\mathcal{E}_{ab}(h^R) = T_{ab}-\mathcal{E}_{ab}(h^P) =: T_{ab}^R,
\end{equation}
which applies over the entire spacetime.
$T_{ab}^R$ is dubbed the \textit{effective source}.

The residual field $h_{ab}^R$ has the properties that (1) its value and the value of its first $n$ derivatives at the particle coincide with those of the regular field $h_{ab}^r$, (2) $h_{ab}^R = h_{ab}^{ret}$ outside the worldtube.
In what is commonly referred to as the \textit{effective source approach} or \textit{puncture scheme}~\cite{Vega:2007mc, Barack:2007jh, Warburton:2013lea}, one solves Eq.~\eqref{eqn:hREqnGlobal} directly for the residual metric perturbation $h_{ab}^R$. 
This is in contrast to the mode-sum regularization approach, where one instead solves for the modes of the physical metric perturbation $h_{ab}^{ret}$ and then extracts physical effects of $h_{ab}^{\cal R}$ by subtracting the modes of $h_{ab}^{\cal P}$ before summing over modes~\cite{Barack2009}. 
We refer the reader to Ref.~\cite{Wardell2015} for a more thorough discussion and comparison of these and other methods.


As stated in the Introduction, the goal of Paper I's puncture scheme is to compute the residual field $h^R_{ab}$ through GHZ reconstruction, leaving the puncture in the Lorenz gauge. This means we solve Eq.~\eqref{eqn:hREqnGlobal} by reconstructing part of $h^R_{ab}$ through the CCK procedure (or more accurately, a CCK-Ori procedure~\cite{Ori2002}) and completing it through the addition of a corrector tensor.
In the following subsections we summarise this procedure, tracking
the regularity of relevant fields 
encountered at each stage, and summarizing these regularity properties at the level of individual $\ell m$ modes in Table~\ref{table:regularity}.

\subsection{Residual Weyl scalar}

The first step is to define a residual Weyl scalar $\psi_0^R$, constructible from $h_{ab}^R$ by applying a certain second-order differential operator $\mathcal{T}^{ab}_0$:
\begin{equation} \label{eqn:Psi0Tabhab}
\psi_0^R := \mathcal{T}^{ab}_0 h_{ab}^R = \psi_0^{ret} - \psi_0^P.
\end{equation}
The puncture $\psi_0^P$ is computed directly from  $h_{ab}^P$,
\begin{equation}
\psi_0^P := \mathcal{T}^{ab}_0 h_{ab}^P.
\end{equation}
Similarly, $\psi_0^{ret}$ is the Weyl scalar associated with $h_{ab}^{ret}$.

$\psi_0^{ret}$ satisfies the Teukolsky equation with a point-particle source,
\begin{equation}\label{eqn:Psi0retEqn}
\mathcal{O} \psi_0^{ret} =  S^{ab}_0 T_{ab},
\end{equation}
where $\mathcal{O}$ and $S^{ab}_0$ are again certain second-order differential operators. $\psi_0^R$, on the other hand, satisfies the Teukolsky equation with an effective source,
\begin{equation}
\label{eqn:Psi0REqn}
\mathcal{O} \psi_0^R = S^{ab}_0 T_{ab}^R = S^{ab}_0 T_{ab} - \mathcal{O} \psi_0^P,
\end{equation}
where we made use of the identity $\mathcal{O} \mathcal{T}^{ab}_0 = S^{cd}_0 \mathcal{E}_{cd}{}^{ab}$~\cite{Wald1978}. To make the index structure clear, here we have defined $\mathcal{E}_{cd}{}^{ab}h_{ab}=\mathcal{E}_{cd}(h)$.
The quantities $\mathcal{T}^{ab}_0 h_{ab}$, $S^{ab}_0 T_{ab}$, and $\mathcal{O}$ are given explicitly in Appendix~\ref{section:GHP_formalism}. 

The Teukolsky equations~\eqref{eqn:Psi0retEqn} and \eqref{eqn:Psi0REqn} are both supplemented with retarded 
boundary conditions at the horizon and null infinity, such that $\psi_0^R$ reduces to $\psi_0^{ret}$ outside the worldtube. In practice, then, there are two routes to calculating $\psi_0^R$: (1) solve Eq.~\eqref{eqn:Psi0REqn} for $\psi_0^R$ directly, or (2) solve Eq.~\eqref{eqn:Psi0retEqn} for the retarded field first and then compute $\psi_0^R = \psi_0^{ret} - \psi_0^P$. We detail these two methods in Sec.~\ref{sec:Weyl}.

\begin{figure} \centering
	\includegraphics[width=\columnwidth]{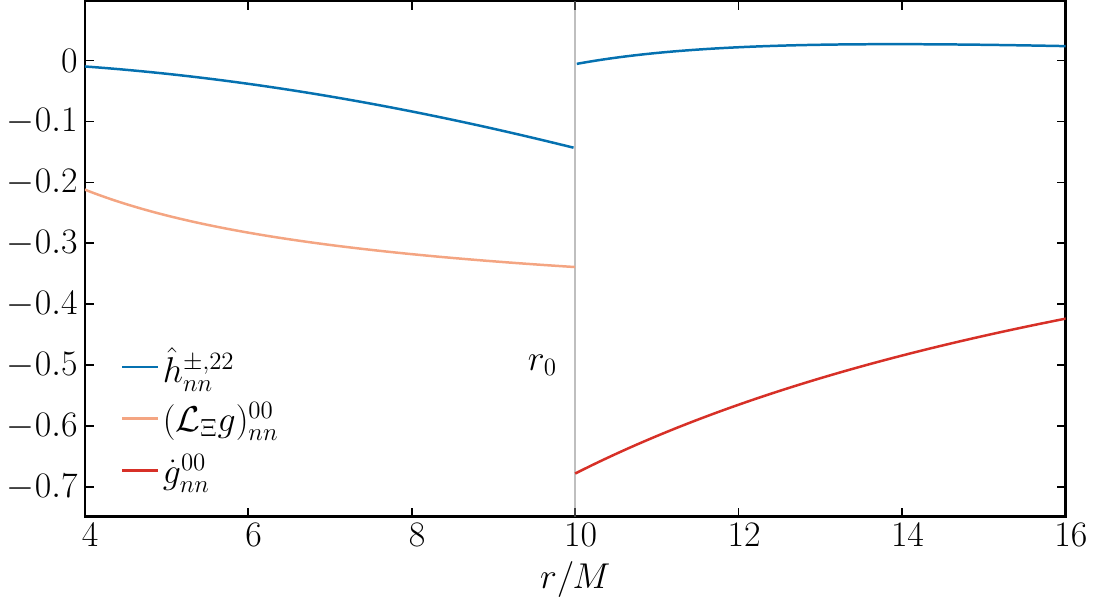}
	\caption{Representative modes of the quantities computed in traditional no-string CCK reconstruction and completion. }
	\label{fig:CCK_Fields}
\end{figure}

Since $T_{ab}$ describes a point-particle source, $S_0^{ab} T_{ab}$ is made up of terms proportional to $\delta(r-r_p(u))$ and its first and second derivatives, where $u$ is a time coordinate (which will later be taken to be retarded time).
The solution  to Eq.~\eqref{eqn:Psi0retEqn} is then a piecewise smooth function,
\begin{equation}\label{eq:psi0ret}
\psi_0^{ret} = \psi_0^- \Theta(r_p(u)-r) + \psi_0^+ \Theta(r-r_p(u)) + \psi_0^\delta \delta(r-r_p(u)),
\end{equation}
where $\psi^\pm_0$ are homogeneous solutions. Note this expression is valid at the level of $\ell m$ modes, which is the context of most of our calculations in this paper.\footnote{The sum of modes, yielding the four-dimensional function, contains no delta function or Heaviside functions but instead has a local power-law singularity supported only on the particle's worldline. See, e.g., Eq.~(170a) in Ref.~\cite{Keidl2006}.} This type of behaviour is displayed in Fig.~\ref{fig:CCK_Fields}, where we see smooth vacuum field\sout{s} modes connected by discontinuous jumps at the particle's orbital radius.

\begin{figure*} \centering
	\includegraphics[width=0.75\textwidth]{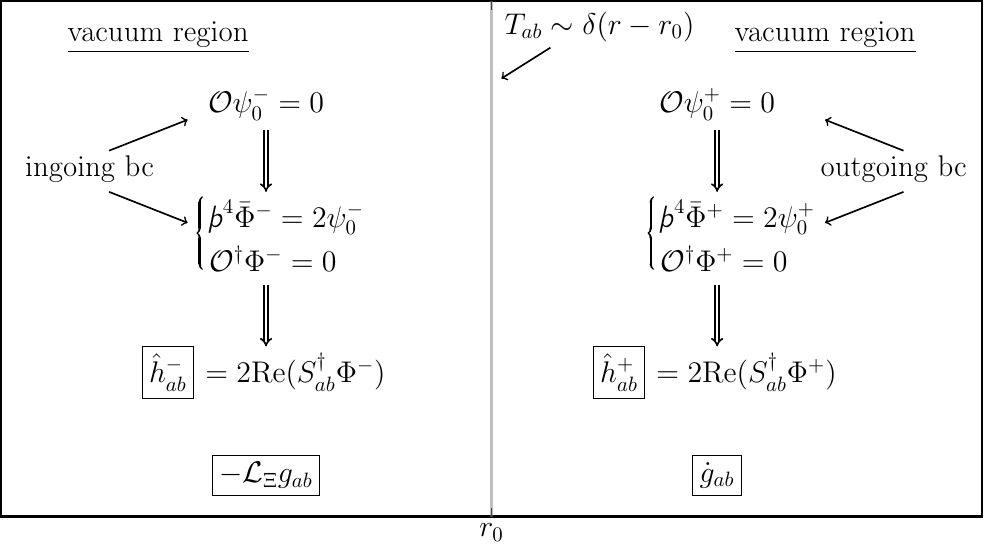}
	\caption{Summary of the no-string CCK reconstruction and completion procedure. The total solution in each region is given by the sum of the boxed quantities.}
	\label{fig:CCK_scheme}
\end{figure*}

\begin{figure*} \centering
\includegraphics[width=0.75\textwidth]{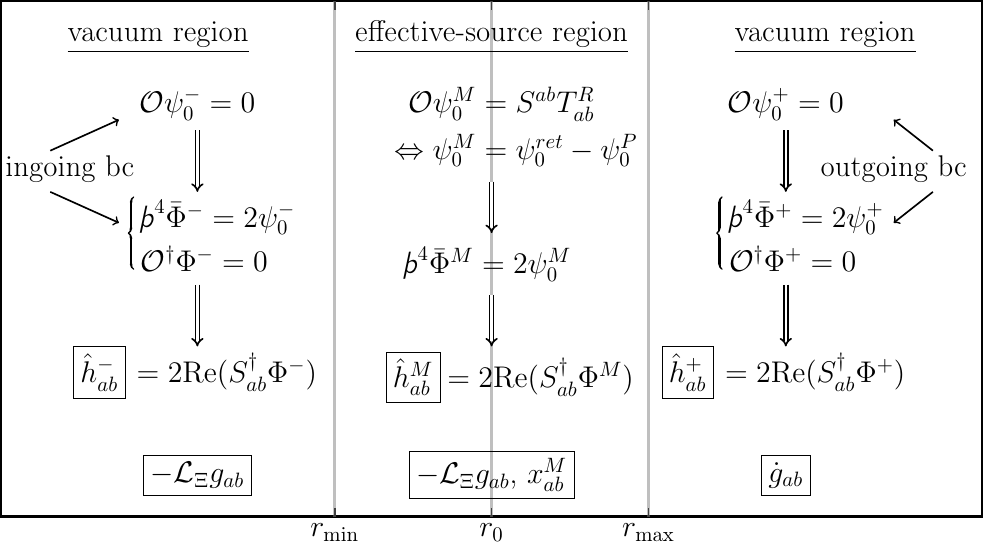}
	\caption{Summary of the GHZ puncture scheme. As in Fig.~\ref{fig:CCK_scheme}, the total solution in each region is given by the sum of the boxed quantities. Note that inside the worldtube, $\psi_0^M = \psi_0^R$.}
	\label{fig:GHZ_puncture_scheme}
\end{figure*}

\begin{figure} \centering
	\includegraphics[width=\columnwidth]{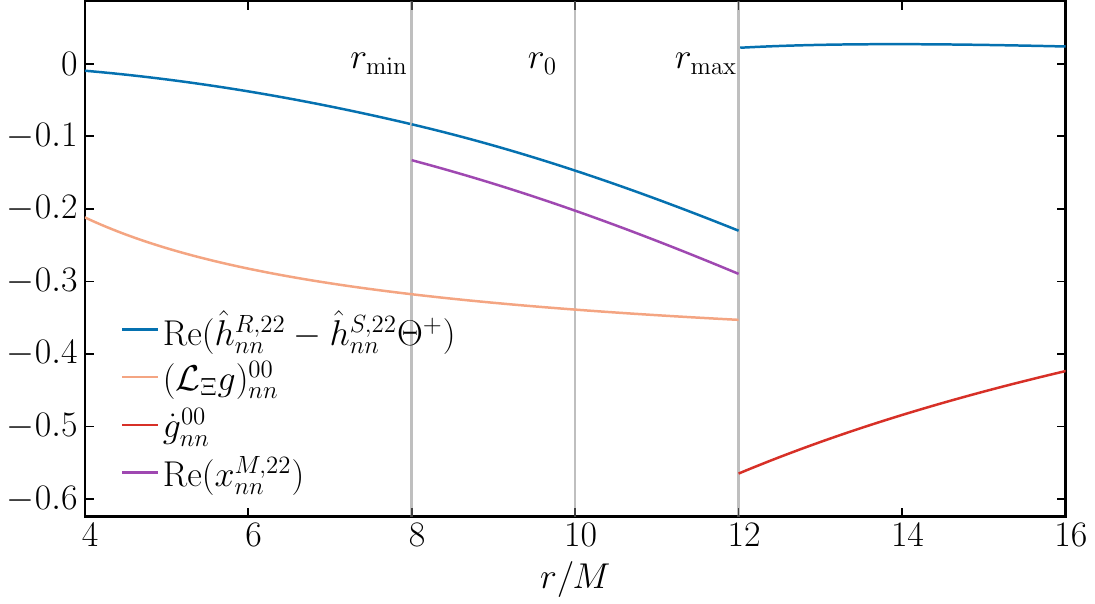}
	\caption{Representative modes of the quantities entering into the solution~\eqref{eqn:habShadowlessGauge} in the GHZ puncture scheme with a puncture of order $n=2$. Note that $\hat{h}^{R,22}_{nn} - \hat{h}^{S,22}_{nn} \Theta^+$ corresponds to $\hat{h}^-_{nn}$  for $r<\rmin$, and $\hat{h}^+_{nn}$ for $r>\rmax$.
    }
	\label{fig:GHZ_Fields}
\end{figure}

Since the residual field $\psi_0^R$ has an extended source that fills the worldtube, it has a more complicated structure. Outside the worldtube, where the puncture vanishes, $\psi_0^R$ reduces to the vacuum solutions $\psi_0^\pm$. Inside the worldtube, (at the level of modes) the retarded field's $\delta$ function and discontinuity are precisely cancelled by $\psi_0^P$, leaving a residual field whose degree of smoothness at the orbital radius is governed by the order of the puncture, as listed in Table~\ref{table:regularity}. At the worldtube boundaries $\rmin$ and $\rmax$, the box window function in the puncture~\eqref{eqn:hP box window} introduces $\delta$ and $\delta'$ functions in $\psi_0^P$ and therefore in $\psi_0^R$ (both in 4D and at the mode level). This type of behavior is displayed in Figs.~\ref{fig:GHZ_puncture_scheme} and \ref{fig:GHZ_Fields}, where we see smooth vacuum fields outside the worldtube, finitely differentiable residual fields in the worldtube interior, and jumps and delta functions at the junctions between these regions. 

\begin{table}[tb] 
    \begin{ruledtabular}
        \begin{tabular}{M{0.15\columnwidth}|
				M{0.12\columnwidth}
				M{0.12\columnwidth}
                M{0.12\columnwidth}
                M{0.12\columnwidth}
                M{0.12\columnwidth}
                M{0.12\columnwidth}}
		$r$ & $h^{P,\ell m}_{ab}$ & ${}_2 \psi^P_{\ell m}$ & ${}_2 \psi^{R}_{\ell m}$ & ${}_{-2} \Phi^{R}_{\ell m}$ & $x_{ab}^{\ell m}$ & $h^{R,\ell m}_{ab}$\\
		\midrule
		$r_0$ & $C^0$ & $\Theta,\delta$ & $C^n$ & $C^{n+4}$ & $C^{n+2}$ & $C^{n+2}$ \\
		$r_{\rm min/max}$ & $\Theta$ & $\Theta,\delta, \delta'$ & $\Theta,\delta,\delta'$ & $C^1$ & $\Theta$ & $\Theta$ \\
	\end{tabular}
    \end{ruledtabular}
\caption{Regularity properties of  relevant quantities at the particle's orbital radius $r_0$ and at the worldtube boundaries $r_{\rm min/max}$. $\Theta$, $\delta$, and $\delta'$ denote Heaviside, Dirac delta, and the first derivative of Dirac delta functions, respectively. $n$ refers to the order of the puncture. 
}
\label{table:regularity}
\end{table}

\subsection{Residual Hertz potential}
\label{sec:summary Hertz}

From $\psi_0^R$, one can obtain a spin-weight $s=-2$ object, called the ingoing radiation gauge (IRG) Hertz potential $\Phi^R$, by solving the radial inversion relation,\footnote{Here we follow the conventions of Paper I. In the conventions of Ref.~\cite{Pound2021}, for example, the factor of 2 is instead a factor of 4.}
\begin{equation}
\label{eqn:PhiREqn}
\pth^4 \bar{\Phi}^R = 2 \psi_0^R,
\end{equation}
where $\bar{\Phi}^R := (\Phi^R)^\star$ and similarly for other Hertz potentials. Throughout, an upper `$\star$' denotes complex conjugation. The operator $\pth$ is a Geroch-Held-Penrose (GHP) derivative along outgoing null curves, making Eq.~\eqref{eqn:PhiREqn} a fourth-order ODE along those curves.
We refer the reader to Appendix~\ref{section:GHP_formalism} for a summary of the GHP formalism. In appropriate retarded coordinates $(u,r)$ and with an appropriate choice of Newman-Penrose tetrad, $\pth$ simply reduces to a radial derivative $\frac{\partial}{\partial r}$. 

The Hertz potential also satisfies the adjoint Teukolsky equation 
\begin{equation}\label{eqn:OdagPhi=eta}
\mathcal{O}^\dagger\Phi^R=\eta^R, 
\end{equation}
where `$\dagger$' on an operator denotes the adjoint, and where the source $\eta^R$ satisfies a transport equation along outgoing null geodesics, which will not be needed here. The support of $\eta^R$ generically extends from $r_{\rm min}$ to $\infty$. However, in practice, we instead solve vacuum equations outside the worldtube, 
\begin{equation}
\mathcal{O}^\dagger\Phi^\pm=0.
\end{equation}
In the region connected to null infinity, we impose purely outgoing boundary conditions at null infinity, corresponding to $\Phi^+$, while in the region connected to the black hole horizon, we impose purely ingoing boundary conditions at the horizon, corresponding to $\Phi^-$. These conditions determine the homogeneous solutions $\Phi^\pm$ in each region, up to an overall constant. The overall constants are then fixed by Eq.~\eqref{eqn:PhiREqn}, which becomes $\pth^4\bar\Phi^\pm=2\psi^\pm_0$ in these regions.\footnote{For $\Phi^-$, this procedure yields the same field as one would obtain by integrating Eq.~\eqref{eqn:PhiREqn} outward along outgoing null rays from the past horizon. However, note that for $\Phi^+$, this procedure differs from integrating Eq.~\eqref{eqn:PhiREqn} inward along radial null rays from future null infinity, which would yield an alternative field
\begin{equation}
\bar\Phi^+_{\rm alt}(u,r) = 2\int^r_\infty\int^{r_4}_\infty\int^{r_3}_\infty\int^{r_2}_\infty \psi^+_0(u,r_1) dr_1 dr_2 dr_3 dr_4.
\end{equation}
Such a field would not satisfy ${\cal O}^\dagger\Phi^+=0$; instead, it would have a noncompact source, ${\cal O}^\dagger\Phi^+_{\rm alt}=\eta^R$, that extends to future null infinity. Unfortunately, Paper I does not distinguish between $\Phi^+$ and $\Phi^+_{\rm alt}$. The mathematics in Paper~I consistently describes $\Phi^+$, but text in Paper~I, particularly text above Eq.~(39) therein, incorrectly conflates $\Phi^+$ with $\Phi^+_{\rm alt}$. Either solution can be used, but $\Phi^+_{\rm alt}$ is acausal, and $\Phi^+$ is the field that is consistent with the traditional no-string solution.} Note that, in these regions, the residual Weyl scalar coincides with the retarded solution, $\psi_0^R = \psi_0^{ret}$, and the corresponding Hertz potential will therefore be labelled by $\Phi^{ret}$. This notation serves to indicate that $\Phi^\pm$ are the Hertz potentials that are constructed in the traditional no-string method of obtaining the retarded metric perturbation sourced by a point particle; we do not mean to suggest that $\Phi^{ret}$ is the retarded solution to a field equation. 

Inside the worldtube, we solve Eq.~\eqref{eqn:PhiREqn} numerically for $\Phi^R$, subject to boundary/jump conditions at $r=\rmin$, which are determined from  Eq.~\eqref{eqn:PhiREqn}. 
We label this internal solution $\Phi^M$ (with `M' indicating the `middle' region or effective `matter' region).
Equation~\eqref{eqn:PhiREqn} also provides jump conditions at $r=\rmax$; see Sec.~\ref{sec:Hertz} for more details. Requiring the Hertz potential to satisfy these jump conditions 
across $r=\rmax$ requires one to add an additional field, $\Phi^S$, to $\Phi^+$ at $r>\rmax$. This additional field, referred to as the \textit{shadow} field, satisfies the homogeneous version of Eq.~\eqref{eqn:PhiREqn} and the inhomogeneous version of Eq.~\eqref{eqn:OdagPhi=eta}. The total residual field can then be written in the form
\begin{equation}\label{eqn:PhiR summary}
\Phi^R =   \Phi^- \Theta^- +  \Phi^M \Theta^M + (\Phi^+ + \Phi^S) \Theta^+.
\end{equation}
In the above, we defined $\Theta^- := \Theta(\rmin-r)$, $\Theta^+ := \Theta(r-\rmax)$, and $\Theta^M := 1 - \Theta^- - \Theta^+$. In particular, note that $\Theta^M(r)=1$ inside the worldtube $\rmin < r < \rmax$. Our use of Heaviside functions here does not indicate discontinuity at the worldtube boundaries: since $\psi^R_0$ contains Dirac $\delta'$ terms there and $\Phi^R$ is four integrals of $\psi^R_0$, the solution~\eqref{eqn:PhiR summary} is $C^1$ at $\rmin$ and $\rmax$.

Although we included it in Eq.~\eqref{eqn:PhiR summary}, $\Phi^S$ is never required (or calculated) in our GHZ puncture scheme. We recall the reason why below. 

\subsection{Residual metric perturbation: Hertz term}

Next, from $\Phi^R$, one can compute a $(0,2)$ tensor via the explicit relation
\begin{equation}
\label{eqn:habRecons}
\hat{h}^R_{ab} = 2\, {\rm Re} \bracket{(S^\dagger_0)_{ab} \Phi^R},
\end{equation}
which satisfies the IRG conditions $\hat{h}_{ab}^R l^a = 0 = g^{ab} \hat{h}^R_{ab}$, where $l^a$ is the principal outgoing null vector, given in Eq.~\eqref{eq:tetrad,Kinn,t} in the Kinnersley tetrad.

Splitting the right-hand side of Eq.~\eqref{eqn:habRecons} into the three domains gives
\begin{equation}
\label{eqn:hhatab2ndForm}
\hat{h}_{ab}^R =  \hat{h}_{ab}^- \Theta^- + \hat h_{ab}^M \Theta^M + (\hat{h}_{ab}^+ + \hat{h}_{ab}^S) \Theta^+,
\end{equation}
where each field is constructed from the corresponding term in Eq.~\eqref{eqn:PhiR summary}. Note that the second-order differential operator $(S_0^\dagger)_{ab}$ did not introduce Dirac delta distributions in $\hat{h}^R_{ab}$ because 
$\Phi^R$ is $C^1$ at the worldtube's boundaries. However, the two derivatives do introduce jump discontinuities there; 
see Table~\ref{table:regularity} for more details.



Like $\Phi^S$, the shadow field $\hat h^S_{ab}$ is never explicitly needed or calculated. The fields $\hat{h}^\pm_{ab}$ satisfy the vacuum linearised Einstein equation ${\cal E}_{ab}(\hat h^\pm)=0$. If we were to reduce the worldtube to zero size, effectively removing the term involving $\Theta^M$, and additionally discard the shadow field $\hat h^S_{ab}$, then Eq.~\eqref{eqn:hhatab2ndForm} would reduce to the traditional form of the no-string CCK metric reconstruction. 

\subsection{Residual metric perturbation: corrector tensor}\label{sec:summary corrector}

The reconstructed field $\hat{h}^R_{ab}$ is not, on its own, a solution to the linearised Einstein equation~\eqref{eqn:hREqnGlobal}. Intuitively, this failure is associated with the fact that if $\hat{h}^R_{ab}$ is in the IRG, then $\mathcal{E}_{ll}(\hat{h}^R)=0$~\cite{Price2006}. It follows that, since $T^R_{ll} \neq 0$, $\hat{h}^R_{ab}$ cannot satisfy~\eqref{eqn:hREqnGlobal}. 
The novel step in the GHZ procedure is to supplement CCK-Ori reconstruction with a \textit{corrector tensor}, $x^R_{ab}$, defined to satisfy the components of the Einstein equations that $\hat{h}_{ab}^R$ cannot,
\begin{equation}
\label{eqn:xabEqn}
\bracket{T_{ab}^R - \mathcal{E}_{ab}(x^R)}l^a = 0.
\end{equation}
As shown by GHZ, the total field $\hat h^R_{ab}+x^R_{ab}$ then satisfies the full set of equations~\eqref{eqn:hREqnGlobal}.

Equation~\eqref{eqn:xabEqn} is a system of four real independent equations. The GHZ scheme adopts the following ansatz for the corrector tensor:
\begin{multline}
\label{eqn:xab}
x^R_{ab} = 2 m_{(a} \bar{m}_{b)} x^R_{m\bar{m}} - 2 l_{(a} \bar{m}_{b)} x^R_{nm} \\
- 2 l_{(a} m_{b)} x^R_{n\bar{m}} + l_a l_b x^R_{nn},
\end{multline}
where $x^R_{n\bar{m}} = (x^R_{nm})^\star$ and we have adopted a Newman-Penrose null tetrad $\{l^\alpha,n^\alpha,m^\alpha,\bar m^\alpha\}$; see, e.g., Eq.~\eqref{eq:tetrad,Kinn,t} for the Kinnersley tetrad. The key feature of this ansatz is its inclusion of a trace component, $x^R_{m\bar m}$; the trace term, which is omitted in CCK-Ori reconstruction, is what allows the corrector tensor to satisfy~\eqref{eqn:xabEqn}. 
By projecting~\eqref{eqn:xabEqn} into the $l^a$, $n^a$ and $m^a$ directions, one obtains a sequence of three ODEs along the integral curves of $l^a$,
\begin{align}
\label{eqn:CorrectorTensorxmmbEqnGHP}
-\rho^2 \pth \bracket{\frac{1}{\rho^2} \pth x^R_{m\bar{m}}} &= T_{ll}^R,\\
\label{eqn:CorrectorTensorxnmEqnGHP}
-\frac{1}{2} \pth \bracket{\rho^2 \pth \bracket{\frac{x^R_{nm}}{\rho^2}}} &= T_{lm}^R + \mathcal{N} x^R_{m\bar{m}},\\
\label{eqn:CorrectorTensorxnnEqnGHP}
-\rho^2 \pth \bracket{\frac{x^R_{nn}}{\rho}} &= T_{ln}^R + \mathcal{U}x^R_{m\bar{m}} + \mathfrak{V}x^R_{nm}+ \bar{\mathfrak{V}}x^R_{n\bar{m}},
\end{align}
where the spin coefficient $\rho$ is given in Eq.~\eqref{eqn:rho} and the operators $\mathcal{N}$, $\mathcal{U}$, $\mathfrak{V}$ and $\bar{\mathfrak{V}}$ are given in Eqs.~\eqref{eqn:N}-\eqref{eqn:Vb}.
This hierarchical system of transport equations can be solved for $x^R_{m\bar{m}}$, $x^R_{nm}$ and $x^R_{nn}$, in that order. 

The transport equations are supplemented with trivial data at the past horizon, implying that $x^R_{ab}$ vanishes for all $r<\rmin$. We then solve the equations within the worldtube 
$\rmin < r < \rmax$, subject to boundary/jump conditions at $r=\rmin$; see Sec.~\ref{sec:Corrector_tensors}. For $r>\rmax$, the transport equations dictate that another shadow field appears, such that the global solution takes the form
\begin{equation}\label{eqn:x summary}
x^R_{ab} = x^M_{ab}\Theta^M + x^S_{ab}\Theta^+.
\end{equation}
Again observe that no delta functions arise at the worldtube boundaries despite the presence of $\delta'$ terms in $T^R_{ab}$; this is due to a fortuitous cancellation on the right-hand side of Eq.~\eqref{eqn:CorrectorTensorxnnEqnGHP}, which we describe below. Also again note that we will not explicitly need or calculate the shadow field.


\subsection{Total residual metric perturbation}\label{sec:total hR}

The total residual metric perturbation is the sum of the reconstructed piece~\eqref{eqn:hhatab2ndForm} and the corrector piece~\eqref{eqn:x summary}:
\begin{equation}
\label{eqn:hab2ndForm}
h_{ab}^{R'} = \hat{h}_{ab}^- \Theta^- + (\hat h_{ab}^M+x^M_{ab}) \Theta^M + (\hat{h}_{ab}^+ + \hat h_{ab}^S + x_{ab}^S) \Theta^+.
\end{equation}
Here we have added a prime on the residual field to indicate that this is not yet in our ultimate choice of gauge. 

To put the residual field in its final form, we will perform a gauge transformation of the shadow field $\hat h_{ab}^S+x_{ab}^S$. There are three reasons for this. First, it is simply unnecessary to calculate $\hat h_{ab}^S+x_{ab}^S$. Second, the source $T^R_{ab}$ has finite differentiability at the particle, and the transport equations~\eqref{eqn:PhiREqn} and \eqref{eqn:CorrectorTensorxmmbEqnGHP}--\eqref{eqn:CorrectorTensorxnnEqnGHP} cause the shadow field to inherit this nonsmoothness at all points along ougoing null geodesics emanating from the particle to future null infinity; this string singularity is increasingly softened as the order of the puncture is increased, but it is still undesirable. Third, the $\ell=0,1$ modes in the shadow field are not in an asymptotically flat gauge, which means, for example, that the orbital frequencies measured at the particle are not those measured by an inertial observer at infinity, and ``invariant'' quantities such as the Detweiler redshift consequently take incorrect values; for discussions of this point, see Refs.~\cite{Shah2015,Barack2018}, for example. 

In fact, since $r>\rmax$ is a vacuum region, the CCK term $\hat h^+_{ab}$ contains almost all of the invariant content there. This follows from the fact that ${\cal T}_0^{ab}\hat h^+_{ab}=\psi_0^+$, while the shadow field contributes nothing to the Weyl scalar: ${\cal T}_0^{ab}(\hat h^S_{ab}+x^S_{ab})=0$. Wald's theorem~\cite{Wald1973} therefore implies that the shadow field can only be composed of a perturbation $\dot{g}_{ab}$ toward another Kerr solution plus a gauge perturbation $\mathcal{L}_\xi g_{ab}$, meaning
\begin{equation}
\hat h_{ab}^S+x_{ab}^S = \dot{g}_{ab} + \mathcal{L}_\xi g_{ab} 
\end{equation}
for some $\dot{g}_{ab}$ and $\xi^a$.

We split $\xi^a$ into an $\ell=0,1$ piece $\Xi^a$ that is strictly linear in time and a piece $\xi_S^a$ that has at most periodic time dependence. We then subtract a gauge perturbation generated by
\begin{equation}\label{eq:discontinuous xi}
    \xi^a = \Xi^a + \xi^a_S\Theta^+.
\end{equation}
In this way, we subtract $\mathcal{L}_\Xi g_{ab}$ from $h_{ab}$ over the entire range of $r$, not only in the region $r>\rmax$. This is necessary because $\Xi^a$ alone alters frequency values, and it must be subtracted globally to ensure that the frequencies throughout spacetime are those measured by an inertial observer at infinity. The final form of the metric perturbation in the resulting ``shadowless gauge'' is hence 
\begin{align}
\label{eqn:habShadowlessGauge}
h_{ab}^R &=  (\hat h_{ab}^- - \mathcal{L}_\Xi g_{ab}) \Theta^- + \bracket{\hat h_{ab}^M+x^M_{ab} - \mathcal{L}_\Xi g_{ab}} \Theta^M\nonumber \\
&\hphantom{:}\quad + \bracket{\hat h_{ab}^+ +  \dot{g}_{ab}} \Theta^+ - 2\xi^S_{(a}r_{b)}\delta(r-\rmax),
\end{align}
where $r_b:=\partial_b r$. The vector field $\Xi^a$ is calculated in Eq.~\eqref{Xi} below. The vector field $\xi^a_S$, which now appears in the Dirac $\delta$ term in Eq.~\eqref{eqn:habShadowlessGauge}, is given in Paper I in terms of $T^R_{ab}$. In this paper we will not explicitly calculate the $\delta$ term in the residual field, but we return to it in the Conclusion.

Note that if one were to shrink the worldtube to zero size, $\rmin \to \rmax \to r_0$, the terms involving $\Theta^M$  would disappear. The field $h_{ab}^R$ would then be precisely the same as the retarded field in the standard no-string construction. See Fig.~\ref{fig:CCK_scheme} for a pictorial summary of the CCK reconstruction and completion, and Fig.~\ref{fig:CCK_Fields} for a plot showing the different quantities involved. 
Similarly, see Figs.~\ref{fig:GHZ_puncture_scheme} and~\ref{fig:GHZ_Fields} for a summary of the corresponding GHZ puncture procedure.

In Table~\ref{table:regularity}, we give a summary of the theoretical expectation of the (ir)regularity of most of the involved quantities, decomposed into Fourier and harmonic modes (raised and lower mode indices have the same meaning); see for example Eqs.~\eqref{eqn:habSphericalDecomposition}, \eqref{eq:psi0_decomp}, \eqref{eqn:Phidecomp}, and \eqref{eqn:xabdecomp}.
The coordinates $(t,r,\theta,\phi)$ are the usual Schwarzschild coordinates, and $u$ is the outgoing null coordinate, see Eqs.~\eqref{eqn:Schwarzschildt} and \eqref{eqn:Schwarzschildu}.
Note that the first two equations show a decomposition into Fourier-$t$ modes, instead of $u$-modes, but this does not affect regularity of the modes.

For the most part, the degree of regularity at the particle can be straightforwardly predicted by, for example, counting the number of derivatives and integrals at each stage of the calculation. The puncture modes $h_{ab}^{P,\ell m}$, given in Lorenz gauge, have a well-known kink at the particle, due to the presence of terms involving $|r-r_0|$. The second-order differential operator $\mathcal{T}^{ab}_0$ therefore introduces a jump and delta function into the modes of $\psi_0^P$ (denoted by ${}_2 \psi^P_{\ell m}$, where the left-subscript ``2'' refers to its spin weight). The regularity of the residual field modes ${}_2 \psi^R_{\ell m} = {}_2 \psi^{ret}_{ \ell m} - {}_2 \psi^P_{\ell m}$ depends on the order of the puncture: the $n=-1$ term in $h^P_{ab}$  removes the $\delta$ function in ${}_2 \psi^{ret}_{\ell m}$, the $n=0$ term removes the jump discontinuity, and so on. For a puncture of order $n \geq 0$, ${}_2 \psi^R_{\ell m}$ is therefore $C^{n}$. The Hertz potential modes ${}_{-2} \Phi^R_{\ell m}$ are obtained by integrating (radially) ${}_2 \psi^R_{\ell m}$ four times, implying it is $C^{n+4}$ at the particle.  The reconstructed field $\hat h_{ab}^{R, \ell m}$, which involves two derivatives of  ${}_{-2} \Phi^R_{\ell m}$, is therefore $C^{n+2}$. 

The regularity of the corrector tensor is slightly less straightforward. The effective stress-energy modes $T_{ab}^{R, \ell m}$ are $C^{n}$ at the particle, analogous to ${}_2\psi^R_{\ell m}$. The second-order radial ODEs~\eqref{eqn:CorrectorTensorxmmbEqnGHP} and \eqref{eqn:CorrectorTensorxnmEqnGHP} then imply that $x^{R,\ell m}_{m\bar{m}}$ and $x^{R,\ell m}_{nm}$ (and $x^{R,\ell m}_{n\bar{m}}$) are $C^{n+2}$. But Eq.~\eqref{eqn:CorrectorTensorxnnEqnGHP} dictates that $x^{R,\ell m}_{nn}$ is obtained via a single integration of the stress-energy tensor. As a result, one might naively think that the modes of $x^R_{nn}$ would only be $C^{n+1}$ instead of $C^{n+2}$. However, it turns out that they are one degree more regular than naively expected due to cancellations on the right-hand side of Eq.~\eqref{eqn:CorrectorTensorxnnEqnGHP}. These cancellations, described in Sec.~\ref{sec:Corrector_tensors}, are also the reason no Dirac $\delta$ terms appear at the worldtube boundaries in Eq.~\eqref{eqn:x summary}.  As a consequence, the modes $x_{ab}^{R,\ell m}$ are all at least $C^{n+2}$ at the particle.


\section{Calculation of the Lorenz-gauge puncture}
\label{sec:Lorenz puncture}

The first step in all puncture schemes is the calculation of the puncture field. Although mode decompositions of singular fields are a standard ingredient in most self-force calculations, they typically make use of approximations which are incompatible with our GHZ-Teukolsky puncture scheme. A similar issue also affects punctures used in dealing with the problem of infinite mode coupling when constructing the source for second-order self-force calculations \cite{Miller:2016hjv}. As such, and since the same puncture was an essential ingredient in producing existing results from second-order self-force \cite{Pound2019,Warburton2021,Wardell:2021fyy}, we defer a full description of the construction of the puncture to a forthcoming paper \cite{PuncturePaper} and give here only the essential details. 

Once the puncture is in hand, the bulk of our calculations in later sections reduce to solving radial differential equations. We do so numerically, making use of the Black Hole Perturbation Toolkit (BHPToolkit) \cite{BHPToolkit} and Mathematica's built-in function \texttt{NDSolve}.

\subsection{Circular orbits in Schwarzschild spacetime}

While our overview in the preceding section remained general, we will from now on focus our attention on a particle in a circular orbit around a Schwarzschild black hole. The background metric $g_{ab}$, in Schwarzschild coordinates $(t,r,\theta,\varphi)$, is
\begin{equation}
\label{eqn:Schwarzschildt}
    ds^2 = -f(r) dt^2 + \frac{dr^2}{f(r)} + r^2 (d\theta^2+\sin^2\theta d\varphi^2),
\end{equation}
where $f(r) := 1-2M/r$.
At leading order in the mass ratio, the particle is a test particle that moves along (timelike) geodesics of the background metric $g_{ab}$. Given the background symmetry, we can assume without loss of generality that the motion is in the equatorial plane $\theta=\pi/2$. Then, for a circular orbit with trajectory $x^a = (t, r_0, \pi/2, \varphi_p(t))$ the four-velocity is given by
\begin{equation}
    u^a = u^t (1,0,0,\Omega_\varphi),
\end{equation}
where $u^t := (1-3M/r_0)^{-1/2}$ and $\Omega_\varphi :=d\varphi_p/dt= \sqrt{M/r_0^3}$. The point-particle stress-energy tensor~\eqref{eqn:covariant Tab} reduces to
\begin{equation}
\label{eqn:Tab}
    T_{ab} = 8 \pi M \frac{u_a u_b}{r_0^2 u^t} \delta(r-r_0) \delta\bracket{\theta -\pi/2} \delta(\varphi - \Omega_\varphi t).
\end{equation}

Given the importance of outgoing null rays in the GHZ procedure, it is useful to adopt outgoing Eddington-Finkelstein coordinates, $(u,r,\theta,\varphi)$, defined by $u := t-r_\star$, where $r_\star$ 
is the usual tortoise coordinate $r_\star = r + 2M \ln \left( r/2M - 1 \right)$. We will refer to Schwarzschild coordinates and Eddington-Finkelstein coordinates as \textit{$t$-slicing} and \textit{$u$-slicing} coordinates, respectively. In $u$ 
slicing, the metric takes the form
\begin{equation}
\label{eqn:Schwarzschildu}
    ds^2 = -f(r) du^2 - 2 dr du + r^2 (d\theta^2+\sin^2\theta d\varphi^2).
\end{equation}

\subsection{Puncture with exact mode decomposition}

The construction of the puncture begins from a 4D, covariant expression for a Lorenz-gauge puncture derived from an approximation to the Detweiler-Whiting singular field \cite{Detweiler:2002mi}. The approximation is obtained as a covariant expansion in 4D distance from the worldline and is given explicitly in Eq.~(4.7) of Ref.~\cite{Heffernan2012}. It includes the leading-order term given in Eq.~\eqref{eq:hs}, but we also keep a further three orders in the expansion so that the puncture agrees with the Detweiler-Whiting singular field through order $(\text{distance})^2$ from the worldline; in our nomenclature this corresponds to an order $n=2$ puncture.

Following the methods in Ref.~\cite{Heffernan2012}, this covariant expression is then converted to a coordinate expression in terms of a rotated coordinate system such that the particle is instantaneously at the north pole. The conversion process involves re-expanding covariant distances in terms of coordinate distances, again preserving terms in the puncture through order $(\text{distance})^2$. We then make the standard choice $\Delta t = 0$, such that the field point and the reference point on the worldline are at the same Schwarzschild time; this has the effect of making all time dependence appear implicitly via the dependence of the puncture on the coordinate location of the worldline. This yields our final, exact\footnote{Note that the puncture is exact both in terms of the rotated, time-dependent coordinate system and in terms of the original, unrotated coordinate system in which the worldline is on an equatorial orbit; the two are related by an exact, time-dependent 3D rotation.} definition for the puncture $h^{\cal P}_{ab}\left(t, r, \theta, \varphi\right)$.

We next follow Ref.~\cite{Wardell:2015ada} and decompose the puncture onto the Barack-Lousto-Sago (BLS) basis of tensor spherical harmonics \cite{Barack:2005nr, Barack:2007tm}, which provide an orthogonal basis for symmetric rank-2 tensors in Schwarzschild spacetime. 
We further decompose the time dependence into Fourier modes, with $\omega$ representing the frequency of the mode (in our case of circular orbits in Schwarzschild spacetime $\omega = m \Omega_\varphi$ and the Fourier transform is discrete). The result is a decomposition
\begin{equation}
 h^{\cal P}_{ab}(t, r, \theta, \varphi) = \sum_{i=1}^{10} \sum_{\ell, m} \frac{a_{i\ell}}{r} h^{\cal P}_{i\ell m}(r) Y^{i\ell m}_{ab} (\theta, \varphi) e^{-i m \varphi_p(t)}
\end{equation}
where $a_{i\ell}$ is an $\ell$-dependent constant, where the infinite sum over $\ell$ starts at $\ell=0$ for $i=1,2,3,6$, at $\ell=1$ for $i=4,5,8,9$, and at $\ell=2$ for $i=7,10$, and where the sum over $m$ is over all integers $-\ell \le m \le \ell$.

To obtain the puncture mode coefficients $h^{\cal P}_{i\ell m}(r)$, we must integrate the 4D puncture field
against a spherical harmonic over spheres of constant $(t,r)$. This is most efficiently done by considering the modes $h^{\cal P}_{i\ell m'}(r)$ with respect to the rotated coordinate system and then applying an exact rotation at the level of modes \cite{Wardell:2015ada}:
\begin{equation}
 h^{\cal P}_{i\ell m}(r) = \sum_{m'} D^{\ell}_{m m'} h^{\cal P}_{i\ell m'}(r)
\end{equation}
where $D^{\ell}_{m m'}$ is the Wigner-D matrix for the time-dependent rotation.

Traditionally, the angular integrals over the sphere are made analytically tractable by making approximations such that the mode-decomposed puncture captures the behaviour of the exact puncture only near the worldline. In particular,
the integrals can be approximated by a power series in $\Delta r := r-r_0$  \cite{Heffernan2012}. Similarly, the use of a rotated coordinate system means that the puncture can be approximated by a small number of $m'$ modes \cite{Wardell:2015ada}. By considering a sufficient number of powers of $\Delta r$ and a sufficient number of $m'$ modes, one can ensure that, when summed over modes and evaluated on the worldline, the mode decomposed puncture and its first few derivatives agree with those of the exact puncture.  The number of derivatives is in one-to-one correspondence with the number of powers of $\Delta r$ and the number of $m'$ modes. Unfortunately, neither of these approximations are valid when using a puncture to address the problem of infinite mode coupling in second-order self-force calculations \cite{Miller:2016hjv}, and they also prove problematic in our context of a GHZ-Teukolsky puncture scheme. One reason for this is that the expansion in powers of $\Delta r$ necessarily introduces a divergence at large $\ell$ at all points away from $\Delta r=0$, implying that the $\ell$-mode sum of the residual field diverges everywhere except at the particle; see the discussion around Eq.~(127) in Ref.~\cite{Toomani2021}. Instead, we work with an exact mode decomposition of the puncture, which we next describe, so that we do not need to be concerned with such issues. 


Fortunately, the approach used in approximate mode decompositions can be adapted to instead produce exact decompositions. A version of this exact mode decomposition approach is detailed in the context of scalar fields in Sec.~IV B of Ref.~\cite{Miller:2016hjv}. Defining the rotated angular coordinates $(\theta',\varphi')$ in which the particle is instantaneously at the north pole ($\theta'=0$), the approach essentially relies on Eq.~(44) of Ref.~\cite{Miller:2016hjv}, which gives a closed form for the integral over $\theta'$ in the case where the integral is against a scalar spherical harmonic with $m'=0$. Integrals for $m' \ne 0$ can be reduced to the $m'=0$ form by integrating by parts $m'$ times. Similarly, integrals against vector and tensor harmonics can be reduced to the same form using identities relating associated Legendre polynomials and their derivatives. This just leaves the integral over $\varphi'$. This can also, in principle, be written in closed form in terms of complete elliptic integrals. In practice, however, the complexity of the expressions meant that we found it more straightforward to evaluate them as 1D numerical integrals.

In summary, our puncture is constructed using a natural extension of the hybrid analytical-numerical mode decomposition scheme developed in Ref.~\cite{Miller:2016hjv}. We also adopt the following practical choices in our implementation:
\begin{enumerate}
    \item We use a second order puncture, accurate through order $(\text{distance})^2$ from the worldline;
    \item We use the angular window function $\mathcal{W}^4_{10}$ from Ref.~\cite{Miller:2016hjv} to smooth undesirable behaviour at the south pole while preserving four orders in the behaviour in the radial direction and modes up to $m'=10$;
    \item We truncate the sum over rotated modes at $m'=10$, which we found sufficient to recover the full unrotated puncture up to approximately machine precision; 
    \item We use the \texttt{NIntegrate} function in \textit{Mathematica} with up to 32 digits of working precision, so that the integral is determined to approximately machine precision.
\end{enumerate}

\section{Calculation of the residual Weyl scalar}
\label{sec:Weyl}

The next step in our scheme is to calculate the residual Weyl scalar $\psi^R_0$.
It satisfies Eq.~\eqref{eqn:Psi0REqn}, which we reproduce here for convenience, along with the corresponding equation for the spin-weight $s=-2$ Weyl scalar $\psi_4^R$:
\begin{align}
\label{eqn:Teukolsky}
\mathcal{O} \psi_0^R = \mathcal{S}_0^{ab} T_{ab}^R, \\
\label{eqn:TeukolskyPsi4}
\mathcal{O}' \psi_4^R = \mathcal{S}_4^{ab} T_{ab}^R.
\end{align}
The second-order differential operators $\mathcal{O}, \mathcal{O}'$, 
$\mathcal{S}_0^{ab}$ 
and $\mathcal{S}_4^{ab}$ 
are explicitly given in Appendix \ref{section:GHP_formalism}. 
In principle, one could calculate the residual IRG Hertz potential from either $\psi^R_0$ or $\psi^R_4$. Following Paper I, we choose to work with $\psi_0$ exclusively, 
but we include the equations satisfied by $\psi_4$ because we will encounter them while reconstructing the Hertz potential. 


The Teukolsky equation is fully separable in a basis of Fourier modes and spin-weighted spheroidal harmonics, even in Kerr spacetime. In Schwarzschild spacetime the spheroidal harmonics reduce to spherical ones. The specific form of the mode ansatz then depends on the choice of tetrad. Choosing the Kinnersley tetrad, decomposing into Fourier-$t$ modes, and specialising to circular orbits in Schwarzschild spacetime, we write the ansatz as
\begingroup
\allowdisplaybreaks
\begin{align}
\label{eq:psi0_decomp}
    \psi_0 =&\; \sum_{\ell=2}^\infty \sum_{m=-\ell}^\ell  \tensor[_2]{\psi}{_\ell_m}(r)  \,_{2}Y_{\ell m}(\theta,\phi) e^{-i m \Omega_\varphi t},\\
    r^4 \psi_4 =&\; \sum_{\ell=2}^\infty \sum_{m=-\ell}^\ell \tensor[_{-2}]{\psi}{_\ell_m}(r) \,_{-2}Y_{\ell m}(\theta,\phi) e^{-i m \Omega_\varphi t},\\
    \mathfrak{T}_0 :=&\; \mathcal{S}_0^{ab} T_{ab} \nonumber\\*
    =&\; -\frac{1}{2 r^2} \sum_{\ell=2}^\infty \sum_{m=-\ell}^\ell \tensor[_2]{\mathfrak{T}}{_\ell_m}(r) \,_{2}Y_{\ell m}(\theta,\phi) e^{-i m \Omega_\varphi t}, \\
     \mathfrak{T}_4 :=&\; r^4 \mathcal{S}_4^{ab} T_{ab}\nonumber\\*
    =&\; -\frac{1}{2 r^2} \sum_{\ell=2}^\infty \sum_{m=-\ell}^\ell \tensor[_{-2}]{\mathfrak{T}}{_\ell_m}(r) \,_{-2}Y_{\ell m}(\theta,\phi) e^{-i m \Omega_\varphi t},
\end{align}
\endgroup
where it is understood that the expansions apply for the retarded, residual, and puncture fields and for the physical and effective sources. 
In the above, the left subscripts are used to keep track of the quantities' spin weights $s=\pm 2$.

We remark that, while the choice of $t$ slicing has traditionally been used when computing homogeneous solutions of the Teukolsky equation, 
our final result for the modes of retarded, puncture and residual fields are all written in $u$-slicing. At the level of modes, the transition from $t$ to $u$ slicing  corresponds to a simple multiplicative factor: the radial coefficients in expansions of the form $\sum_{\ell m} R^{[u]}_{\ell m}(r)\,{}_s Y_{\ell m}e^{-im\Omega_\varphi u}$ and $\sum_{\ell m} R^{[t]}_{\ell m}(r)\,{}_s Y_{\ell m}e^{-im\Omega_\varphi t}$ are related by
\begin{equation}
\label{eqn:tTouSlicing}
R_{\ell m}^{[u]}(r) = R_{\ell m}^{[t]}(r) e^{-im \Omega_\varphi r_\star}.
\end{equation}
This applies to the mode decompositions of all fields.

The 
radial coefficients $\tensor[_s]{\psi}{_\ell_m}(r)$ satisfy 
the radial Teukolsky equation:
\begin{align}
\label{eqn:RadialTeukolsky}
&\,_{s}\Box_{\ell m} \tensor[_s]{\psi}{_\ell_m} := \Big[ \Delta^{-s} \frac{d}{dr} \bracket{\Delta^{s+1} \frac{d}{dr}} \nonumber \\ 
      &+ \frac{K^2-2 i s (r-M)K}{\Delta} + 4ism\Omega_\varphi r- \tensor[_s]{\lambda}{_\ell_m} \Big] \tensor[_s]{\psi}{_\ell_m} = \tensor[_s]{\mathfrak{T}}{_\ell_m},
\end{align}
where $\Delta := r^2 f(r)$ and $K := m r^2 \Omega_\varphi$. 
In Schwarzschild spacetime, the eigenvalues $ \tensor[_s]{\lambda}{_\ell_m}$ are given explicitly as
\begin{equation}
 \tensor[_s]{\lambda}{_\ell_m} = \ell (\ell+1) - s (s+1).
\end{equation}


In the following subsections, we calculate $\psi^R_0$ at the level of its individual modes ${}_2\psi^R_{\ell m}$. As explained below Eq.~\eqref{eqn:Psi0REqn}, we can find these modes in two different ways: either by first calculating $\psi^{ret}_0$ and subtracting $\psi^P_{0}$ or by directly solving the field equation~\eqref{eqn:Teukolsky} for $\psi^R_0$. We refer to the first as the subtraction method and the second as the effective-source method. In either approach, the key input is the puncture field modes ${}_2\psi^P_{\ell m}$. We describe our calculation of ${}_2\psi^P_{\ell m}$ before turning to our implementation of the two methods.

\subsection{Puncture field}

It is straightforward to compute $\psi^P_0$ from $h_{ab}^P$ using the definition of the perturbed Weyl scalar in terms of the metric perturbation, $\psi^P_0 := \mathcal{T}^{ab}_0 h_{ab}^P$, given in Eq.~\eqref{eqn:T0abhab}.
As described in Sec.~\ref{sec:Lorenz puncture}, the puncture is computed using the BLS tensor spherical harmonics~\cite{Barack:2005nr, Barack:2007tm}. Here we express the modes of $\psi^P_0$, defined as in Eq.~\eqref{eq:psi0_decomp}, directly in terms of the BLS modes $h^{P}_{i\ell m}$.

We start by writing the modes of $\mathcal{T}^{ab}_0 h_{ab}^P$ in terms of the modes $h^{P,\ell m}_{ab}$ in a spin-weighted harmonic decomposition,
\begin{align}
\label{eqn:habSphericalDecomposition}
h_{ab}^P &= \sum_{\ell=|s|}^\infty \sum_{m=-\ell}^\ell h_{ab}^{P, \ell m}(r)  \,_{s}Y_{\ell m}(\theta,\varphi)e^{-im\Omega_\varphi t},
\end{align}
where $s$ denotes the spin weight of the given tetrad component (e.g., $s=0$ for $h^{P,\ell m}_{ll}$, $s=-1$ for $h^{P,\ell m}_{l\bar m}$, etc.); see Appendix~\ref{section:GHP_formalism}. We suppress the spin weight $s$ of the modes ${h}_{ab}^{P, \ell m}(t,r)$ for notational simplicity. 
We then re-express the modes ${h}_{ab}^{P, \ell m}$ in terms of the BLS modes ${h}^{P}_{i\ell m}$ using the conversions in Appendix \ref{section:TetradToBLS}. (Note the mode indices $\ell,m$ have the same meaning whether they are subscripts or superscripts.) Our result for ${}_2{\psi}_{\ell m}^P$ is
\begin{align}
\label{eqn:habToPsi0}
{}_2 {\psi}_{\ell m}^P &= - \frac{\sqrt{(\ell-1) \ell (\ell+1) (\ell+2)}}{4 r^3 f^2} \bracket{{h}^{P}_{1\ell m} + {h}^{P}_{2\ell m}}  \nonumber \\
&-\frac{ \bracket{f^2 \partial_r^2 -(\partial_{r}f) \partial_t +2 f \partial_t \partial_r +\partial_t^2} \bracket{{h}^{P}_{7\ell m} -i {h}^{P}_{10,\ell m}}}{4 r f^2 \sqrt{(\ell-1)\ell(\ell+1)(\ell+2)}} \nonumber\\
&+ \frac{\sqrt{(\ell-1)(\ell+2)}}{4r^2f^2 \sqrt{\ell (\ell+1)}} \bracket{-\partial_{r}f +f \partial_r+\partial_t} \nonumber\\
&\times\bracket{{h}^{P}_{4\ell m} + {h}^{P}_{5\ell m} -i ({h}^{P}_{8\ell m} + {h}^{P}_{9\ell m})},
\end{align}
where it is understood that $\partial_t = -i m \Omega_{\varphi}$. The same relationship holds between ${}_2 {\psi}_{\ell m}^R$ and $h^{R}_{i\ell m}$ and between ${}_2 {\psi}_{\ell m}^{ret}$ and $h^{ret}_{i\ell m}$.

Now recall that $h_{ab}^P = h_{ab}^\mathcal{P} \Theta^M$. The sharp window function $\Theta^M$ causes a jump in the puncture $h_{ab}^P$ at the worldtube's edges, which then translates into radial $\delta$ and $\delta'$ distributions for ${}_2 \psi^P_{\ell m}$. Consequently, the Weyl scalar  modes ${}_2 {\psi}^P_{\ell m}$ are schematically of the form
\begin{multline}
\label{eqn:PsiSchematicForm}
{}_2 {\psi}^P_{\ell m} = {}_2 {\psi}^{smooth}_{\ell m}(r) \Theta^M(r)  + \psi_{0,\ell m}^{\delta, P} \delta(r-r_0) \\
({}_2 {\psi}^{\delta,P}_{\ell m})^\pm \delta^M(r)
+ ({}_2 {\psi}^{\delta',P}_{\ell m})^\pm {\delta'}^M(r)
\end{multline}
for some smooth ${\psi}^{smooth}_{\ell m}$ and constants $ \psi_{0,\ell m}^{\delta, P}$, $({}_2 {\psi}^{\delta,P}_{\ell m})^\pm$, and $({}_2 {\psi}^{\delta',P}_{\ell m})^\pm$. In the above, we defined for convenience,
\begin{align}
\delta^M(r) &:= {\Theta'}^M = \delta(r-\rmin) - \delta(r-\rmax),\\
{\delta'}^M(r) &:= {\Theta''}^M = \delta'(r-\rmin) - \delta'(r-\rmax).
\end{align}
The $\pm$ superscripts indicate that these quantities are to be evaluated at $\rmin$ or $\rmax$, depending on which $\delta$ function they are associated with.
This form is obtained by making use of the distributional identities
\begin{align}
F(x) \delta(x-x_0) &= F(x_0) \delta(x-x_0), \\
F(x) \delta'(x-x_0) &= F(x_0) \delta'(x-x_0)\nonumber\\
&\quad -F'(x_0)\delta(x-x_0) .
\end{align}
Doing so, one finds the coefficients ${}_2 {\psi}^{\delta,P}_{\ell m}$ and ${}_2 {\psi}^{\delta',P}_{\ell m}$ are explicitly given by
\begin{align}
{}_2 {\psi}^{\delta,P}_{\ell m} &= -\frac{\bracket{f+f r \partial_r + 2r \partial_t} \bracket{{h}^{\mathcal{P}}_{7\ell m}-i {h}^{\mathcal{P}}_{10,\ell m}}}{4 f r^2 \sqrt{(\ell-1)\ell(\ell+1)(\ell+2)}}\\
&\quad+\frac{(\ell-1)(\ell+2) \bracket{{h}^{\mathcal{P}}_{4\ell m}+{h}^{\mathcal{P}}_{5\ell m}-i({h}^{\mathcal{P}}_{8\ell m}+{h}^{\mathcal{P}}_{9\ell m})}}{4 f r^2 \sqrt{(\ell-1)\ell(\ell+1)(\ell+2)}}, \nonumber \\
{}_2 {\psi}^{\delta',P}_{\ell m}  &= -\frac{{h}^{\mathcal{P}}_{7\ell m} -i {h}^{\mathcal{P}}_{10,\ell m}}{4r\sqrt{(\ell-1)\ell(\ell+1)(\ell+2)}},
\end{align}
where $\partial_t$ is understood as $-im\Omega_\varphi$. 
Finally, we use Eq.~\eqref{eqn:tTouSlicing} to convert to $u$-slicing.

\subsection{Subtraction method}

In the subtraction method, we calculate ${}_2\psi^R_{\ell m}$ from  ${}_2\psi^{ret}_{\ell m}$ using
\begin{equation}
    {}_2\psi^R_{\ell m} = {}_2\psi^{ret}_{\ell m}-{}_2\psi^P_{\ell m}.
\end{equation}

In the remainder of this section we write formulae for generic $s$ rather than specializing to $s=2$. This is because, while the residual and puncture fields are only computed for $s=+2$, solutions to the Teukolsky equation for $s=-2$ will also play a role when computing the ($s=-2$) Hertz potential in later sections.

The retarded field modes are obtained via the method of variation of parameters. Specifically, we make use of the Teukolsky package from the BHPToolkit to compute two linearly independent solutions to the homogeneous radial Teukolsky equation \eqref{eqn:RadialTeukolsky}, referred to as the ``in'' and ``up'' solutions. When combined with the Fourier time factor, the ``in'' solutions represent purely ingoing waves at the future horizon. 
The ``up'' solutions, combined with the Fourier factor, instead represent 
purely outgoing waves at future null infinity.
Their asymptotic behaviours are 
\begin{align}
{}_s \psi^{in}_{\ell m}(r) &\sim \Delta^{-s} e^{-im\Omega_\varphi r_\star}, \quad r \to 2M, \\ 
{}_s \psi^{up}_{\ell m}(r) &\sim r^{-(1+2s)} e^{im\Omega_\varphi r_\star}, \quad r \to \infty.
\end{align}

The inhomogeneous retarded solution is then written in terms of these homogeneous solutions, ${}_s \psi_0^{in}$ and ${}_s \psi_0^{up}$, as
\begin{equation}\label{eqn:psi-ret variation of parameters}
    {}_s \psi_{\ell m}^{ret} = {}_s C^{in}_{\ell m}(r) {}_s \psi_{\ell m}^{in}(r) + {}_s C^{up}_{\ell m}(r) {}_s \psi_{\ell m}^{up}(r),
\end{equation}
where the weighting coefficients are given by
\begin{align}
    {}_s C^{in}_{\ell m}(r) &= \int_r^\infty \frac{{}_s \psi^{up}_{\ell m}(r')}{{}_s W(r') \Delta(r^{\prime})} \tensor[_s]{\mathfrak{T}}{_\ell_m}(r') dr', \\
    {}_s C^{up}_{\ell m}(r) &= \int_{2M}^r \frac{{}_s \psi^{in}_{\ell m}(r')}{{}_s W(r') \Delta(r^{\prime})} \tensor[_s]{\mathfrak{T}}{_\ell_m}(r') dr',
\end{align}
with
\begin{equation}
    {}_s W(r) := {}_s \psi^{in}_{\ell m}(r) \frac{d {}_s \psi^{up}_{\ell m}}{dr} - {}_s \psi^{up}_{\ell m} \frac{d {}_s \psi^{in}_{\ell m}(r)}{dr}.
\label{eq:Wronsk}
\end{equation}
Since the retarded field is sourced by a point-particle stress-energy, ${}_s \mathfrak{T}_{\ell m}$ is a linear combination of $\delta(r-r_0)$ and its first and second derivatives. The integrals can therefore be evaluated explicitly in terms of the homogeneous solutions (and their derivatives) on the  worldline.
The integrals are constants for $r>r_0$ and $r<r_0$, with ${}_s C^{in}_{\ell m} = 0$ when $r>r_0$ and ${}_s C^{up}_{\ell m} = 0$ when $r<r_0$.
The modes of the retarded solution therefore split into two solutions inside and outside the particle's orbit:
\begin{equation}
\label{eqn:PsiRet}
{}_s \psi_{\ell m}^{ret}(r) = 
\begin{cases} 
      {}_s \psi^-_{\ell m} := {}_s C^{-}_{\ell m}\, {}_s \psi_{\ell m}^{in}(r), & r < r_0, \\
     {}_s \psi^+_{\ell m} := {}_s C^{+}_{\ell m}\, {}_s \psi_{\ell m}^{up}(r), & r > r_0,
\end{cases}
\end{equation}
where the constants
\begin{align}
    {}_s C^{-}_{\ell m} &:= {}_s C^{in}_{\ell m}(r<r_0)=\int_{2M}^\infty \frac{{}_s \psi^{up}_{\ell m}(r')}{{}_s W(r') \Delta(r^{\prime})} \tensor[_s]{\mathfrak{T}}{_\ell_m}(r') dr', \\
    {}_s C^{+}_{\ell m} &:= {}_s C^{up}_{\ell m}(r>r_0)=\int_{2M}^\infty \frac{{}_s \psi^{in}_{\ell m}(r')}{{}_s W(r') \Delta(r^{\prime})} \tensor[_s]{\mathfrak{T}}{_\ell_m}(r') dr'
    \label{eq:Cin/up}
\end{align}
can be explicitly evaluated in terms of the coefficients of $\delta$, $\delta'$, and $\delta''$ in $\tensor[_s]{\mathfrak{T}}{_\ell_m}$. As before, we modify these  expression by including the appropriate exponential factor to bring the mode solution from $t$- to $u$-slicing using Eq.~\eqref{eqn:tTouSlicing}.

In addition to ${}_s \psi^\pm_{\ell m}$, the retarded field modes contain a term proportional to $\delta(r-r_0)$, as shown in Eq.~\eqref{eq:psi0ret}. The coefficient of the delta function can be obtained straightforwardly by evaluating the weighting coefficients $ {}_s C^{in}_{\ell m}(r)$ and  ${}_s C^{up}_{\ell m}(r)$; the result is necessarily identical to the coefficient $\psi_0^{\delta, P}$ appearing in Eq.~\eqref{eqn:PsiSchematicForm}. Since the delta functions therefore cancel out in ${}_2\psi^R_{\ell m}$, we will not need the explicit value of $\psi_0^{\delta, P}$ here.
\begin{figure}[t!] 
    \includegraphics[width=\columnwidth]{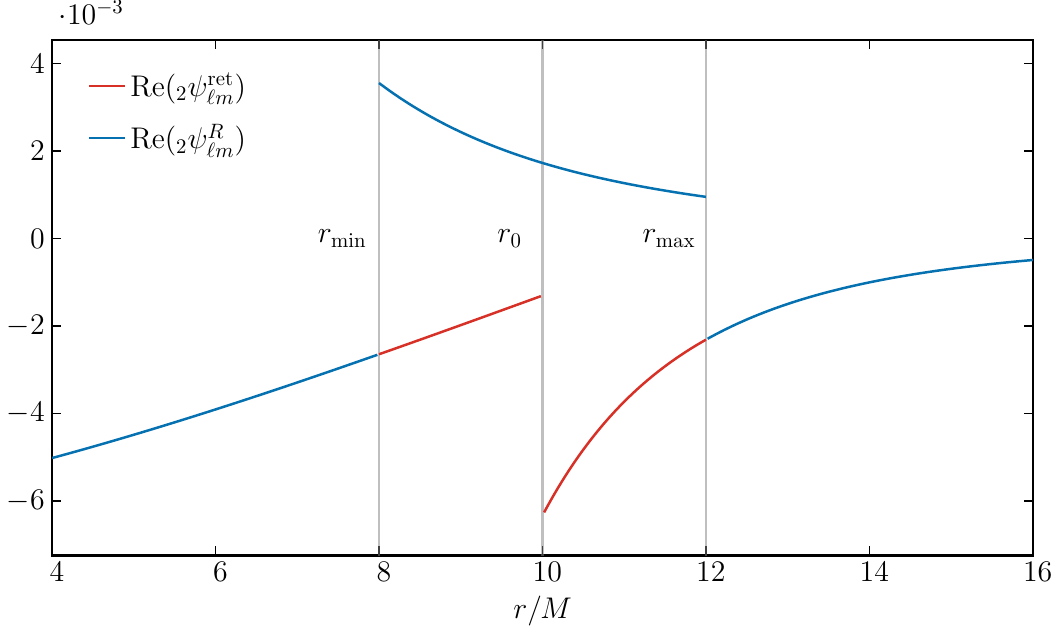}
	\caption{The $\ell=m=2$ mode of the retarded and residual Weyl scalars, $\psi_0^{ret}$ and $\psi_0^R$. The retarded field mode is discontinuous at the particle's orbital radius $r_0$.
    $_2\psi_{\ell m}^{ret}$ also contains a $\delta$ function at $r_0$, although not visible in the plot. The residual field is continuous at $r_0$ but discontinuous on the worldtube's edges, $r_{\rm min/max}$, and additionally contains a $\delta$ and $\delta'$ there, a consequence of our choice of sharp window function.}
	\label{fig:Psi0Fields}
\end{figure}


\subsection{Effective-source method}

In the effective-source approach, we directly solve Eq.~\eqref{eqn:Teukolsky} for the residual field.  
At the level of modes, 
the equation becomes
\begin{align}
    \,_{2}\Box_{\ell m} \,_{2}{\psi^{R}_{\ell m}} 
    &= \,_{2}{\mathfrak{T}_{\ell m}} - 
    \,_{2}\Box_{\ell m} {}_{2}{\psi^{P}_{\ell m}} =: {}_{2}{S^{\rm eff}_{\ell m}}.
    \label{eq:effective_source_field_eqn}
\end{align}
Therefore, solving this alternative field equation with an effective source, $\,_{2}S^{\rm eff}_{\ell m}$, will yield the residual field without the need for mode-by-mode subtraction of the puncture field from the retarded field.  
Furthermore, the puncture field, $\psi^{P}_{\ell m}$, is defined in such a manner that doing this cancellation at the level of the field equation will leave us with a source that does not involve Dirac delta distributions.

We solve  Eq.~\eqref{eq:effective_source_field_eqn} using a \emph{worldtube} approach, based on the similar scheme implemented in~\cite{Leather:2023dzj}.
Schematically, the residual radial Teukolsky function obeys Eq.~\eqref{eq:effective_source_field_eqn} inside the worldtube; outside the worldtube the residual is identical to the retarded field and thus is a solution to the homogeneous Teukolsky equation.
Quantitatively, we adopt the following ansatz for the residual field,
    \begin{equation}
    \,_{2}{\psi^{R}_{\ell m}} = 
    \begin{cases}
        {}_{2}a_{\ell m}^{in}\,_{2}{\psi^{in}_{\ell m}} & r <  \rmin, \\[.8em]
	\begin{array}{l}
            {}_{2}b_{\ell m}^{up}\,_{2}{\psi^{up}_{\ell m}} \\
            \qquad + {}_{2}b_{\ell m}^{in}\,_{2}{\psi^{in}_{\ell m}}
            + \,_{2}{\psi^{\rm inh}_{\ell m}}\end{array} &
	\rmin < r < \rmax,\\[1.2em]
	{}_{2}a_{\ell m}^{up}\,_{2}{\psi^{up}_{\ell m}} & r >  \rmax.
    \end{cases}
    \label{eq:residual_field_ansatz}
    \end{equation}
Here $\,_{2}{\psi^{\rm inh}_{\ell m}}$ is the particular inhomogeneous solution found from the standard variation of parameters approach, in analogy with Eq.~\eqref{eqn:psi-ret variation of parameters}:
\begin{equation}
    {}_2 \psi_{\ell m}^{inh} = {}_2 {\cal C}^{in}_{\ell m}(r) {}_2 \psi_{\ell m}^{in}(r) + {}_2 {\cal C}^{up}_{\ell m}(r) {}_2 \psi_{\ell m}^{up}(r),
\end{equation}
with
\begin{align}
    \,_2{\cal C}^{in}_{\ell m}(r) &= \int^{\rmax}_{r} \frac{{}_2 \psi^{up}_{\ell m}(r^{\prime}) \,_2{S^{\rm eff}_{\ell m}}(r^{\prime})}{_2W(r')\Delta(r^{\prime})} dr^{\prime},\\
    \,_2{\cal C}^{up}_{\ell m}(r) &= \int^{r}_{\rmin} \frac{{}_2 \psi^{in}_{\ell m}(r^{\prime}) \,_2{S^{\rm eff}_{\ell m}}(r^{\prime})}{_2W(r')\Delta(r^{\prime})} dr^{\prime}.
\end{align}
Here, the unknown coefficients ${}_2a_{\ell m}^{in/up}$ and ${}_2b_{\ell m}^{in/up}$ are constrained by demanding continuity 
of ${}_2\psi^{ret}_{\ell m} = {}_2\psi^{R}_{\ell m} + {}_2\psi^{P}_{\ell m}$ and $\dfrac{d{}_2\psi^{ret}_{\ell m}}{dr}$ at the 
worldtube boundaries, yielding
\begin{align}
	{}_{2}a_{\ell m}^{up} =& \frac{1}{{}_{2}\psi^{up}_{\ell m}(\rmax)}\big\{{}_{2}\psi^{up}_{\ell m}(\rmax)[{}_{2}b_{\ell m}^{up} + {{}_{2}\cal C}_{\ell m}^{up}(\rmax) ]\nonumber \\
	& + {}_{2}b_{\ell m}^{in}\,{}_{2}\psi^{in}_{\ell m}(\rmax)+ {}_{2}\psi^{P}_{\ell m}(\rmax)\big\}, \\
	{}_{2}a_{\ell m}^{in} =& \frac{1}{{}_{2}\psi^{in}_{\ell m}(\rmin)}\big\{{}_{2}b_{\ell m}^{up}\,{}_{2}\psi^{up}_{\ell m}(\rmin)  \nonumber \\
	  & + {}_{2}\psi^{in}_{\ell m}(\rmin)[{}_{2}b_{\ell m}^{in} + {}_{2}C_{\ell m}^{in}(\rmin)] + {}_{2}\psi^{P}_{\ell m}(\rmin)\big\}.
	\label{eq:a_coefficients}
\end{align}
Similarly, the coefficients ${}_{2}b^{in/up}_{\ell m}$ are given by
\begin{align}
	\begin{split}
            {}_{2}b^{up}_{\ell m} &= \left. \frac{W[{}_{2}\psi^{P}_{\ell m}(r),
            {}_{2}\psi^{in}_{\ell m}(r)]}{W[{}_{2}\psi^{in}_{\ell m}(r),{}_{2}\psi^{up}_{\ell m}(r)]}\right|_{r=\rmin} , \\
            {}_{2}b^{in}_{\ell m} &= \left. \frac{W[{}_{2}\psi^{P}_{\ell m}(r),
            {}_{2}\psi^{up}_{\ell m}(r)]}{W[{}_{2}\psi^{up}_{\ell m}(r),{}_{2}\psi^{in}_{\ell m}(r)]}\right|_{r=\rmax},
	\label{eq:b_coefficients}
	\end{split}	
\end{align}
where $W[\psi_{1}, \psi_{2}] := \psi_{1}\frac{d\psi_{2}}{dr} - \psi_{2}\frac{d\psi_{1}}{dr} $ is the usual Wronskian. We note that since the residual field must reduce to the retarded field outside the worldtube, the coefficients $_2a^{up/in}_{\ell m}$ are necessarily found to be equal to the coefficients ${}_2C^\pm_{\ell m}$ in the retarded solution~\eqref{eqn:PsiRet}.

The placement of the worldtube boundaries should have no effect on our results on the whole, but practically we find the constraints $|r_0 - r_{\rm min/max}| \lesssim 3M$ and $\rmin > 3M$ ensure an optimal level of error.

\subsection{Behaviour of the modes of the retarded, puncture and residual fields}

Due to the point-particle source, 
the modes of the retarded and puncture fields are discontinuous at $r=r_0$. See, for example, Fig.~\ref{fig:Psi0Fields} for the jump 
in the $\ell=m=2$ mode. They also contain a radial delta function $\delta(r-r_0)$, as alluded to earlier. In the residual field modes ${}_2 \psi^R_{\ell m}(r):= {}_2 \psi^{ret}_{\ell m}(r) - {}_2 \psi^P_{\ell m}(r)$, the distributional content and discontinuities cancel out. More precisely, the jumps in the retarded and puncture field modes across $r=r_0$ are such that 
${}_2 \psi^R_{\ell m}$ is $C^{n}$ there, 
where $n$ denotes the order of the puncture.
\begin{figure}
   \includegraphics[clip, width=1.\columnwidth]{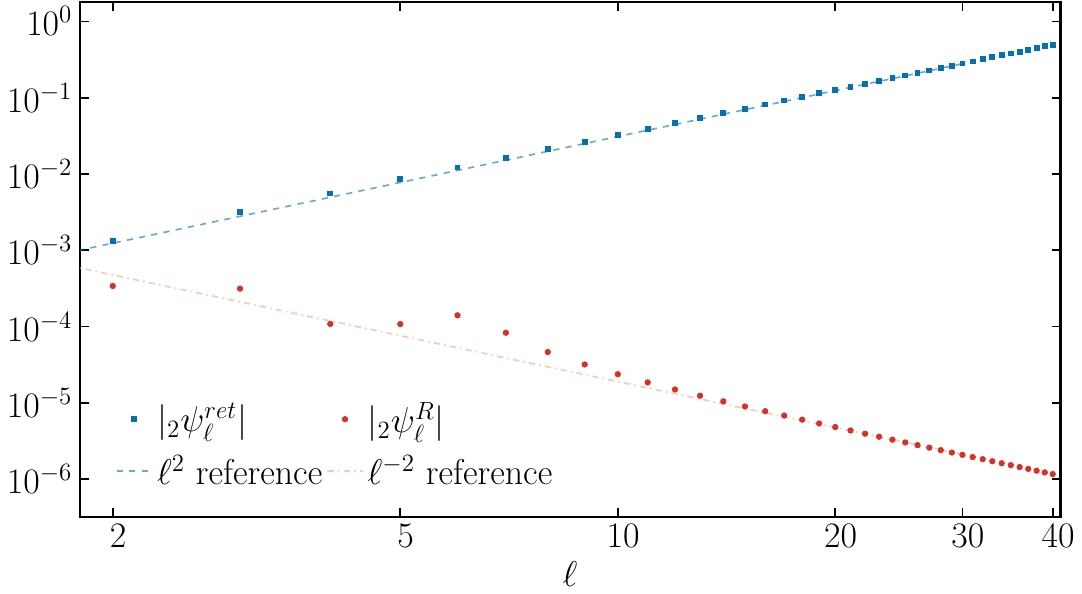}%
\hspace{1cm}

   \includegraphics[clip, width=1.\columnwidth]{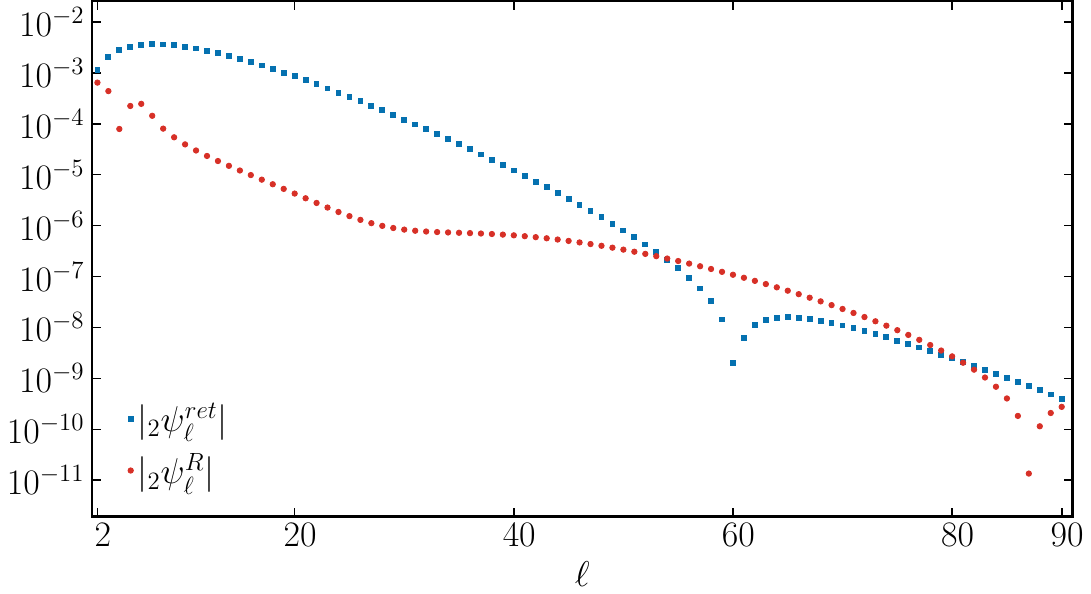}%
\caption{$\ell$-mode behaviour of $\psi^{ret}_0$ and $\psi^R_0$ at the particle, $r=r_0=10M, \theta=\pi/2$ (top, log-log grid), and away from the particle, $r=8M, \theta=\pi/2$ (bottom, semilog grid). Note the much larger range in $\ell$ in the bottom panel.}
\label{fig:Psi0Ret_lmode}
\end{figure}

The sum over the modes of $\psi^{ret}_0$ does not converge at the particle, as expected for a divergent quantity. On the other hand, the sum of modes of $ \psi^R_0$ converges algebraically at the particle, as one would expect for a finitely differentiable function. For the second-order puncture used here, the terms in the sum fall off asymptotically as $\ell^{-2}$ for large $\ell$. This is shown in the top panel of Fig.~\ref{fig:Psi0Ret_lmode}, where we plot the $\ell$-modes
\begin{equation}\label{eqn:psi_ell def}
    {}_2\psi_{\ell}(r,\theta) := \sum_{m=-\ell}^\ell {}_2\psi_{\ell m}(r) e^{i m \Omega_\varphi r_\star} {}_2 Y_{\ell m}(\theta,0),
\end{equation}
evaluated at the particle's position $r=r_0$, $\theta=\pi/2$, for the retarded and residual field. Note that the exponential factor arises because ${}_2 \psi_{\ell m}$ is in $u$ slicing, and we have used ${}_2Y_{\ell m}(\theta,\varphi)e^{-im\Omega_\varphi t}={}_2Y_{\ell m}(\theta,0)e^{im(\varphi-\Omega_\varphi t)}$ and set $\varphi=\Omega_\varphi t$. 

In the bottom panel of Fig~\ref{fig:Psi0Ret_lmode}, we show a similar plot, but where the $\ell$-modes are evaluated at $r=r_0-2M$, away from the particle. In principle, away from $r=r_0$ the fields are smooth functions on the sphere, and we should expect the mode-sum to converge exponentially. As is clear from the plot, exponential behaviour only sets in for the retarded field modes above $\ell\gtrsim 60$, and for the residual field modes it is not apparent even at $\ell=90$. This is in contrast to the clean polynomial behaviour that is apparent at the particle for much smaller $\ell$, as shown in the upper plot.
Upon investigating this feature 
more closely, we can remark that: (1) the poor behaviour of the residual field modes worsens the further away the puncture is evaluated from the particle, with a particularly poor behaviour for $r<r_0$ and for small $r_0$; (2) the order of the puncture strongly affects this behaviour; for example, when computing the residual field using a zeroth-order puncture (instead of a second-order one as is otherwise the case throughout the rest of this work) at $r=r_0-2M=8M$, the residual field has a zero crossing at around $\ell \simeq 41$, whereas it does not for a second-order puncture; see again Fig.~\ref{fig:Psi0Ret_lmode}, where the residual field only crosses through zero around $\ell \simeq 87$. 
This poor behaviour away from the particle is expected to be particularly significant for second-order calculations.
\section{Calculation of the residual Hertz potential}
\label{sec:Hertz}

Generically, the Hertz potential $\Phi$ is a solution to a fourth-order linear differential equation, sourced by either $\psi_0$ or $\psi_4$. There are in fact four such equations, which come in pairs called the angular and radial inversion formulae, one pair each for the cases sourced by $\psi_0$ and $\psi_4$. The choice of which formula is used, and which Weyl scalar one is working 
with, determines whether the ensuing reconstructed metric will be in the IRG or outgoing radiation gauge (ORG) 
\cite{Pound2021}. In the present case, we will work with the radial inversion relation, given in Eq.~\eqref{eqn:PhiREqn}, in order to 
reconstruct the metric perturbation in the IRG from $\psi_0$.

When solving the radial inversion relation, it is convenient to work with the complex conjugate of the Hertz potential, $\bar{\Phi} := \Phi^\star$.
We can decompose $\Phi$ and $\bar{\Phi}$ in spin-weighted spherical harmonics,
\begin{align}
\label{eqn:Phidecomp}
    \Phi &= \sum_{\ell=2}^\infty \sum_{m=-\ell}^\ell {}_{-2} \Phi_{\ell m}(r) \, \tensor[_{-2}]{Y}{_\ell_m}(\theta,\varphi) e^{-im \Omega_\varphi u},\\
    \bar{\Phi} &= \sum_{\ell=2}^\infty \sum_{m=-\ell}^\ell {}_2 \bar{\Phi}_{\ell m}(r) \, \tensor[_{2}]{Y}{_\ell_m}(\theta,\varphi) e^{-im \Omega_\varphi u}. \label{eqn:PhiDecomposition}
\end{align}
Note $\bar{\Phi}$ has spin weight $+2$, motivating the notation ${}_2 \bar{\Phi}_{\ell m}(r)$, but it is not a solution to the $s=+2$ Teukolsky equation. These expansions are used for the residual Hertz potential as well as for the no-string vacuum potentials outside the worldtube. 
The well-known identity
\begin{equation}
\label{eqn:SlmProperty}
_s{Y}^\star_{\ell m} = (-1)^{m+s} \tensor[_{-s}]{Y}{_\ell_{-m}}
\end{equation}
implies that the modes of $\Phi$ and $\bar{\Phi}$ are simply related by
\begin{equation}
{}_2 \bar{\Phi}_{\ell m} = (-1)^m ({}_{-2} \Phi_{l -m})^\star.
\end{equation}
(Note that the quantity on the left represents a mode of a complex-conjugated field, while the quantity on the right represents the complex conjugate of a mode coefficient.) The inversion relation~\eqref{eqn:PhiREqn} is then given by
\begin{equation}\label{eq:inv rln mode}
	\frac{d^4}{dr^4} {}_2 \bar{\Phi}_{\ell m}^R = 2 {}_2 \psi^R_{\ell m},
\end{equation}
where ${}_2 \psi^R_{\ell m}$ is the mode coefficient in $u$ slicing.

As explained in Sec.~\ref{sec:summary Hertz}, we only directly solve Eq.~\eqref{eq:inv rln mode} inside the worldtube. Outside the worldtube, we solve the vacuum adjoint Teukolsky equation ${\cal O}^\dagger\Phi^\pm=0$ for the no-string potentials $\Phi^\pm$. In analogy with $_2\psi^\pm_{\ell m}$, the modes $_{-2}\Phi^\pm_{\ell m}$  are proportional to ``in'' and ``up'' vacuum solutions to the $s=-2$ radial Teukolsky equation. As reviewed in Appendix~\ref{App:static}, the constants of proportionality are determined by imposing Eq.~\eqref{eq:inv rln mode}.

For non-static modes (in this case, modes where $m \neq 0$), the final expressions (see again Sec.~\ref{sec:summary Hertz}), in $u$ slicing, are given by
\begin{align}
    {}_2 \bar{\Phi}_{\ell m}^{ret}(r) &= 2 A^+_{\ell m} {}_2 C^+_{\ell m} {}_{-2} \psi^{up}_{\ell m}(r) e^{i m \Omega_\varphi r_\star}, \quad r>\rmax,\label{eqn:PhiPlus} \\
    {}_2 \bar{\Phi}_{\ell m}^{ret}(r) &= 2 A^-_{\ell m} {}_2 C^-_{\ell m} {}_{-2} \psi^{in}_{\ell m}(r) e^{i m \Omega_\varphi r_\star}, \quad r<\rmin, \label{eqn:PhiMinus}
\end{align}
where ${}_2 C^\pm_{\ell m}$ are the coefficients in Eq.~\eqref{eqn:PsiRet}, and
\begin{align}
    A^+_{\ell m} &:= \frac{16 \omega^4}{p},\\
    A^-_{\ell m} &:= \frac{1}{(w+4iM) (w^2+4M^2) w},\\
    p &:= {}_{-2} \lambda^2_{\ell m} ({}_{-2} \lambda^2_{\ell m} +2)^2 + 144 \omega^2 M^2,\label{eq:p}\\ 
    w &:= 8 \omega M^2.
\end{align}
Here we have corrected transcription errors that appeared in Eq.~(53) of Paper I.

For static modes, 
they instead read
\begin{equation}\label{eq:PhiRet,static}
\begin{split}
{}_2\bar\Phi^{ret}_{\ell 0}(r)&=\frac{2 M^4}{\left(\ell-1\right)_4}{}_2 C^{+}_{\ell 0}{}_{-2}\psi_{\ell 0}^{up}(r),\quad r > \rmax,\\
{}_2\bar\Phi^{ret}_{\ell 0}(r)&=\frac{2 M^4}{\left(\ell-1\right)_4}{}_2 C^{-}_{\ell 0} {}_{-2}\psi_{\ell 0}^{in}(r),\quad r < \rmin. 
\end{split}
\end{equation}
Note that the functional form of the above formulae is sensitive to the choice of normalisation of the basis functions (although the actual numerical values of ${}_2 \bar{\Phi}_{l0}^{ret}$, of course, are not); see Appendix~\ref{App:static} for details. To our knowledge, Eq.~\eqref{eq:PhiRet,static} appears here for the first time.

Moving now to solving for the Hertz potential inside the worldtube, we solve the ODE~\eqref{eq:inv rln mode} with the following boundary conditions on the worldtube boundary $r=\rmin$:
\begin{equation}
    \lim_{r \to \rmin^+} \partial_r^n {}_2 \bar{\Phi}^R_{\ell m} = \partial_r^n {}_2 \bar{\Phi}_{\ell m}^{ret} +  [\partial_r^n {}_2 \bar{\Phi}_{\ell m}^R],
\end{equation}
for $n=0,1,2,3$. In the above, $[F] := (\lim_{r \to r^+_{\rm min}}F(r)) - (\lim_{r \to r^-_{\rm min}}F(r))$
denotes the jump of the quantity $F(r)$ across $\rmin$.
In other words, we demand that the jump of the Hertz potential (and its first three derivatives) across the worldtube at $\rmin$ matches the jump induced by the puncture.

The jumps $[\partial_r^n {}_2 \bar{\Phi}_{\ell m}^R]$ can be easily computed in terms of the coefficients of the delta functions in $\psi^R_0$. Specifically, recalling the schematic form of $\psi^R_0$ in Eq.~\eqref{eqn:PsiSchematicForm}, in a neighbourhood of $r=\rmin$ 
we have
\begin{equation}
    {}_2 \psi^R_{\ell m} = {}_2 A^o_{\ell m} \delta'(r-\rmin) + {}_2 B^o_{\ell m} \delta(r-\rmin),
\end{equation}
plus a piecewise smooth function.
Note that ${}_2 A^o_{\ell m}$ and ${}_2 B^o_{\ell m}$ are constants. 
By integrating the inversion relation in a neighbourhood of $\rmin$, we find these $\delta$ functions then imply that 
\begin{align}
[{}_2 \bar{\Phi}_{\ell m}^R] &= [\pth {}_2 \bar{\Phi}_{\ell m}^R] = 0,\\ 
[\pth^2 {}_2 \bar{\Phi}_{\ell m}^R] &= 2 \, {}_2 A^o_{\ell m}, \\
[\pth^3 {}_2 \bar{\Phi}_{\ell m}^R] &= 2 \, {}_2 B^o_{\ell m}, 
\end{align}
as given in \cite{Toomani2021} (with a change in sign convention).

\begin{figure}
\centering
  \centering
  \includegraphics[width=\linewidth]{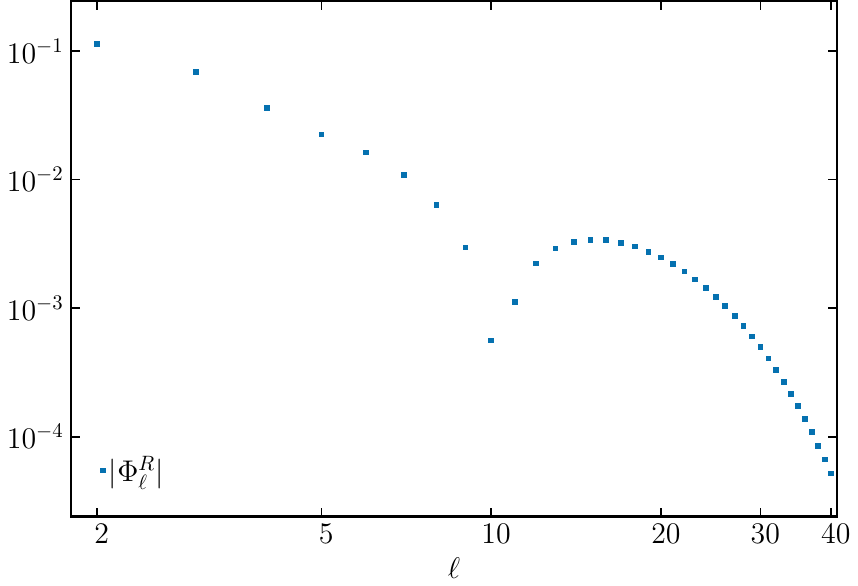}
\caption{Large-$\ell$ behavior of the modes of $\Phi^R$ at the particle, $r=r_0=10M$, $\rmin = 8M$ and $\rmax = 12M$.}
\label{fig:HertzR_lmode}
\end{figure}

In Fig.~\ref{fig:HertzR_lmode}, we show the large-$\ell$ behavior of the modes of the residual Hertz potential, ${}_{-2} \Phi^R_{\ell m}$, at the particle, $r_0=10M$. Here, in analogy with Eq.~\eqref{eqn:psi_ell def}, we define the $\ell$ modes as
\begin{equation}\label{eqn:Hertz_ell def}
    \Phi^R_{\ell}(r,\theta) := \sum_{m=-\ell}^\ell {}_{-2}\Phi^R_{\ell m}(r)e^{i m \Omega_\varphi r^\star} {}_{-2}Y_{\ell m}(\theta,0)
\end{equation}
and evaluate at $r=r_0$, $\theta=\pi/2$. We would naively expect a power-law behaviour that is dependent on the order of the puncture, but a clear power law cannot be identified for $\ell_{max}=40$.

We do not display a corresponding plot for the retarded field. But we note that, unlike the residual field, the retarded field exhibits clear power-law convergence at the particle. More precisely, if the fields $\Phi^\pm$ are evaluated at the particle (as in a traditional no-string solution), then after summing over the $m$-modes, the large-$\ell$ behaviour of the Hertz potential modes are given by ${}_{-2} \Phi_{\ell}^{ret} \sim \ell^{-2}$. This contrasts with the sum of the $\ell$-modes of $\psi_0^{ret}$, which diverges at the particle like $\ell^2$; as expected, the Hertz potential is four orders more regular than the Weyl scalar. 

\section{Residual metric perturbation: Hertz term}
\label{sec:metric_reconstruction}

From the Hertz potential, most of the residual metric perturbation is reconstructed via Eq.~\eqref{eqn:habRecons}.
The reconstructed metric perturbation is in the IRG since $\hat{h}^R_{ab} \ell^b = 0 = g^{ab} \hat{h}^R_{ab}$. The only non-trivial components are
\begin{align}
   \hat{h}^R_{mm} &= \bracket{(S^\dagger_0)_{\bar{m}\bar{m}} \Phi^R}^\star, \\
   \hat{h}^R_{nm} &= \bracket{(S^\dagger_0)_{n\bar{m}} \Phi^R}^\star, \\
   \hat{h}^R_{nn} &= (S^\dagger_0)_{nn} \Phi^R + \bracket{(S^\dagger_0)_{nn} \Phi^R}^\star. 
\end{align}
Decomposing the left- and right-hand sides in spherical harmonics and $u$ slicing as given in Eqs.~\eqref{eqn:habSphericalDecomposition} and \eqref{eqn:PhiDecomposition}, we find the modes are explicitly given by
\begin{align}
   \hat{h}^{\ell m}_{mm} &= \bracket{\frac{2 {}_{-2} \bar{ \Phi}_{\ell m}'(r)}{r}-{}_{-2} \bar{\Phi}_{\ell m}''(r)}, \\
   \hat{h}^{\ell m}_{nm} &= 
   -\frac{1}{r^2} \sqrt{\frac{(\ell-1)(\ell+2)}{2}} \nonumber\\
   &\quad\times\bracket{-2 {}_{-2} \bar{\Phi}_{\ell m}(r)+r {}_{-2} \bar{\Phi}_{\ell m}'(r) }, \\
   \hat{h}^{\ell m}_{nn} &= -\frac{\sqrt{(\ell-1)\ell(\ell+1)(\ell+2)}}{2r^2}  \nonumber\\
   & \quad \times\bracket{{}_{-2} \bar{\Phi}_{\ell m}(r) + (-1)^m {}_{-2} \bar{\Phi}^\star_{\ell -m}(r)}, \\
   \hat{h}^{\ell m}_{\bar{m} \bar{m}} &= (-1)^m \bracket{\hat{h}^{\ell -m}_{mm}}^\star, \\
   \hat{h}^{\ell m}_{n \bar{m}} &= (-1)^{m+1} \bracket{\hat{h}^{\ell -m}_{nm}}^\star.
\end{align}
\section{Residual metric perturbation: corrector tensor}
\label{sec:Corrector_tensors}

The generic form of the corrector tensor $x^R_{ab}$ is given in Eq.~\eqref{eqn:xab}.
In particular, there are only three non-trivial pieces (one  of which is complex): $x^R_{nn}$, $x^R_{nm}$, 
and $x^R_{m\overline{m}}$, along with $x^R_{n\overline{m}} := (x^R_{nm})^\star$. These three pieces satisfy the hierarchical system of equations \eqref{eqn:CorrectorTensorxmmbEqnGHP}--\eqref{eqn:CorrectorTensorxnnEqnGHP}. In the Kinnersley tetrad and $u$ slicing coordinates, these equations become radial ODEs given by
\begin{align}
    -(\partial_r^2+2 r^{-1} \partial_r)x^R_{m\overline{m}} &= T_{ll}^R, \label{eqn:xmmb}\\
    -\bracket{ \frac{1}{2} \partial_r^2 + r^{-1} \partial_r - r^{-2}} x_{nm}^R &= T_{lm}^R + \mathcal{N}x^R_{m\overline{m}}, \label{eqn:xnm}\\
    \bracket{r^{-1} \partial_r + r^{-2}} x_{nn}^R &= T_{ln}^R+\mathcal{U}x^R_{m\overline{m}} + \mathfrak{V}x_{nm} \nonumber\\
    &\quad + \overline{\mathfrak{V}}x^R_{n\overline{m}} \label{eqn:xnn},
\end{align}
where the operators $\mathcal{N}$, $\mathcal{U}$, $\mathfrak{V}$ and $\overline{\mathfrak{V}}$ are given in Eqs.~\eqref{eqn:N}--\eqref{eqn:Vb}.

As usual, one can decompose $x^R_{ab}$ and $T_{ab}^R$ into spin-weighted spherical harmonic and Fourier $u$-modes,
\begin{align}
\label{eqn:xabdecomp}
x^R_{ab} &= \sum_{\ell=|s|}^\infty \sum_{m=-\ell}^\ell x_{ab}^{R,\ell m}(r) \, \tensor[_s]{Y}{_\ell_m}(\theta,\phi) e^{-i m \Omega_\varphi u}, \\
T^R_{ab} &= \sum_{\ell=|s|}^\infty \sum_{m=-\ell}^\ell T_{ab}^{R,\ell m}(r) \, \tensor[_s]{Y}{_\ell_m}(\theta,\phi) e^{-i m \Omega_\varphi u},
\end{align}
where $s$ is the spin weight of the tetrad component ($s=0$ for $x^R_{nn}$, $s=1$ for $x^R_{nm}$, etc.).  
When applied directly to the radial modes, the operators $\mathcal{N}$, $\mathcal{U}$, $\mathfrak{V}$ and $\overline{\mathfrak{V}}$ are explicitly given by
\begin{align}
    \mathcal{N} &= -\frac{\sqrt{(\ell-s) (\ell+s+1)}}{2 \sqrt{2} r} \frac{\partial}{\partial r},\\
    \mathcal{U} &= \frac{1}{{2 r^2}}\biggr[-\left(\ell^2+\ell-4 i m r \Omega_\varphi -2\right) + \\
    & \left(2 i m r^2 \Omega_\varphi -6 M+4 r\right) \frac{\partial}{\partial r}+ r (r-2 M) \frac{\partial^2}{\partial r^2} \biggl],\\
    \mathfrak{V} &= \frac{\sqrt{(\ell-s+1) (\ell+s)} \left(3 + r \frac{\partial}{\partial r} \right)}{2 \sqrt{2} r^2},\\
    \overline{\mathfrak{V}} &= -\mathfrak{V}.
\end{align}

At the level of the radial modes, the usual property of the spin-weighted spherical harmonics then implies that $x_{n\overline{m}}^{R,\ell m} = (-1)^{m+1} \bracket{x_{nm}^{R,\ell-m}}^\star$. Furthermore, one can check explicitly that $x_{ab}^{R,\ell-m} = (-1)^\ell \bracket{x_{ab}^{R,\ell m}}^\star$. Combining both relations, one finds a simple relationship between the radial modes of $x^R_{nm}$ and $x^R_{n\overline{m}}$, namely 
\begin{equation}
x_{n\overline{m}}^{R,\ell m} =  (-1)^{\ell+m+1} x_{nm}^{R,\ell m}.
\end{equation}
From now on in this section, we will suppress mode indices $\ell m$ for brevity.

The form of the solution for the corrector tensor is given in Eq.~\eqref{eqn:x summary}. In that solution, we only require the piece inside the worldtube, denoted $x^M_{ab}$, which we obtain by solving the ODEs \eqref{eqn:xmmb}-\eqref{eqn:xnn} numerically.
Recall that the puncture $h_{ab}^P = h_{ab}^\mathcal{P} \Theta^M$. Since the effective stress energy $T_{ab}^R := T_{ab}-\mathcal{E}_{ab}(h^P) \sim \partial^2 h^P$, it follows that $T_{ab}^R$ contains both $\delta$ and $\delta'$ distributions on the worldtube's boundaries. This distributional content results in jump conditions on the corrector tensor across the worldtube.
Since we are only interested in computing the corrector tensor inside the worldtube, we only need to account for the distributional content at $r=\rmin$.

Therefore, with the distributional content at $r = \rmax$ ignored, each of the terms appearing on the right-hand side of \eqref{eqn:xmmb}-\eqref{eqn:xnn} are of the schematic form,
\begin{align}
    T_{ab}^R &= (\cdots) \Theta(r-\rmin) + \delta T^-_{ab} \delta(r-\rmin)\nonumber\\
    &\quad +\Delta T^-_{ab} \delta'(r-\rmin),
    \label{eqn:TabR_distr} \\
    \mathcal{N}x^R_{m\overline{m}} &= (\cdots) \Theta(r-\rmin) + \frac{1}{2}\eth^- x^-_{m\overline{m}} \delta(r-\rmin),\label{eq:Nxmmb}\\
    \mathfrak{V}x^R_{nm} &= (\cdots) \Theta(r-\rmin) + \frac{1}{2} {\eth'}^- x_{nm}^- \delta(r-\rmin), \\
    \overline{\mathfrak{V}}x^R_{n\overline{m}} &= (\cdots) \Theta(r-\rmin) + \frac{1}{2} \eth^- x_{n\overline{m}}^- \delta(r-\rmin), \\
    \mathcal{U}x^R_{m\overline{m}} &= (\cdots) \Theta(r-\rmin) +\frac{f^-}{2} {x'}_{m\overline{m}}^- \delta(r-\rmin) \nonumber \\ 
    &\quad +\bracket{i m \Omega_\varphi +\frac{2 f^-}{\rmin}} x_{m\overline{m}}^-\delta(r-\rmin)\nonumber\\
    &\quad + \frac{f^-}{2} x_{m\overline{m}}^- \delta'(r-\rmin), \label{eqn:Uxmmb_distr}
\end{align}
where a prime on $x_{ab}$ denotes a radial derivative and a `$-$' superscript on a quantity indicates evaluation of that quantity at $\rmin$. In the case of discontinuous quantities, such as the corrector tensor itself, the superscript denotes the limit from above, as in 
\begin{equation}
x^-_{ab}:=\displaystyle\lim_{r\to (\rmin)^+}x^R_{ab}=\displaystyle\lim_{r\to (\rmin)^+}x^M_{ab}.
\end{equation}
In the above, the operators $\edth$ and $\edth'$ are to be understood as the operator acting directly at the level of radial modes, in which case they simply reduce to a multiplicative factor
\begin{align}
\label{eqn:edthRadialModes}
\edth = -\frac{\sqrt{(\ell-s) (\ell+s+1)}}{\sqrt{2} r}, \\
\edth' = \frac{\sqrt{(\ell+s) (\ell-s+1)}}{\sqrt{2} r}.
\end{align}

There are multiple equivalent ways of translating the $\delta$ and $\delta'$ source terms into boundary/jump conditions for the corrector tensors.
One way is to use the explicit integral solution of the ODE system, given in Eqs.~(59)-(61) of Ref.~\cite{Toomani2021}. Only the distributional content will contribute in computing $x^R_{ab}$ (and its first derivative) at $\rmin$.

Alternatively, one can work with the ODE system itself.
Returning to the ODE for $x^R_{m\overline{m}}$, Eq.~\eqref{eqn:xmmb}, we require the corrector tensors to vanish outside the worldtube for $r<\rmin$.
The corrector tensor $x^R_{ab}$ then contains a Heaviside function at $r=\rmin$. Explicitly writing it out in Eq.~\eqref{eqn:xmmb} for example, and using the standard identity $F(r)\delta'(r-\rmin)=F^-\delta'(r-\rmin)-\partial_rF^-\delta(r-\rmin)$, we find the left-hand side reads
\begin{align}
   & (\cdots) \Theta(r-\rmin) - \frac{2}{r}\bracket{x^M_{m\overline{m}} + r\partial_r x^M_{m\overline{m}}} \delta(r-\rmin) \nonumber \\
   & \qquad - x^M_{m\overline{m}} \delta'(r-\rmin) \\
    &= (\cdots) \Theta(r-\rmin) - \bracket{\frac{2}{\rmin}x_{m\overline{m}}^- + {x'}_{m\overline{m}}^-} \delta(r-\rmin) \nonumber \\
    & \qquad - x_{m\overline{m}}^- \delta'(r-\rmin).
\end{align}
Equating the coefficients in front of $\delta(r-\rmin)$ and $\delta'(r-\rmin)$ in this expression to the corresponding coefficients in Eq.~\eqref{eqn:TabR_distr} gives
\begin{align}
    x_{m\overline{m}}^- &= -\deltapT_{ll}^-, \label{eq:xmmb- val}\\
    {x'}_{m\overline{m}}^- &= \frac{2}{\rmin} \deltapT_{ll}^-  - \deltaT_{ll}^-.\label{eq:xmmb'-}
\end{align}

We can then proceed in a similar fashion for the jump conditions of $x^R_{nm}$ and $x^R_{nn}$.
The left-hand side of Eq.~\eqref{eqn:xnm} reads
\begin{align}
    & (\cdots) \Theta(r-\rmin)  - \bracket{\frac{x^M_{nm}}{r} + \partial_r x^M_{nm}} \delta(r-\rmin) \nonumber \\
    & \qquad -\frac{1}{2} x^M_{nm} \delta'(r-\rmin) \label{eq:xnm eq1} \\
    &= (\cdots) \Theta(r-\rmin) - \bracket{\frac{x_{nm}^-}{\rmin} + \frac{1}{2}{x'}_{nm}^-} \delta(r-\rmin) \nonumber \\
    & \qquad - \frac{1}{2} x_{nm}^- \delta'(r-\rmin).\label{eq:xnm eq2}
\end{align}
The right-hand side is
\begin{align}
    T_{lm}^R + \mathcal{N}x_{m\overline{m}} &= (\cdots) \Theta(r-\rmin) \nonumber \\
    &\quad + \bracket{\deltaT_{lm}^- + \frac{1}{2} \eth^- x_{m\overline{m}}^-} \delta(r-\rmin) \nonumber \\
    &\quad + \deltapT_{lm}^- \delta'(r-\rmin).
\end{align}
Equating the $\delta$ and $\delta'$ on both sides gives
\begin{align}
    x_{nm}^- &= -2 \deltapT_{lm}^-, \\
    {x'}_{nm}^- &= -\eth^- x_{m\overline{m}}^- + \frac{4}{\rmin} \deltapT_{lm}^- - 2\deltaT_{lm}^-.
\end{align}

Finally, turning our attention to \eqref{eqn:xnn}, the left-hand side is simply
\begin{equation}
    (\cdots) \Theta(r-\rmin) + \frac{x_{nn}^-}{\rmin} \delta(r-\rmin).
\end{equation}
The right-hand side reads
\begin{align}
    &(\cdots) \Theta(r-\rmin) + \nonumber \\
    & \left\{\deltaT_{ln}^- + \bracket{i m \Omega_{\varphi}+\frac{2f^-}{\rmin}} x_{m\overline{m}}^- + \frac{f^-}{2} {x'}_{m\overline{m}}^- \right. \nonumber \\
    & \left. + \frac{\sqrt{\ell (\ell+1)}}{2\sqrt{2} \rmin} (x_{nm}^- - x_{n\overline{m}}^-) \right\} \delta(r-\rmin) \nonumber \\
    & + \bracket{\deltapT_{ln}^- + \frac{f^-}{2} x_{m\overline{m}}^-} \delta'(r-\rmin).
    \label{eq:Bmxnmb RHS}
\end{align}
Equating both sides, one can check that the coefficient in front of $\delta'(r-\rmin)$ vanishes, $\deltapT_{ln}^- + \frac{f^-}{2} x_{m\overline{m}}^- = 0$, and we are left with
\begin{align} \label{eq:xnn-}
    x_{nn}^- &= \rmin \biggl[\deltaT_{ln}^- + \bracket{i m \Omega_{\varphi}+\frac{2f^-}{\rmin}} x_{m\overline{m}}^-   \nonumber \\
    &\qquad +\frac{f^-}{2} {x'}_{m\overline{m}}^- + \frac{\sqrt{\ell (\ell+1)}}{2\sqrt{2} \rmin} (x_{nm}^- - x_{n\overline{m}}^-) \biggr].
\end{align}

The cancellation of the $\delta'$ term in the last equation was previously alluded to in Sec.~\ref{sec:summary corrector}. It occurs because $x^-_{m\bar m}$ can be expressed in terms of $T_{ll}$ using Eq.~\eqref{eq:xmmb- val}. The would-be $\delta'$ is then given by $\Delta T_{ln}^- -\frac{f^-}{2} \Delta T_{ll}^-$. These two terms cancel because the $\delta'$ in $T^R_{ll}$ arises from $\partial_r \partial_r h^P_{m\bar{m}}$ while the $\delta'$ in  $T^R_{ln}$ arises from $\frac{f}{2}\partial_r \partial_r h^P_{m\bar{m}}$; these are the only terms in $T^R_{ll}$ and $T^R_{ln}$ involving two radial derivatives. The same cancellation is responsible for all components of $x^{R,\ell m}_{ab}$ being $C^{n+2}$ rather than $C^{n+1}$ at the particle, as also alluded to in Sec.~\ref{sec:summary corrector}.
\section{Gauge correction and completion of the metric perturbation}
\label{sec:Gauge}

There are two final ingredients in the metric perturbation~\eqref{eqn:habShadowlessGauge}: the perturbation~$\dot g_{ab}$ toward another Kerr solution, which appears for $r>\rmax$, and the gauge perturbation $-{\cal L}_\Xi g_{ab}$, which appears for $r<\rmax$.

In the traditional no-string reconstruction procedure, $\dot g_{ab}$ appears instead for all $r>r_0$, and $-{\cal L}_\Xi g_{ab}$ appears for all $r<r_0$. Finding these two contributions was referred to as the \textit{completion problem}~\cite{Merlin2016,vanDeMeent2017,Bini:2019xwn}, referencing the fact that the CCK reconstructed metric perturbation was incomplete.
In the Schwarzschild case, the CCK reconstructed (retarded) metric only corresponds to the $\ell>1$ tensor spherical-harmonic modes of the solution to the linearized Einstein equations with a point-particle source, and the completion problem reduced to solving the $\ell=0$ and $\ell=1$ pieces of the Einstein equations.

Our main conclusion in this section is that $\dot g_{ab}$ and the gauge vector $\Xi^a$ in the GHZ puncture scheme are identical to their values in the traditional no-string retarded-field solution. This should be intuitively reasonable because we expect our solution outside the worldtube to be identical to the traditional no-string solution.

We begin with the perturbation $\dot g_{ab}$. As shown in Paper I (extending Ref.~\cite{vanDeMeent2017}), $\dot g_{ab}$ can be written simply in terms of the total mass and angular momentum contained within (and on the boundaries of) the worldtube. The nontrivial components are
\begin{align}
    \dot{g}_{nn} &= \frac{2}{r} \dot M, \\
    \dot{g}_{nm} &= -i \frac{\sqrt{2}}{r^2} \dot J \sin \theta,
\end{align}
with\footnote{Here we correct the expressions in Paper I by a factor $-1/(8\pi)$. The minus sign arises because we use a future-directed surface element; the common convention~\cite{Poisson:2009pwt}, deriving the surface element from the restriction of the 4D Levi-Civita tensor, instead has a future-directed unit normal and a past-directed surface element.}
\begin{align}
    \dot M &= \frac{1}{8 \pi}\int_{\Sigma_t} T^R_{ab}t^a d\Sigma^b,\label{eq:Mdot int}\\
    \dot J &= -\frac{1}{8 \pi}\int_{\Sigma_t} T^R_{ab}\varphi^a d\Sigma^b,\label{eq:Jdot int}
\end{align}
where $\Sigma_t$ is a hypersurface of constant $t$, $d\Sigma^b = f^{-1}t^b r^2 \sin\theta\, dr\, d\theta\, d\varphi$ is the future-directed surface element on the hypersurface, and $t^a$ and $\varphi^a$ are the timelike and axial Killing vectors.
In the case of a point-particle source $T_{ab}$ (rather than the extended, effective source $T^R_{ab}$), the integrals evaluate to the specific orbital energy ${\cal E}_0$ and angular momentum ${\cal L}_0$,
\begin{align}
    \dot M = {\cal E}_0 &:= -g_{ab} u^a t^b = \frac{1-\frac{2M}{r_0}}{\sqrt{1-\frac{3M}{r_0}}}, \\
    \dot J = {\cal L}_0 &:= g_{ab} u^a \varphi^b = \sqrt{\frac{Mr_0}{1-\frac{3M}{r_0}}}.
\end{align}
As argued in Paper I, these values also remain unchanged in our puncture scheme because the puncture's contribution to $T^R_{ab}$ contributes nothing to the integrals~\eqref{eq:Mdot int} and \eqref{eq:Jdot int}. The puncture's contribution vanishes because Stokes' theorem can be used to express $\int_{\Sigma_t}{\cal E}_{ab}(h^P)t^ad\Sigma_b$ as an integral over the 2D boundary of $\Sigma_t$; since $h^P_{ab}$ vanishes outside the worldtube, the boundary integral likewise vanishes.

We now turn to the gauge perturbation $-{\cal L}_\Xi g_{ab}$, which is generated by the linear-in-time vector $\Xi^a$. Traditionally, in a no-string reconstruction, this perturbation can be understood to enforce continuity conditions on the stationary axially symmetric piece of the metric perturbation, thereby ensuring that the coordinate frequency $\Omega_\varphi$ has the same meaning inside and outside the orbit. 
This is crucial for obtaining the correct values of quasi-invariant quantities such as the Detweiler redshift~\cite{Detweiler2008}. As shown in Paper I, it also serves to prevent Dirac delta distributions whose coefficient grows linearly with time. 

Following Paper I, we write $\Xi^a$ as\footnote{Our expression differs from Paper I by an overall sign to account for the change in metric signature.}
\begin{equation}\label{Xi}
    \Xi^a = -u \bracket{\alpha t^a + \beta \varphi^a},
\end{equation}
where $\alpha$ and $\beta$ are constants given by the $(\ell,m)=(0,0)$ and $(\ell,m)=(1,0)$ modes of the Held quantities $a^o$ and $c^o$,
\begin{align}
    \alpha &= \frac{\braket{a_{00}^o}}{\sqrt{4 \pi}},\\
    \beta &= -i\sqrt{\frac{3}{4\pi}} \braket{c_{10}^o},
\end{align}
where 
\begin{equation}
   \braket{F} := \lim_{T \to \infty} \frac{1}{2T} \int_{-T}^T F du
\end{equation}
is an infinite time average; in our case, this average simply eliminates $m\neq0$ modes.
In the above, the Held scalars $a^o$ and $c^o$ are given by the following integrals:
\begin{align}
    a^o &= -\frac{1}{2} \int_{\rho_{\rm min}^-}^{\rho_{\rm max}^+} d\rho_2 \int_{\rho_{\rm min}^-}^{\rho_2} d\rho_1 \frac{T_{ll}^R}{\rho_1^4} \nonumber \\
    &\quad + \frac{\rho_{\rm max}}{2} \int_{\rho_{\rm min}^-}^{\rho_{\rm max}^+} d\rho_1 \frac{T_{ll}^R}{\rho_1^4},\\
    b^o &= -\frac{1}{2} \int_{\rho_{\rm min}^-}^{\rho_{\rm max}^+} d\rho_1 \frac{T_{ll}^R}{\rho_1^4},\\
    c^o &= \frac{1}{3} \left[\rho_{max} \eth^+ b^o + 2 \int_{\rho_{\rm min}^-}^{\rho_{max}^+} \frac{d\rho}{\rho^2} \bracket{T_{lm}^R + \mathcal{N}x^R_{m\bar{m}}}\right],
\end{align}
where $\rho = -1/r$. The superscript $\pm$ over $\rho_{\rm min}$ and $\rho_{\rm max}$ indicate that the support of the $\delta$ functions are included in the integration domain; so the integrals run from $\rho^-_{\rm min} = \rho_{\rm min} - 0^+$ to $\rho^+_{\rm max} = \rho_{\rm max} + 0^-$. $\eth^+$ is simply equal, at the level of modes, to Eq.~\eqref{eqn:edthRadialModes} evaluated at $\rmax$.
We refer to Paper I for the derivation of these results in a mostly negative signature.

One observes that these integrals are very similar to the integral form of the corrector tensors; see Eqs. (59)--(61) in Paper I for these expressions in a mostly negative signature. For example, the first integral in the expression of $a^o$ precisely equals that of $-x_{m \bar{m}}^R$. The only notable difference is that the integrals for the scalars $a^o$, $b^o$ and $c^o$ include the distributional contribution at $\rmax$, whereas the integrals for the corrector tensor do not (while they do contain the distributional contributions at $\rmin$ via its boundary conditions).
Therefore, the first integral in the expression of $a^o$ is simply the quantity $-x_{m \bar{m}}^R$ evaluated at the worldtube boundary $r=\rmax$, plus the distributional contribution at $\rmax$. In other words, we can write
\begin{align}
\int_{\rho_{\rm min}^-}^{\rho_{\rm max}^+} d\rho_2 \int_{\rho_{\rm min}^-}^{\rho_2} d\rho_1 \frac{T_{ll}^R}{\rho_1^4} &= \int_{\rho_{\rm min}^-}^{\rho_{\rm max}^-} d\rho_2 \int_{\rho_{\rm min}^-}^{\rho_2} d\rho_1 \frac{T_{ll}^R}{\rho_1^4} \nonumber \\
&+ \int_{\rho_{\rm min}^-}^{\rho_{\rm max}^+} d\rho_2 \int_{\rho_{\rm min}^-}^{\rho_2} d\rho_1 \frac{T_{ll}^R}{\rho_1^4}, \\
&= -x_{m \bar{m}} (\rmax) \nonumber \\
&+ \int_{\rho_{\rm max}^-}^{\rho_{\rm max}^+} d\rho_2 \int_{\rho_{\rm min}^-}^{\rho_2} d\rho_1 \frac{T_{ll}^R}{\rho_1^4}
\end{align}
and similarly for $b^o$ and $c^o$.

The remaining integrals from $\rho^-_{\rm max}$ to $\rho^+_{\rm max}$ can be easily evaluated exactly since only the $\delta$ and $\delta'$ terms in $T^R_{ll}$ at $r=\rmax$ will contribute (which were ignored in Eq.~\eqref{eqn:TabR_distr}). Denoting $\delta T^+_{ab}$ and $\Delta T^+_{ab}$ as the coefficients of $\delta(r-\rmax)$ and $\delta'(r-\rmax)$ of $T_{ab}^R$ respectively, and a $+$ superscript indicates evaluation at $r=\rmax$, the final expressions for the Held scalars $a^o$, $b^o$ and $c^o$ are
\begin{align}
    a^o &= -\frac{1}{2} 
    \left[-x^M_{m\bar{m}}(\rmax) + a_1\right] \nonumber \\
    &\quad + \frac{\rho_{\rm max}}{2} 
    \left[-r^2_{\rm max} \partial_r x^M_{m\bar{m}}(\rmax) + a_2\right], \\
    b^o &= -\frac{1}{2} 
    \left[-r^2_{\rm max} \partial_r x^M_{m\bar{m}}(\rmax) + a_2 \right], \\
    c^o &=  -\frac{1}{3 \rmax} \eth^+ b^o + \frac{1}{3} \Bigl[-\partial_r x_{nm}^M(\rmax) \nonumber\\
    &\qquad\qquad\qquad\qquad - \frac{2}{\rmax} x_{nm}^M(\rmax) + 2 c_1\Bigr].
\end{align}
where
\begin{align}
a_1 &:= \Delta T^+_{ll}, \\ 
a_2 &:= \rmax (\rmax \delta T^+_{ll} - 2  \Delta T^+_{ll}), \\
c_1 &:= \delta T^+_{lm} + \frac{\sqrt{\ell (\ell+1)}x^M_{m \bar{m}}(\rmax)}{2\sqrt{2}}.
\end{align}

We find by inspection that $\braket{b^o_{10}} = 0$ because the corresponding $\ell m$ mode of the stress energy vanishes. Furthermore, the constants $\alpha$ and $\beta$ are independent of the worldtube size. In particular, we find numerically that their values are the same as those computed from a point-particle stress-energy~\cite{Shah2015},
\begin{equation}
    \alpha = -\frac{1}{r_0 \sqrt{1-\frac{3M}{r_0}}}, \quad \beta = -\frac{2 \Omega_\varphi}{r_0 \sqrt{1-\frac{3M}{r_0}}}.
\end{equation}
\section{Analysis of the residual field: Detweiler redshift and softened string}
\label{sec:Results}

A strong consistency check of the GHZ procedure is to recover the known value of the Detweiler redshift.
In the first subsection below, we will first calculate the redshift via the CCK reconstruction, before moving to the results using the GHZ infrastructure in the next subsection.
In the following, for a tensor $F_{ab}$, we will define $F_{uu} := u^a u^b F_{ab}$ as the contraction with the $4$-velocity $u^a$, with the understanding that this quantity is to be evaluated in the limit to the particle. In particular, we will identify the redshift with the quantity $h_{uu}$.

\subsection{Traditional method: mode-sum regularization in the no-string gauge}

Perhaps the most straightforward way to compute the redshift at the particle is the following:
\begin{enumerate}
\item First, compute the \textit{retarded} no-string reconstructed metric perturbation, $\hat{h}_{ab}^{ret} = 2 {\rm Re} \bracket{(S^\dagger_0)_{ab} \Phi^{ret}}$. 
\item Subtract the puncture from the retarded field, $\hat{h}_{ab}^{ret}-h_{ab}^P$. Note that, while $\hat{h}_{ab}^{ret}$ is in the (ingoing) radiation gauge, the puncture is in Lorenz gauge.
\item Complete the metric reconstruction by adding the $\ell=0,1$ modes. These correspond to $+\dot{g}_{ab}$ to the right of the particle, $r>r_0$, and $-\mathcal{L}_\Xi g_{ab}$ to the left of the particle $r<r_0$; see again Fig.~\ref{fig:CCK_scheme}.
\begin{align}
h_{ab}^{R,N} &= \hat{h}_{ab}^{ret}-h_{ab}^P - \mathcal{L}_\Xi g_{ab} \Theta(r_0-r) \nonumber\\
&\quad+ \dot{g}_{ab} \Theta(r-r_0).
\end{align}
Here the $N$ indicates that this residual field is in the traditional no-string radiation gauge~\cite{Pound2013}. It is therefore related to ours by a gauge transformation~\cite{Shah2015}.
\item Contract this quantity with the four-velocity and evaluate at the particle.
\end{enumerate}

Note that, since the retarded Hertz potential has been computed separately to the left and right of the particle following the no-string gauge prescription, it is not continuous across the particle, and thus neither is the resulting reconstructed metric,  $\hat{h}^{ret}_{ab}$. Making it continuous across the particle would require the introduction of the shadow field $\Phi^S$.



We have two different formulae for the redshift, depending on whether one wishes to take the limit to the particle from ``the left'' or ``the right'':
\begin{align}
h^R_{uu} &= \sum_{\ell = 0}^\infty \bracket{\hat{h}_{uu}^{ret, \ell} - h_{uu}^{P,\ell}} - \mathcal{L}_\Xi g_{uu}, \quad r \to r_0^-, \\
h^R_{uu} &= \sum_{\ell = 0}^\infty \bracket{\hat{h}_{uu}^{ret,\ell} - h_{uu}^{P,\ell}} + \dot{g}_{uu}, \quad r \to r_0^+.
\end{align}
We refer to the redshift calculated in this manner as $h^{\rm CCK}_{uu}$. In these expressions, the sum is defined in analogy with Eq.~\eqref{eqn:psi_ell def}:
\begin{multline}
    \sum_{\ell = 0}^\infty \bracket{\hat{h}_{uu}^{ret, \ell} - h_{uu}^{P,\ell}} \\
    = \sum_{\ell = |s|}^\infty\sum_{m=-\ell}^\ell\bracket{\hat{h}_{ab}^{ret, \ell m} - h_{ab}^{P,\ell m}}u^a u^b \\
    \times e^{im\Omega_\varphi r_{\star}}{}_sY_{\ell m}(\pi/2,0),
\end{multline}
where $s$ is the spin weight of each tetrad component $h_{ab}$ ($s=0$ for $h_{nn}$, $s=1$ for $h_{nm}$, etc.). Since $\hat{h}_{ab}^{ret}$ is obtained from a CCK metric reconstruction procedure, its $\ell = 0$ and $1$ modes vanish, while the Lorenz-gauge $h^{P,\ell m}_{ab}$ includes these low modes. Since the redshift is a gauge-invariant quantity, it turns out that both formulae are equivalent.
In fact, one can check analytically that $\mathcal{L}_\Xi g_{uu} = - \dot{g}_{uu}$.

In practice, the $\ell$-modes of the reconstructed metric are computed to some finite $\ell = \ell_{\rm max}$.
Since the $\ell$-mode contributions to the redshift scale according to some power law,  the contributions  beyond $\ell > \ell_{\rm max}$ can be approximately accounted for by fitting for this power-law tail.

To benchmark our results, we checked that our calculation of $h^{\rm CCK}_{uu}$ agrees with the known values~\cite{Dolan:2014} within a relative error $\sim 10^{-8}$.

\subsection{GHZ puncture scheme results and comparison}

In the GHZ puncture scheme, the final residual field is given by Eq.~\eqref{eqn:habShadowlessGauge}.
In particular, only three quantities enter the calculation of the redshift:
\begin{equation}
h^R_{uu} = u^a u^b \bracket{\hat h_{ab}^M + x^M_{ab} -  \mathcal{L}_\Xi g_{ab}}|_{r=r_0}. 
\end{equation}
We refer to the redshift calculated in this manner as $h^{\rm GHZ}_{uu}$. 

We again use a power-law tail fit to account for the contribution of large-$\ell$ modes. Concretely, we numerically compute contributions up to $\ell_{\rm max}=40$ and use a power-law fit for the contributions with $\ell>\ell_{max}$. 

\begin{figure*} \centering
	\includegraphics[width=0.75\textwidth]{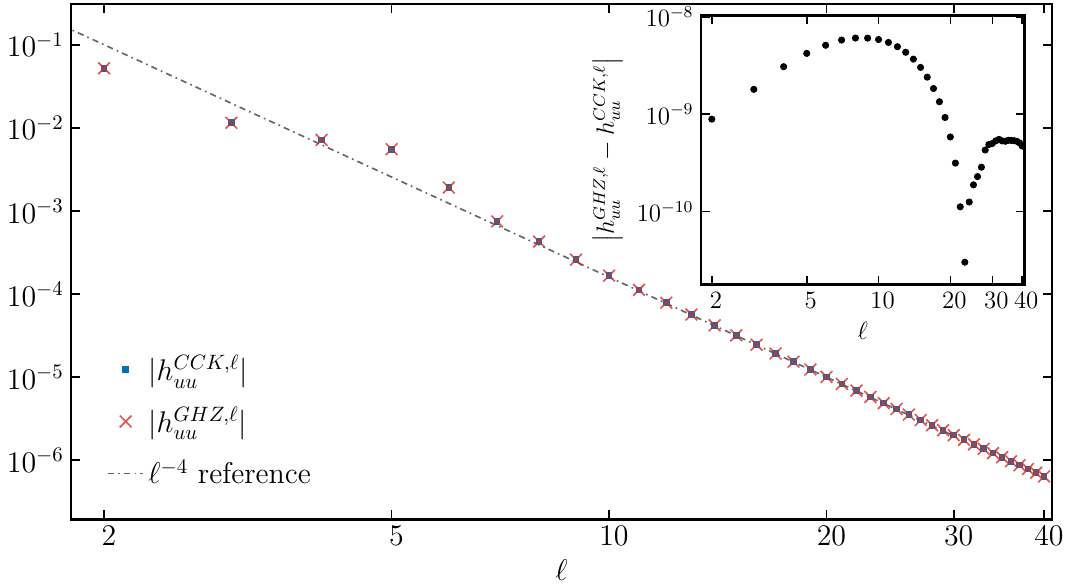}
	\caption{$\ell$-mode contributions of the Detweiler redshift at $r_0=10M$, computed either using the traditional CCK (blue squares), or the new GHZ (red crosses) method. Inset: absolute difference between the two methods.}
	\label{fig:huuR_CCK_GHZ_lmode}
\end{figure*}

In Fig.~\ref{fig:huuR_CCK_GHZ_lmode}, we plot the individual $\ell$-modes of both $h_{uu}^{\rm CCK}$ and $h_{uu}^{\rm GHZ}$ for $r_0/M=10$ and worldtube size $\rmin = 8M$, $\rmax = 12M$. We find good agreement for both methods mode by mode. Table \ref{table:huu} compares $h_{uu}^{\rm GHZ}$ to $h_{uu}^{\rm CCK}$ for a range of $r_0/M$, again showing good agreement. The listed values include a  power-law tail fit in all instances; typically, this improves the relative error by a factor $\sim 10^{-2}$.

Finally, we also verified that, within numerical errors, $h_{uu}^{\rm GHZ}$ does not depend on the worldtube size, although its individual contributions $\hat h_{uu}^M$ and $x^M_{ab}$ do.

\begin{table}\renewcommand{\arraystretch}{1.2} \centering
    \begin{ruledtabular}
	\begin{tabular}{c|
        M{0.27\columnwidth}
        M{0.27\columnwidth}
        M{0.27\columnwidth}}
	$r_0$ & $h_{uu}^{\rm GHZ}$ & $h_{uu}^{\rm CCK}$ & $|h_{uu}^{\rm GHZ}/h_{uu}^{\rm CCK}-1|$ \\
	\midrule
	8   & $-0.2809995(3)$ & $-0.2809995(6)$ & $\sim 1\times 10^{-7}$ \\
	9   & $-0.2439048(6)$ & $-0.2439048(4)$ & $\sim 1\times 10^{-7}$ \\
        10  & $-0.2160628(2)$ & $-0.2160628(8)$ & $\sim 3\times 10^{-7}$ \\
        12  & $-0.1765575(0)$ & $-0.1765575(8)$ & $\sim 5\times 10^{-7}$ \\
        50  & $-0.0404192(5)$ & $-0.0404192(9)$ & $\sim 1\times 10^{-6}$ \\
	\end{tabular}
    \end{ruledtabular}
	\caption{Comparison between the redshift $h_{uu}^{\rm GHZ}$ as calculated from our GHZ puncture scheme and as calculated from traditional no-string CCK reconstruction and completion for different orbital radii $r_0/M$. 
    In each case, the $\ell$-modes of $h_{uu}^{\rm GHZ}$ were computed up to $\ell_{max}=40$ and the contribution of higher modes was included using a power-law fit.}
	\label{table:huu}
\end{table}

\subsection{Softened string singularity}

In the typical no-string CCK reconstruction, the reconstructed metric is highly singular at the particle's orbital radius, possessing both a jump discontinuity and a delta function there. Alternatively, in a half-string reconstruction (as would be obtained by the GHZ scheme without a puncture~\cite{Toomani2021}), there is a strong singularity emanating along null rays from the particle to infinity or from the particle to the horizon.

By using a puncture scheme, we soften this singular behavior. In particular the residual reconstructed metric perturbation is $C^n$ at the particle (in four dimensions; $C^{n+2}$ at the level of modes).
Nonetheless, there is a residual, softened string singularity that is confined inside the worldtube, in the region $r_0 < r < \rmax$. Specifically, the presence of this softened string is due to the fact that that effective stress-energy tensor $T_{ab}^R$ is $C^{n-2}$ at the particle, as is $\psi^R_{0}$.
In the GHZ reconstruction, these functions are integrated along integral curves of $l^a$ to obtain the Hertz potential $\Phi^R$ and the corrector tensors $x^R_{ab}$. As a result, the singularity at $s_0=0$ is propagated along $l^a$. In general, the integration increases the singularity’s dimension by one but reduces its strength by one as well.
Consequently, the corrector tensors, the Hertz potential, and, by extension, the reconstructed metric emerge as $C^k$ functions of coordinates along this string. The value of $k$ is contingent upon the field under consideration and is determined by the order of the puncture. Paper I estimated that $h^R_{ab}$ is at worst $C^{n-3}$ but also observed this is overly pessimistic in known cases. The strength of the string singularity in the retarded half-string solution is $\sim1/\varrho^2$ (i.e., $C^{-3}$), where $\varrho$ is the distance from the string~\cite{Pound2013}. If the singularity is weakened by one order with each additional order of the puncture, we can expect $h^R_{ab}$ to be $C^{n-1}$ at the string for an $n$th-order puncture. This would suggest $h^R_{ab}$ should be $C^1$ for our $n=2$ puncture, with the local behavior $\sim \varrho^2\ln\varrho$.

We can numerically investigate the degree of regularity of our reconstructed metric $h_{ab}^R$ by integrating it over the 2-sphere, $S^2$, with radius $r_0 < r_s < \rmax$.
Specifically, we consider the $L^2$-norm,
\begin{align}
||h_{ab}^R||_{L^2}(r) &:= \int_{S^2} |h_{ab}^R|^2 \sin \theta d\theta d\phi \nonumber\\
&\hphantom{:}= \sum_{\ell \geq |s|} \sum_{m=-\ell}^\ell |h_{ab}^{R,\ell m}(r)|^2,
\end{align}
where the last equality follows from the orthonormality of the (spin-weighted) spherical harmonics. Standard methods~\cite{Orszag:1974} show that if $h^R_{ab}$ has the expected string singularity $\sim\varrho^2\ln\varrho$, then 
\begin{equation}\label{eq:string decay}
    \sum_{m=-\ell}^\ell |h_{ab}^{R,\ell m}(r)|^2\leq C(r)\frac{(2\ell+1)}{[\ell(\ell+1)]^2}\sim \ell^{-3} 
\end{equation}
for some $\ell$-independent $C(r)$.

\begin{figure} \centering
	\includegraphics[width=\columnwidth]{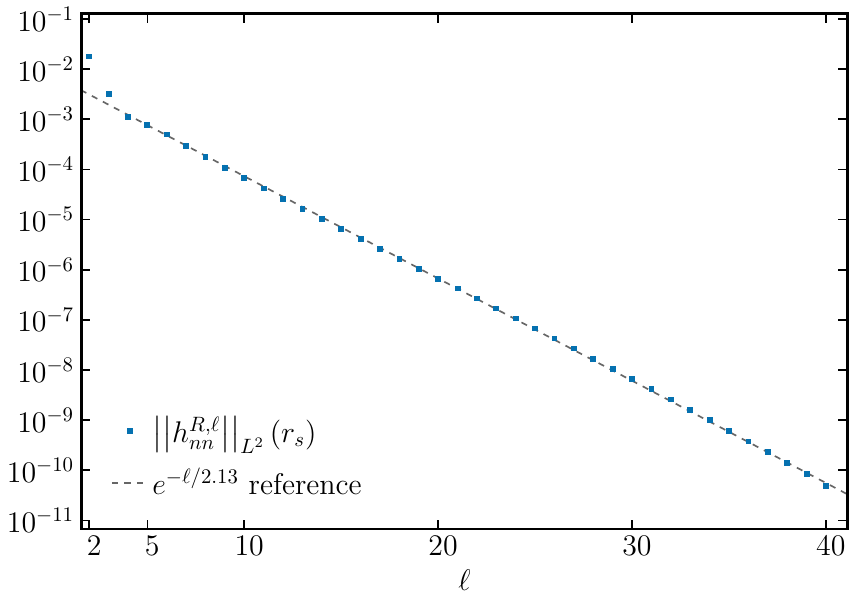}
	\caption{$\ell$-mode contributions to the $L^2$ norm of $h_{nn}^R$ on the 2-sphere $S^2$ centered at radius $r_s = (r_0+\rmax)/2$, with $r_0=10M$ and $\rmax = 12M$. Away from the particle, $h_{nn}^R$ can be written as a sum of smooth and $C^k$ pieces (the softened string), which are expected to converge exponentially and polynomially with $\ell$, respectively. At least up to $\ell=40$, we only observe exponential behaviour, suggesting that the magnitude of the softened string is small.}
	\label{fig:hSLnn_lBehaviour_L2Norm}
\end{figure}

In Fig.~\ref{fig:hSLnn_lBehaviour_L2Norm}, we show the $\ell$-mode contributions of $||h_{nn}^R||_{L^2}(r_s)$, for $r_s := (r_0+\rmax)/2$. Other tetrad components show qualitatively the same behaviour. We find, surprisingly, an exponential decay with $\ell$, instead of a polynomial behaviour as one would have expected.
The appearance of an exponential behaviour itself is expected, since the reconstructed metric contains smooth pieces, but one would expect these pieces to be subdominant compared to the $C^k$ pieces which only decay algebraically with $\ell$. As shown in Fig.~\ref{fig:hSLnn_lBehaviour_L2Norm}, we do not find any hint of polynomial behaviour up to $\ell=40$, suggesting that the amplitude of the softened string is rather small. Quantitatively, if our estimate~\eqref{eq:string decay} is correct, our numerical results suggest $C(r_s)\lesssim 0.00023$. If this turns out to be a general feature of the GHZ reconstruction, the presence of the softened string may not be an important consideration for accurate second-order calculations.

\section{Conclusion}
\label{sec:conclusion}

We have presented the first implementation of Paper~I's GHZ-Teukolsky puncture scheme in an astrophysically relevant scenario, going beyond Paper~I's simple proof-of-concept implementation in flat spacetime. Our calculations, for a particle in circular orbit in Schwarzschild spacetime, also mark the first concrete use of GHZ reconstruction for a nontrivial physical system. 
As a crucial test of our implementation, we have recovered the known values of the Detweiler redshift~\cite{Dolan:2014,Detweiler2008}, for different values of the orbital radius $r_0/M$, and checked that these values are independent of the worldtube size.

The ingredient in the GHZ reconstruction that extends the CCK-Ori reconstruction is the inclusion of a corrector tensor. This tensor obeys a simple set of semi-decoupled first- and second-order ODEs that can be solved very efficiently. The most time-consuming aspect of our implementation is numerically computing the exact mode decomposition of the puncture, and to a lesser extent, computing the solutions to the Teukolsky equation.

The initial aim of Paper I's scheme was to compute the first-order metric perturbation in a sufficiently regular gauge as input for second-order calculations. The residual first-order metric perturbation we obtain is smooth everywhere, except at the particle and on a string from the particle to $r=\rmax$, where the regularity of the first-order metric perturbation depends on the order of the puncture. This finite differentiability leads to polynomial, rather than exponential, decay of mode coefficients. However, for the highest-order punctures in the literature, such as the one we use here, this effect of the field's non-smoothness is mild; at the string, it is numerically undetectable up to $\ell=40$.

Our particular implementation of the puncture scheme also gives rise to another type of singularity: jump discontinuities and delta functions at the worldtube boundaries. These occur for two reasons. First, our use of a box window function in our puncture causes a gauge discontinuity because the retarded field outside the worldtube is in a radiation gauge while the retarded field $h^R_{ab}+h^P_{ab}$ inside the worldtube is in a mixed gauge (with $h^R_{ab}$ in an IRG and $h^P_{ab}$ in the Lorenz gauge). Second, our use of a discontinuous gauge transformation to eliminate the ``shadow field'' outside the worldtube leads to both discontinuities and delta functions at $r=\rmax$; see Eq.~\eqref{eqn:habShadowlessGauge}. All of these features can be eliminated by replacing Heaviside functions with smooth window functions that smoothly taper the puncture and the gauge vector to zero. 

However, such tapering is likely not needed in the case of the delta function (and jump) created by the discontinuous gauge vector~\eqref{eq:discontinuous xi}. Even though the distributions at first order will formally lead to ill-defined products of distributions in the quadratic source term at second order, one can sidestep that problem by directly deriving jump conditions for the second-order fields across the worldtube boundaries. In a smooth gauge, there is no jump in the retarded field across the boundary. Under a gauge transformation generated by the vector field $\xi^a_S$, the second-order metric perturbation changes by an amount $\Delta h^{(2)}_{ab} = \frac{1}{2}{\cal L}_{\xi_S}^2g_{ab}-{\cal L}_{\xi_S}h^{(1)}_{ab}$~\cite{Pound2021}, where we have added superscript labels to indicate the perturbative order of the fields, and where $h^{(1)}_{ab}$ is in the gauge containing the shadow field. Combining these facts, we find that the jump caused by the discontinuous gauge vector~\eqref{eq:discontinuous xi} is simply $\bigl[h^{(2)}_{ab}\bigr]=\Delta h^{(2)}_{ab}\bigr|_{\rmax}$. Note that computing this jump is only possible because $\xi^a_S$ is known explicitly~\cite{Toomani2021}.

Our results therefore demonstrate that the GHZ puncture scheme is a viable path toward second-order self-force calculations in Kerr spacetime. In the remainder of this section, we compare it to several alternatives.

\subsection{Roads to second-order self-force in Kerr}

As mentioned in the Introduction, there are now several possible avenues to second-order  self-force calculations in a Kerr background:
\begin{enumerate}
    \item Traditional no-string CCK reconstruction and completion equipped with suitable regularization.
    \item The GHZ puncture scheme demonstrated here.
    \item A Lorenz-gauge metric reconstruction procedure following Dolan et al.~\cite{Dolan:2021ijg,Dolan:2023enf}.
    \item Directly solving the coupled Lorenz-gauge Einstein equations in an $m$-mode decomposition, following Osburn and Nishimura~\cite{Osburn2022}.
\end{enumerate}
All of these are methods of obtaining first-order solutions that are sufficiently regular to use as input at second order. There are then also several options for solving the second-order field equations, whether starting from Teukolsky equations~\cite{Spiers:2023cip} or directly solving the second-order Einstein equations.

The no-string option would likely be the simplest, as it mostly reduces to solving two vacuum Teukolsky equations (for the Weyl scalar and the Hertz potential). Since the no-string retarded metric perturbation contains jump discontinuities and delta functions, one would have to appropriately regularize it if using it within a second-order quadratic source. As we have pointed out regarding delta functions at our worldtube boundary, this is not necessarily an intractable obstacle. However, little work has been done to this end.

The GHZ puncture scheme is perhaps the second simplest option, since it only adds a small number of simple radial ODEs. However, its significant downside, relative to the alternatives, is that it requires the exact mode decomposition of the puncture. At least as of this writing, this means numerically integrating the puncture over spheres of constant $r$ on a grid of $r$ values within the worldtube. This procedure contrasts with earlier puncture schemes in which the puncture modes are obtained analytically as an expansion in powers of $r-r_0$. As discussed in Paper I, such expansions introduce divergences at large $\ell$ at all points away from the particle. Since the residual and puncture fields of the first-order metric perturbation are in different gauges, their $\ell$-mode-sum divergences do not readily cancel each other in the total, retarded field (as they would in a Lorenz-gauge puncture scheme). An exact mode decomposition of the puncture circumvents this by never introducing spurious large-$\ell$ divergences in the first place, but it is  substantially more expensive than other steps in the calculation. Currently, this exact mode decomposition is also the only missing ingredient for implementing the GHZ puncture scheme in Kerr; all other tools for such a calculation are ready at hand.

However, this drawback of the GHZ puncture scheme might ultimately be immaterial. The only extant method of computing the second-order quadratic source requires the exact puncture modes in any case~\cite{Miller:2016hjv}. In that sense, the exact mode decomposition does not represent an additional cost for the GHZ puncture scheme.

The Lorenz-gauge reconstruction method of Dolan et al.~is moderately more complicated than the GHZ scheme, in that it requires solving several additional Teukolsky equations, one of which has a noncompact source. To date, it has also been more limited than the GHZ procedure because it has been restricted to vacuum reconstruction; although it has been applied to point-particle sources on circular orbits in Kerr~\cite{Dolan:2023enf}, several key steps in the calculation assumed vacuum away from the particle's worldline. However, this restriction is not fundamental, and an extension to generic sources should soon be available~\cite{Wardell:2023-talk,Wardell:2023-inprep}. Even without such an extension, the existing results for the first-order metric perturbation for circular orbits could serve as an immediate starting point for second-order calculations in Kerr.  

Directly solving the Lorenz-gauge Einstein equations in an $m$-mode puncture scheme, as proposed by Osburn and Nishimura, is numerically the most complicated option but conceptually the simplest. This approach avoids a full $\ell m$ mode decomposition because the Lorenz-gauge Einstein equations are not  separable in such a decomposition. Instead, one solves coupled two-dimensional elliptic PDEs in $r$ and $\theta$ for the metric perturbation components. As of this writing, work with this method has nearly completed the calculation of the first-order Lorenz-gauge metric perturbation for circular orbits in Kerr~\cite{OsburnCapra2023}, complementing the $\ell m$-mode results of Dolan et al.

An $m$-mode scheme is generally more expensive than a full separation of variables because it involves PDEs instead of ODEs. However, we note that both the GHZ and Lorenz-gauge reconstruction schemes might actually benefit from an $m$-mode implementation. One recurring difficulty when working in $\ell m$ modes is the poor convergence of the $\ell$-mode sum close to the particle, which leads to major challenges when constructing the quadratic source at second order~\cite{Miller:2016hjv}. While slow convergence might be partly alleviated using a puncture scheme with a high-order puncture~\cite{Bourg2024,Macedo2024}, we find that some residual quantities do not exhibit a ``clean'' power-law convergence, as shown in Fig.~\ref{fig:HertzR_lmode}. This prevents us from extrapolating higher mode numbers by fitting, which in turn impacts the final accuracy of the model. All of these poor convergence properties might be at least partially bypassed by an $m$-mode scheme~\cite{Macedo2024}.


\subsection{Further applications}


As was also mentioned in the Introduction (and in Paper I), a Teukolsky puncture scheme could have additional benefits beyond yielding sufficiently regular fields for second-order calculations. This is true regardless of whether one uses GHZ or Lorenz-gauge reconstruction. One such benefit is more rapid convergence of numerical approximations. In the case of eccentric orbits, this could resolve Gibbs phenomena simply by working with smoother fields. The standard method of overcoming Gibbs phenomena has been the method of extended homogeneous solutions, but that method becomes increasingly expensive for more eccentric orbits due to large cancellations (of more than 30 digits in some cases~\cite{vandeMeent:2017bcc}). A puncture scheme could be a fruitful alternative.

The emergence of non-vacuum reconstruction methods also opens up the possibility of obtaining second-order metric perturbations from solutions to the second-order Teukolsky equation~\cite{Green2019,Spiers:2023cip}. While it is likely that only the second-order Weyl scalar is needed for first-post-adiabatic waveform generation~\cite{Spiers:2023cip}, the complete metric perturbation would be needed to calculate second-order conservative effects, which enter the waveform at second post-adiabatic order. The GHZ puncture scheme represents a viable method for that purpose. Although the GHZ reconstruction was originally derived for \textit{bounded} sources, it necessarily yields a particular solution even for unbounded ones, such as the quadratic source at second order.

In fact, formally, little changes in the GHZ procedure at second order. The GHZ procedure divides into three steps: (1) solving the Teukolsky equation for a Weyl scalar, (2) solving an inversion relation for the Hertz potential and applying derivatives to obtain a reconstructed metric perturbation, and (3) completing the metric perturbation by solving radial ODEs for the corrector tensor. Each of these steps can be carried out at second order. However, in practice, the puncture scheme from Paper I relies on having vacuum regions outside the worldtube, which allows easy use of ready-at-hand vacuum solutions, easy algebraic inversion to obtain the Hertz potential, and easy removal of the shadow field. At second order, one would instead have to directly solve inhomogeneous equations over the whole spacetime at each step. The major challenge then lies in dealing with the unbounded second-order source. In the Lorenz gauge, this noncompact source leads to infrared divergences, in turn necessitating alternative formulations of the small-mass-ratio expansion near the horizon and future null infinity~\cite{Pound:2015wva}. Such problems could be exacerbated in a GHZ solution in the IRG, which is not an asymptotically flat gauge even at first order. However, as recently outlined by Spiers et al.~\cite{Spiers:2023cip}, problems in the infrared might be entirely bypassed by transforming to a well-behaved Bondi-type gauge at first order before proceeding to second order. The resulting asymptotically well-behaved second-order source would then also enable use of compactified hyperboloidal slices at second order, substantially alleviating the expense of solving the inhomogeneous second-order equations with noncompact sources~\cite{PanossoMacedo:2022fdi, DaSilva:2023xif,PanossoMacedo:2023qzp}.

\section*{Acknowledgement}

We thank Stefan Hollands, Vahid Toomani, and Maarten van de Meent for helpful discussions. AP acknowledges the support of a Royal Society University Research Fellowship and the ERC Consolidator/UKRI Frontier Research Grant GWModels (selected by the ERC and funded by UKRI [grant number EP/Y008251/1]). AP and PB acknowledge the support of a Royal Society University Research Fellowship Enhancement Award. PB acknowledges the support of the Dutch Research Council (NWO) (project name: Resonating with the new gravitational-wave era, project number: OCENW.M.21.119). This work makes use of the Black Hole Perturbation Toolkit.

\appendix

\section{GHP formalism}
\label{section:GHP_formalism}

In the GHP formalism, each quantity has a well-defined  \textit{type} $\{p,q\}$ which expresses how the quantity transforms under spin and boost transformation. The group of such transformations is isomorphic to multiplication by a complex number $\lambda$, and an object $f$ of type $\{p,q\}$ transforms as $f \to \lambda^p \bar{\lambda}^q f$. We will use the notation $f \  \stackrel{o}{=} \{p,q\}$. Only objects of the same type can be added together, which provides a useful consistency check on any equations.
Note that the product of an object of type $\{p_1,q_1\}$ with an object of type $\{p_2,q_2\}$ produces an object of type $\{p_1+p_2,q_1+q_2\}$.

As part of the GHP formalism, as in the Newman-Penrose formalism, one adopts a double-null tetrad basis of the form $(l^a,n^a,m^a,\bar{m}^a)$. The vectors are normalised such that
\begin{equation}
l^a n_a = -1, \qquad m^a \bar{m}_a = 1,
\end{equation}
and all other product combinations are zero. The metric in this basis therefore reads
\begin{equation}\label{eq:metric NP}
g_{ab} = -2 l_{(a} n_{b)} + 2 m_{(a} \bar{m}_{b)}.
\end{equation}
We remark here that the above choice of normalisation corresponds to mostly positive signature for the background metric $g_{ab}$.
The tetrad legs have the following GHP type:
\begin{align}
l &\stackrel{o}{=} \{1,1\}, \qquad n \stackrel{o}{=} \{-1,-1\}, \\
m &\stackrel{o}{=} \{1,-1\}, \qquad \bar{m} \stackrel{o}{=} \{-1,1\}.
\end{align}
All other GHP objects' type can be deduced by counting the factors of the tetrad vectors in their definition. In particular,
\begin{align}
h_{ll}  \stackrel{o}{=} \{2,2\}, \; h_{ln}  \stackrel{o}{=} \{0,0\}, \; h_{lm}  \stackrel{o}{=} \{2,0\}, \; h_{l \bar{m}}  \stackrel{o}{=} \{0,2\}, \\
h_{nn} \stackrel{o}{=} \{-2,-2\}, \; h_{nm}  \stackrel{o}{=} \{0,-2\},  \; h_{n\bar{m}}  \stackrel{o}{=} \{-2,0\}, \\
h_{mm} \stackrel{o}{=} \{2,-2\}, \; h_{m\bar{m}}  \stackrel{o}{=} \{0,0\}, \; h_{\bar{m}\bar{m}}  \stackrel{o}{=} \{-2,2\}.
\end{align}

From the tetrad vectors $(l^a,n^a,m^a,\bar{m}^a)$, one can next introduce the \textit{spin coefficients}, defined to be the 12 directional derivatives of the tetrad vectors. Of these, eight have a well-defined GHP type, and are given by
\begin{align}
\kappa &= -l^\mu m^\nu \nabla_\mu l_\nu \stackrel{o}{=} \{3,1\}, \label{eqn:kappa} \\
\kappa' &= -n^\mu \bar{m}^\nu \nabla_\mu n_\nu \stackrel{o}{=} \{-3,-1\}, \\
\sigma &= -m^\mu m^\nu \nabla_\mu l_\nu \stackrel{o}{=} \{3,-1\}, \\ 
\sigma' &= -\bar{m}^\mu \bar{m}^\nu \nabla_\mu n_\nu \stackrel{o}{=} \{-3,1\}, \\ 
\rho &= -\bar{m}^\mu m^\nu \nabla_\mu l_\nu \stackrel{o}{=} \{1,1\}, \label{eqn:rho} \\
\rho' &= -m^\mu \bar{m}^\nu \nabla_\mu n_\nu \stackrel{o}{=} \{-1,-1\}, \\
\tau &= -n^\mu m^\nu \nabla_\mu l_\nu \stackrel{o}{=} \{1,-1\}, \\
\tau' &= -l^\mu \bar{m}^\nu \nabla_\mu n_\nu \stackrel{o}{=} \{-1,1\}. \label{eqn:taup}
\end{align}

The remaining four spin coefficients do not have a well-defined GHP type by themselves, but instead appear in the definition of four GHP derivative operators. When acting on an object of GHP type $\{p,q\}$, they are given by
\begin{align}
\pth &= l^\mu \nabla_\mu -p \epsilon - q \bar{\epsilon} \stackrel{o}{=} \{1,1\}, \\
\pth' &= n^\mu \nabla_\mu +p \epsilon' + q \bar{\epsilon}' \stackrel{o}{=} \{-1,-1\}, \\
\eth &= m^\mu \nabla_\mu -p \beta + q \bar{\beta}' \stackrel{o}{=} \{1,-1\}, \\
\eth' &= \bar{m}^\mu \nabla_\mu +p \beta' - q \bar{\beta} \stackrel{o}{=} \{-1,1\},
\end{align}
where
\begin{align}
\beta &= \frac{1}{2} \bracket{m^\mu \bar{m}^\nu \nabla_\mu m_\nu - m^\mu n^\nu \nabla_\mu l_\nu},\\
\beta' &= \frac{1}{2} \bracket{\bar{m}^\mu m^\nu \nabla_\mu \bar{m}_\nu - \bar{m}^\mu l^\nu \nabla_\mu n_\nu},\\
\epsilon &= \frac{1}{2} \bracket{l^\mu \bar{m}^\nu \nabla_\mu m_\nu - l^\mu n^\nu \nabla_\mu l_\nu},\\
\epsilon' &=  \frac{1}{2} \bracket{n^\mu m^\nu \nabla_\mu \bar{m}_\nu - n^\mu l^\nu \nabla_\mu n_\nu}.
\end{align}

The definition of the (linearized) Weyl scalars $\psi_0$ and $\psi_4$ can be found for example in Eq.~(54) of Ref.~\cite{Pound2021}, which uses a mostly positive  convention:
\begin{align}
    \psi_0 &:= C_{abcd} l^a m^b l^c m^d = \mathcal{T}_0^{ab} h_{ab}, \\
    \psi_4 &:= C_{abcd} n^a \bar{m}^b n^c \bar{m}^d = \mathcal{T}_4^{ab} h_{ab},
\end{align}
where $C_{abcd}$ is the linearized Weyl tensor. In a mostly negative sign convention, such as in \cite{Toomani2021}, an additional minus sign is added to the above definition to ensure that the resulting Teukolsky equations are independent of sign convention.
The operators $\mathcal{T}_0^{ab}$ and $\mathcal{T}_4^{ab}$ are explicitly given by
\begin{align}
\label{eqn:T0abhab}
    \mathcal{T}_0^{ab} h_{ab} &= -\frac{1}{2} \left\{ \eth^2 h_{ll} + (\pth-\rho) (\pth-\rho) h_{mm} \right. \nonumber \\
    	&\quad \left. - \left[(\pth-\rho) \eth + \eth (\pth-2 \rho)\right] h_{lm} \right\}, \\
\label{eqn:T4abhab}
   \mathcal{T}_4^{ab} h_{ab} &= -\frac{1}{2} \left\{ {\eth'}^2 h_{nn} + (\pth'-\rho') (\pth'-\rho') h_{\bar{m} \bar{m}} \right. \nonumber \\
    	&\quad \left. - \left[(\pth'-\rho') \eth' + \eth' (\pth' - 2 \rho')\right] h_{n \bar{m}} \right\}.
\end{align}

The source operators $\mathcal{S}_0^{ab}$ and $\mathcal{S}_4^{ab}$ in the Teukolsky equations, \eqref{eqn:Teukolsky}-\eqref{eqn:TeukolskyPsi4}, are given by
\begin{align}
\label{eqn:S0abTab}
    \mathcal{S}_0^{ab} T_{ab} &:= \eth \bigl[(\pth - 2 \rho) T_{lm} - \eth T_{ll}\bigr] \\
    &\quad + (\pth-5\rho) \bigl[\eth T_{lm} - (\pth-\rho) T_{mm}\bigr], \nonumber \\
\label{eqn:S4abTab}
    \mathcal{S}_4^{ab} T_{ab} &:= \eth' \bigl[(\pth' - 2 \rho') T_{n \bar{m}} - \eth' T_{nn}\bigr] \nonumber \\
    &\quad + (\pth'-5\rho') \bigl[\eth' T_{n \bar{m}} - (\pth'-\rho') T_{\bar{m} \bar{m}}\bigr].
\end{align}
Their adjoints are also important as they directly appear when reconstructing the metric perturbation from the Hertz potential, \eqref{eqn:habRecons}. The adjoints are
\begin{align}
\label{eqn:S0abAdj}
    (S_0^\dagger)_{ab} &= -l_a l_b \eth^2 - m_a m_b (\pth-\rho) (\pth+3 \rho) \nonumber \\
    &\quad + l_{(a} m_{b)} [\pth \eth - \eth (\pth + 3 \rho)], \\
\label{eqn:S4abAdj}
    (S_4^\dagger)_{ab} &= -n_a n_b {\eth'}^2 - \bar{m}_a \bar{m}_b (\pth'-\rho') (\pth'+3 \rho') \nonumber \\
    &\quad + n_{(a} \bar{m}_{b)} [\pth' \eth' - \eth' (\pth' + 3 \rho')].
\end{align}
The operators on the left-hand sides of the Teukolsky equation are
\begin{align}
\label{eqn:TeukolskyO}
    \mathcal{O} &:= 2 \bracket{(\pth - 4 \rho - \bar{\rho}) (\pth' - \rho') - \eth \eth' -3 \psi_2}, \\
\label{eqn:TeukolskyOp}
    \mathcal{O}' &= 2 \bracket{(\pth' - 4 \rho' - \bar{\rho}') (\pth - \rho) - \eth' \eth -3 \psi_2},
\end{align}
where, in the Kinnerlsey tetrad, $\psi_2 = M \rho^3$. We also require the adjoint of ${\cal O}$, which is related to ${\cal O}'$ by
\begin{equation}\label{eqn:TeukolskyOadj}
   {\cal O}^\dagger = \rho^{-4} \mathcal{O}'\rho^{4}.
\end{equation}

Finally, the operators appearing as source terms in the ODEs for the corrector tensor, \eqref{eqn:CorrectorTensorxmmbEqnGHP}-\eqref{eqn:CorrectorTensorxnnEqnGHP}, are given by
\begin{align}
    \label{eqn:N}
    \mathcal{N} &= \frac{1}{2} \bracket{\pth - \rho} \eth,\\
    \mathcal{U} &= \frac{1}{2}\left[\eth' \eth + \eth \eth'-2\Psi_2 + (\pth'-2\rho') \rho + (\pth-2\rho)\rho' \right. \nonumber\\
    &\quad \left. +\rho (3\pth'-2\rho')+\rho' (3\pth-2\rho)-2\pth' \pth+2\rho\rho' \right],\\
    \mathfrak{V} &= \frac{1}{2} \bracket{\pth-4\rho} \eth',\\
    \overline{\mathfrak{V}} &= \frac{1}{2} \bracket{\pth-4\rho} \eth.
    \label{eqn:Vb}
\end{align}

So far, the discussion has been kept general. We now specialise to the Kinnersley tetrad. In $t$-slicing coordinates, the Kinnersley tetrad reads
\begin{align} \label{eq:tetrad,Kinn,t}
    l^\alpha &= \frac{1}{f} \biggl\{1,f,0,0  \biggr\},\\
    n^\alpha &= \frac{1}{2} \biggl\{1,-f,0,0  \biggr\},\\
    m^\alpha &= \frac{1}{\sqrt{2}\, r} \left\{0,0,1,\frac{i}{\sin \theta}  \right\}.
\end{align}
In $u$-slicing coordinates, the Kinnersley tetrad reads instead,
\begin{align}
    l^\alpha &= \biggl\{0,1,0,0  \biggr\},\\
    n^\alpha &= \biggl\{1,-\frac{f}{2},0,0  \biggr\},\\
    m^\alpha &= \frac{1}{\sqrt{2}\, r} \left\{0,0,1,\frac{i}{\sin \theta}  \right\}.
\end{align}



The scalar quantities \eqref{eqn:kappa}-\eqref{eqn:taup} in the Kinnersley tetrad are
\begin{align}
    \beta &= \beta' = \frac{\cot \theta}{2 \sqrt{2}r},\\
    \epsilon &=  0,\\
    \epsilon' &=  -\frac{M}{2r^2},\\
    \kappa' &= \kappa = 0,\\
    \sigma' &= \sigma = 0,\\
    \rho &= -\frac{1}{r},\\
    \rho' &= -\frac{f}{2r},\\
    \tau' &= \tau = 0.
\end{align}

In $t$-slicing in the Kinnersley tetrad, the radial GHP operators are
\begin{align}
    \pth &= \frac{1}{f} \bracket{\pfrac{}{t} + f\pfrac{}{r}},\\
    \pth' &= \frac{1}{2} \bracket{\pfrac{}{t} - f\pfrac{}{r}} - \frac{p+q}{2}\frac{M}{r^2},
\end{align}
while in $u$-slicing, they are given by
\begin{align}
    \pth &= \pfrac{}{r},\\
    \pth' &= \bracket{\pfrac{}{u} - \frac{f}{2}\pfrac{}{r}} - \frac{p+q}{2}\frac{M}{r^2}.
\end{align}

The angular operators in the Kinnersley tetrad are (in both slicings)
\begin{align}
    \eth &= -\frac{1}{r} \prescript{}{-s}{\mathcal{L}^\dagger},\\
    \eth' &= -\frac{1}{r} \prescript{}{s}{\mathcal{L}},\\
    \prescript{}{s}{\mathcal{L}} &= -\frac{1}{\sqrt{2}} \bracket{\pfrac{}{\theta} - \frac{i}{\sin \theta}\pfrac{}{\varphi}+s\cot \theta},
\end{align}
where $s := (p-q)/2$ is the spin weight of the quantity that the operator acts on. The operators $\prescript{}{-s}{\mathcal{L}^\dagger}$ and $\prescript{}{s}{\mathcal{L}}$ serve to, respectively, raise and lower the spin-weight of the harmonics they act on,
\begin{align}
\prescript{}{-s}{\mathcal{L}^\dagger} \prescript{}{s}{Y_{lm}} &= \sqrt{\frac{(\ell-s) (\ell+s+1)}{2}} \prescript{}{s+1}{Y_{lm}}, \\
\prescript{}{s}{\mathcal{L}} \prescript{}{s}{Y_{lm}} &= - \sqrt{\frac{(\ell+s) (\ell-s+1)}{2}} \prescript{}{s-1}{Y_{lm}}.
\end{align}

\section{Tetrad to BLS basis}
\label{section:TetradToBLS}

Below, we give the formulas relating the modes of the (not trace-reversed) metric perturbation in the Kinnersley tetrad basis to its BLS quantities $h_{i\ell m}$, $i=1,\ldots 10$. See Table 1, page 29 in Ref.~\cite{Pound2021} for comparison. With $\ell m$ mode indices suppressed, the relations are
\begingroup
\allowdisplaybreaks
\begin{align}
    h_{ll} &= \frac{h_1+h_2}{r f^2},\\
    h_{ln} &= \frac{h_3}{2r},\\
    h_{lm} &= -\frac{h_4+h_5-i(h_8+h_9)}{2rf\sqrt{2\ell(\ell+1)}},\\
    h_{l\bar{m}} &= \frac{h_4+h_5+i(h_8+h_9)}{2rf\sqrt{2\ell(\ell+1)}},\\
    h_{nn} &= \frac{h_1-h_2}{4r},\\
    h_{nm} &= \frac{-h_4+h_5+i(h_8-h_9)}{4r\sqrt{2\ell(\ell+1)}},\\
    h_{n\bar{m}} &= \frac{h_4-h_5+i(h_8-h_9)}{4r\sqrt{2\ell(\ell+1)}},\\
    h_{mm} &= \frac{h_7-ih_{10}}{2r\sqrt{(\ell-1)\ell(\ell+1)(\ell+2)}},\\
    h_{m\bar{m}} &= \frac{h_6}{2r},\\
    h_{\bar{m}\bar{m}} &= \frac{h_7+ih_{10}}{2r\sqrt{(\ell-1)\ell(\ell+1)(\ell+2)}}.
\end{align}
\endgroup

\section{Modes of the no-string Hertz potential}\label{App:static}

The modes of the no-string Hertz potentials $\Phi^\pm$ are obtained by solving the vacuum adjoint Teukolsky equation ${\cal O}^\dagger\Phi^\pm=0$ in the two regions outside the worldtube ($r<r_{\rm min}$ for $\Phi^-$ and $r>r_{\rm max}$ for $\Phi^+$). We obtain these as solutions to the spin-weight $-2$ Teukolsky equation, using the relationship ${\cal O}^\dagger\Phi = r^4{\cal O}'(r^{-4}\Phi)$; cf. Eq.~\eqref{eqn:TeukolskyOadj}. That relationship implies $_{-2}\Phi_{\ell m}$ (when multiplied by $e^{-im\Omega_\varphi r_\star}$ to account for our use of $u$ slicing) satisfies the radial Teukolsky equation~\eqref{eqn:RadialTeukolsky}.
Recall that $\psi^{ret}_0$ is a solution to the $s=2$ Teukolsky equation, which is written as a linear combination of so-called ``in'' and ``up'' basis functions as in Eq.~\eqref{eqn:PsiRet}. In analogy with ${}_2\psi^\pm_{\ell m}$ in Eq.~\eqref{eqn:PsiRet}, $_{-2}\Phi^\pm_{\ell m}$ are proportional to the $s=-2$ ``in'' and ``up'' basis solutions:
\begin{align}
_{-2}\Phi^+_{\ell m} &= {}_{-2}a^+_{\ell m} {}_{-2}\psi^{up}_{\ell m}(r)e^{im\Omega_\varphi r_*} \\
_{-2}\Phi^-_{\ell m} &= {}_{-2}a^-_{\ell m} {}_{-2}\psi^{in}_{\ell m}(r)e^{im\Omega_\varphi r_*},
\end{align}
where the exponential accounts for our use of $u$ slicing. 

Following Ori~\cite{Ori2002}, we can determine the coefficients ${}_{-2}a^\pm_{\ell m}$ by evaluating the ODE~\eqref{eq:inv rln mode} in a neighbourhood of the horizon or null infinity, allowing us to relate ${}_{-2}a^\pm_{\ell m}$ to the coefficients ${}_{2}C^\pm_{\ell m}$ appearing in Eq.~\eqref{eqn:PsiRet}.
In half of cases, doing this by directly expanding Eq.~\eqref{eq:inv rln mode} requires analytically computing the basis solutions to fourth or fifth order in the $1/r$ or $(r-2M)$ expansion, essentially because Eq.~\eqref{eq:inv rln mode} is a fourth-order ODE. In these cases, the procedure is facilitated by using the Teukolsky-Starobinsky identities to relate the $s=\pm 2$ basis solutions to each other. 

Ori only carried out this procedure for $\omega\neq0$, but the same method can be used in the static case, as detailed in Ref.~\cite{van2015metric}. However, there are some typographical errors in 
that paper (which we detail at the end of this appendix). We therefore will derive Eq.~\eqref{eq:PhiRet,static} for the static modes of the Hertz potential here.



Let us first start with generic modes (i.e., not necessarily static).
In $t$-slicing and for the retarded fields, Eq.~\eqref{eq:inv rln mode} is re-written as
\begin{equation}\label{eq:inv rln mode,t-sl}
D_{m}^4{}_2\bar\Phi^{ret}_{\ell m}(r)=2 {}_2\psi^{ret}_{\ell m}(r),
\end{equation}
where
\begin{equation}
D_{m}:= \frac{d}{dr}-i\frac{m\Omega_{\varphi}}{f(r)}.
\end{equation}
The Hertz potential $\Phi^{ret}$ satisfies the spin $s=-2$ Teukolsky equation, thus its radial modes $_{-2}\Phi^{ret}_{\ell m}(r)$ satisfy the spin $s=-2$ radial Teukolsky equation.
The Teukolsky-Starobinsky identities~\cite{Starobinskil:1974nkd,teukolsky1974perturbations}
yield
\begin{equation}\label{eq:T-S}
p\, {}_2\bar\Phi^{ret}_{\ell m}(r)=
\Delta^2 \left(D^\star_{m}\right)^4\Delta^2 \left(D_{m}\right)^4{}_2\bar\Phi^{ret}_{\ell m}(r),
\end{equation}
where $p$ is given in Eq.~\eqref{eq:p}. Thus, applying the operator 
$\Delta^2 \left(D^\star_{m}\right)^4\Delta^2$ 
to Eq.~\eqref{eq:inv rln mode,t-sl}, we obtain
\begin{equation}\label{eq:T-S,Phi-psi}
p\, {}_2\bar\Phi^{ret}_{\ell m}(r)=2\Delta^2 \left(D^\star_{m}\right)^4\Delta^2{}_2\psi^{ret}_{\ell m}(r).
\end{equation}

To relate the coefficients $_{-2}a^-_{\ell m}$ in $_{-2}\Phi^-_{\ell m}$ to the coefficients $_2C^-_{\ell m}$ in $_{2}\psi^-_{\ell m}$, we can simply substitute the leading asymptotic near-horizon behavior of the ``in'' solutions in Eq.~\eqref{eq:inv rln mode,t-sl}. Doing the same for the ``up'' solutions near infinity would require four orders in $1/r$; we then instead use Eq.~\eqref{eq:T-S,Phi-psi}, which requires only the leading terms in the asymptotic forms of $_s\psi^{up}_{\ell m}$ to determine the coefficient $_{-2}a^+_{\ell m}$. The immediate outcome of these calculations is shown in Eqs.~\eqref{eqn:PhiPlus} and~\eqref{eqn:PhiMinus}. 

Let us turn to the static case, $m=0$. In this case, obviously, there is no difference between $u$-slicing and $t$-slicing. The radial Teukolsky solutions are easy to find in closed form: 
\begin{align}\label{eq:psi^in,m=0}
{}_s\psi^{in}_{\ell 0}(r)&=
\left(\ell+s+1\right)_{-2s}\frac{\Gamma(1+s)}{\Gamma(1-s)}\left(\frac{r}{2M}-1\right)^{-s}\times
\nonumber
\\ &{}_2F_1\left(-\ell,\ell+1,1-s,1-\frac{r}{2M}\right),\quad \text{if}\ s>0,
\\
{}_s\psi^{in}_{\ell 0}(r)&=
2^{-2s}
\left(\frac{r}{2M}-1\right)^{-s}\times
\nonumber
\\ &{}_2F_1\left(-\ell,\ell+1,1-s,1-\frac{r}{2M}\right),\quad \text{if}\ s\leq 0,
\\
{}_s\psi^{up}_{\ell 0}(r)&=
2^{-s-\ell-1}
\left(\frac{r}{2M}-1\right)^{-s}\left(\frac{r}{2M}\right)^{-\ell-1}
\times
\label{eq:psi^up,m=0}
\\&
{}_2F_1\left(\ell+1,\ell+1-s,2\ell+2,\frac{2M}{r}\right),\quad \forall s.
\nonumber
\end{align}
The specific choices of  hypergeometric functions for the solutions are made so that ${}_s\psi^{in}_{\ell 0}$ is regular at $r=2M$  and ${}_s\psi^{up}_{\ell 0}$ is regular at $r=\infty$;
here we have used the normalization choices made in the BHPToolkit. Note that any normalization choices are without loss of generality since the normalizations are taken care of by the Wronskian in Eqs.~\eqref{eq:Wronsk}--\eqref{eq:Cin/up}, which is equal to 
\begin{align}
   {}_sW(r)&=-\frac{M^{1+2s}2^{\ell+|s|+1} \Gamma \left(\ell+\frac{3}{2}\right) \left(|s|\right)! \Delta^{-1-s}}{\sqrt{\pi }\,
   (\ell+|s|)!}
\end{align}
for the static solutions in Eqs.~\eqref{eq:psi^in,m=0}-\eqref{eq:psi^up,m=0} for any spin $s$.
It is easy to check that
\begin{align}\label{eq:rln psi2-psi-2}
2\Delta^2 \left(D_{0}\right)^4\Delta^2{}_2\psi^{in/up}_{\ell 0}(r)=
2 M^4 \, p^{1/2}\,{}_{-2}\psi^{in/up}_{\ell 0}(r)
\end{align}
with $p=\left(\left(\ell-1\right)_4\right)^2$.
Since, from Eq.~\eqref{eqn:PsiRet},
\begin{equation}
{}_2\psi_{\ell 0}^{ret}(r) = 
\begin{cases} 
      {}_2C^{-}_{\ell0}\, {}_2\psi_{\ell 0}^{in}(r) & r < \rmin, \\
     {}_2C^{+}_{\ell 0}\, {}_2\psi_{\ell 0}^{up}(r)& r > \rmax,
\end{cases}
\end{equation}
it follows from Eq.~\eqref{eq:T-S,Phi-psi} and using Eq.~\eqref{eq:rln psi2-psi-2} (together with $D^\star_{0}=D_{0}$) 
that
\begin{equation}\label{eq:BarPhi-Psi-2}
\begin{split}
{}_2\bar\Phi^{ret}_{\ell 0}(r)&=\frac{2 M^4}{\left(\ell-1\right)_4}{}_2 C^{+}_{\ell 0}{}_{-2}\psi_{\ell 0}^{up}(r),\quad r > \rmax,\\
{}_2\bar\Phi^{ret}_{\ell 0}(r)&=\frac{2 M^4}{\left(\ell-1\right)_4}{}_2 C^{-}_{\ell 0}{}_{-2}\psi_{\ell 0}^{in}(r),\quad r < \rmin. \end{split}
\end{equation}
This is Eq.~\eqref{eq:PhiRet,static}.

We finish by detailing some typographical errors in the static-mode section (Sec.~III.B)
and the section for the modes of the Hertz potential (Sec.~III.C)
of Ref.~\cite{van2015metric}\footnote{We note that some typographical errors in the Phys.~Rev.~D version of Ref.~\cite{van2015metric} were corrected in its arXiv version and so we do not include those here. That is,  we only include here the typographical errors which we believe are still present in both the Phys.~Rev.~D and current arXiv versions.}. In the following, all equation numbers correspond to Ref.~\cite{van2015metric}. 
\begin{enumerate}
\item
In Eq.~(54) there is a factor $1/2!$ missing in front of $\left(3-s\right)_{s-2}$. Furthermore, the coefficient of the ${}_sB_{lmn}$ is denoted by ``divergent terms" but these terms are not actually divergent for spin $s=-1,-2$. 

\item
In Eq.~(55), the term with ${}_sB_{lmn}$ inside the parenthesis should instead read:
\begin{equation}
\dots+(-1)^s{}_sB_{lmn}\frac{\sqrt{\pi}\left(l+\frac{3}{2}\right)_{s-\frac{1}{2}}}{2^{2l+2}}+\dots
\end{equation}



\item
In Eq.~(62),
the signs of $s$ in the two Pochhammer functions in the numerators  are wrong;
also, the `+' after ${}_sB_{lm}$ should not be there.

\item
The boundary conditions should be reversed on one side of Eq.~(79), so that it reads:
\begin{equation}
\Delta^s \,{_{s}\bar R_{lmn}^{\pm}} = \,{_{-s}R_{lmn}^{\not{\pm}}}.
\end{equation}


\item

On the LHS of Eq.~(81), the $\Delta^2$ term should be outside the operator $D_{mn}^4$.



\end{enumerate}

\nocite{*}
\bibliography{bibfile}

\begin{thebibliography}{85}%
\makeatletter
\providecommand \@ifxundefined [1]{%
 \@ifx{#1\undefined}
}%
\providecommand \@ifnum [1]{%
 \ifnum #1\expandafter \@firstoftwo
 \else \expandafter \@secondoftwo
 \fi
}%
\providecommand \@ifx [1]{%
 \ifx #1\expandafter \@firstoftwo
 \else \expandafter \@secondoftwo
 \fi
}%
\providecommand \natexlab [1]{#1}%
\providecommand \enquote  [1]{``#1''}%
\providecommand \bibnamefont  [1]{#1}%
\providecommand \bibfnamefont [1]{#1}%
\providecommand \citenamefont [1]{#1}%
\providecommand \href@noop [0]{\@secondoftwo}%
\providecommand \href [0]{\begingroup \@sanitize@url \@href}%
\providecommand \@href[1]{\@@startlink{#1}\@@href}%
\providecommand \@@href[1]{\endgroup#1\@@endlink}%
\providecommand \@sanitize@url [0]{\catcode `\\12\catcode `\$12\catcode
  `\&12\catcode `\#12\catcode `\^12\catcode `\_12\catcode `\%12\relax}%
\providecommand \@@startlink[1]{}%
\providecommand \@@endlink[0]{}%
\providecommand \url  [0]{\begingroup\@sanitize@url \@url }%
\providecommand \@url [1]{\endgroup\@href {#1}{\urlprefix }}%
\providecommand \urlprefix  [0]{URL }%
\providecommand \Eprint [0]{\href }%
\providecommand \doibase [0]{http://dx.doi.org/}%
\providecommand \selectlanguage [0]{\@gobble}%
\providecommand \bibinfo  [0]{\@secondoftwo}%
\providecommand \bibfield  [0]{\@secondoftwo}%
\providecommand \translation [1]{[#1]}%
\providecommand \BibitemOpen [0]{}%
\providecommand \bibitemStop [0]{}%
\providecommand \bibitemNoStop [0]{.\EOS\space}%
\providecommand \EOS [0]{\spacefactor3000\relax}%
\providecommand \BibitemShut  [1]{\csname bibitem#1\endcsname}%
\let\auto@bib@innerbib\@empty
\bibitem [{\citenamefont {Abbott}\ \emph {et~al.}(2016)\citenamefont {Abbott}
  \emph {et~al.}}]{LIGOScientific2016}%
  \BibitemOpen
  \bibfield  {author} {\bibinfo {author} {\bibfnamefont {B.~P.}\ \bibnamefont
  {Abbott}} \emph {et~al.} (\bibinfo {collaboration} {LIGO Scientific,
  Virgo}),\ }\bibfield  {title} {\enquote {\bibinfo {title} {{Observation of
  Gravitational Waves from a Binary Black Hole Merger}},}\ }\href {\doibase
  10.1103/PhysRevLett.116.061102} {\bibfield  {journal} {\bibinfo  {journal}
  {Phys. Rev. Lett.}\ }\textbf {\bibinfo {volume} {116}},\ \bibinfo {pages}
  {061102} (\bibinfo {year} {2016})},\ \Eprint
  {http://arxiv.org/abs/1602.03837} {arXiv:1602.03837 [gr-qc]} \BibitemShut
  {NoStop}%
\bibitem [{\citenamefont {Audley}\ \emph {et~al.}(2017)\citenamefont {Audley}
  \emph {et~al.}}]{LISA}%
  \BibitemOpen
  \bibfield  {author} {\bibinfo {author} {\bibfnamefont {Heather}\ \bibnamefont
  {Audley}} \emph {et~al.} (\bibinfo {collaboration} {LISA}),\ }\bibfield
  {title} {\enquote {\bibinfo {title} {{Laser Interferometer Space Antenna}},}\
  }\href@noop {} {\  (\bibinfo {year} {2017})},\ \Eprint
  {http://arxiv.org/abs/1702.00786} {arXiv:1702.00786 [astro-ph.IM]}
  \BibitemShut {NoStop}%
\bibitem [{LIS()}]{LISA-LaunchDate}%
  \BibitemOpen
  \href@noop {} {\enquote {\bibinfo {title} {{LISA} mission summary},}\
  }\bibinfo {note}
  {\url{https://sci.esa.int/web/lisa/-/61367-mission-summary}}\BibitemShut
  {NoStop}%
\bibitem [{\citenamefont {Babak}\ \emph {et~al.}(2017)\citenamefont {Babak},
  \citenamefont {Gair}, \citenamefont {Sesana}, \citenamefont {Barausse},
  \citenamefont {Sopuerta}, \citenamefont {Berry}, \citenamefont {Berti},
  \citenamefont {Amaro-Seoane}, \citenamefont {Petiteau},\ and\ \citenamefont
  {Klein}}]{Babak2017}%
  \BibitemOpen
  \bibfield  {author} {\bibinfo {author} {\bibfnamefont {Stanislav}\
  \bibnamefont {Babak}}, \bibinfo {author} {\bibfnamefont {Jonathan}\
  \bibnamefont {Gair}}, \bibinfo {author} {\bibfnamefont {Alberto}\
  \bibnamefont {Sesana}}, \bibinfo {author} {\bibfnamefont {Enrico}\
  \bibnamefont {Barausse}}, \bibinfo {author} {\bibfnamefont {Carlos~F.}\
  \bibnamefont {Sopuerta}}, \bibinfo {author} {\bibfnamefont {Christopher
  P.~L.}\ \bibnamefont {Berry}}, \bibinfo {author} {\bibfnamefont {Emanuele}\
  \bibnamefont {Berti}}, \bibinfo {author} {\bibfnamefont {Pau}\ \bibnamefont
  {Amaro-Seoane}}, \bibinfo {author} {\bibfnamefont {Antoine}\ \bibnamefont
  {Petiteau}}, \ and\ \bibinfo {author} {\bibfnamefont {Antoine}\ \bibnamefont
  {Klein}},\ }\bibfield  {title} {\enquote {\bibinfo {title} {{Science with the
  space-based interferometer LISA. V: Extreme mass-ratio inspirals}},}\ }\href
  {\doibase 10.1103/PhysRevD.95.103012} {\bibfield  {journal} {\bibinfo
  {journal} {Phys. Rev. D}\ }\textbf {\bibinfo {volume} {95}},\ \bibinfo
  {pages} {103012} (\bibinfo {year} {2017})},\ \Eprint
  {http://arxiv.org/abs/1703.09722} {arXiv:1703.09722 [gr-qc]} \BibitemShut
  {NoStop}%
\bibitem [{\citenamefont {Barausse}\ \emph {et~al.}(2020)\citenamefont
  {Barausse} \emph {et~al.}}]{Barausse2020}%
  \BibitemOpen
  \bibfield  {author} {\bibinfo {author} {\bibfnamefont {Enrico}\ \bibnamefont
  {Barausse}} \emph {et~al.},\ }\bibfield  {title} {\enquote {\bibinfo {title}
  {{Prospects for Fundamental Physics with LISA}},}\ }\href {\doibase
  10.1007/s10714-020-02691-1} {\bibfield  {journal} {\bibinfo  {journal} {Gen.
  Rel. Grav.}\ }\textbf {\bibinfo {volume} {52}},\ \bibinfo {pages} {81}
  (\bibinfo {year} {2020})},\ \Eprint {http://arxiv.org/abs/2001.09793}
  {arXiv:2001.09793 [gr-qc]} \BibitemShut {NoStop}%
\bibitem [{\citenamefont {Barack}\ and\ \citenamefont
  {Pound}(2019)}]{Barack2018}%
  \BibitemOpen
  \bibfield  {author} {\bibinfo {author} {\bibfnamefont {Leor}\ \bibnamefont
  {Barack}}\ and\ \bibinfo {author} {\bibfnamefont {Adam}\ \bibnamefont
  {Pound}},\ }\bibfield  {title} {\enquote {\bibinfo {title} {{Self-force and
  radiation reaction in general relativity}},}\ }\href {\doibase
  10.1088/1361-6633/aae552} {\bibfield  {journal} {\bibinfo  {journal} {Rept.
  Prog. Phys.}\ }\textbf {\bibinfo {volume} {82}},\ \bibinfo {pages} {016904}
  (\bibinfo {year} {2019})},\ \Eprint {http://arxiv.org/abs/1805.10385}
  {arXiv:1805.10385 [gr-qc]} \BibitemShut {NoStop}%
\bibitem [{\citenamefont {Pound}\ and\ \citenamefont
  {Wardell}(2021)}]{Pound2021}%
  \BibitemOpen
  \bibfield  {author} {\bibinfo {author} {\bibfnamefont {Adam}\ \bibnamefont
  {Pound}}\ and\ \bibinfo {author} {\bibfnamefont {Barry}\ \bibnamefont
  {Wardell}},\ }\bibfield  {title} {\enquote {\bibinfo {title} {{Black hole
  perturbation theory and gravitational self-force}},}\ }\href {\doibase
  10.1007/978-981-15-4702-7\_38-1} {\  (\bibinfo {year} {2021}),\
  10.1007/978-981-15-4702-7\_38-1},\ \Eprint {http://arxiv.org/abs/2101.04592}
  {arXiv:2101.04592 [gr-qc]} \BibitemShut {NoStop}%
\bibitem [{\citenamefont {Wardell}\ \emph {et~al.}(2023)\citenamefont
  {Wardell}, \citenamefont {Pound}, \citenamefont {Warburton}, \citenamefont
  {Miller}, \citenamefont {Durkan},\ and\ \citenamefont
  {Le~Tiec}}]{Wardell:2021fyy}%
  \BibitemOpen
  \bibfield  {author} {\bibinfo {author} {\bibfnamefont {Barry}\ \bibnamefont
  {Wardell}}, \bibinfo {author} {\bibfnamefont {Adam}\ \bibnamefont {Pound}},
  \bibinfo {author} {\bibfnamefont {Niels}\ \bibnamefont {Warburton}}, \bibinfo
  {author} {\bibfnamefont {Jeremy}\ \bibnamefont {Miller}}, \bibinfo {author}
  {\bibfnamefont {Leanne}\ \bibnamefont {Durkan}}, \ and\ \bibinfo {author}
  {\bibfnamefont {Alexandre}\ \bibnamefont {Le~Tiec}},\ }\bibfield  {title}
  {\enquote {\bibinfo {title} {{Gravitational Waveforms for Compact Binaries
  from Second-Order Self-Force Theory}},}\ }\href {\doibase
  10.1103/PhysRevLett.130.241402} {\bibfield  {journal} {\bibinfo  {journal}
  {Phys. Rev. Lett.}\ }\textbf {\bibinfo {volume} {130}},\ \bibinfo {pages}
  {241402} (\bibinfo {year} {2023})},\ \Eprint
  {http://arxiv.org/abs/2112.12265} {arXiv:2112.12265 [gr-qc]} \BibitemShut
  {NoStop}%
\bibitem [{\citenamefont {Albertini}\ \emph {et~al.}(2022)\citenamefont
  {Albertini}, \citenamefont {Nagar}, \citenamefont {Pound}, \citenamefont
  {Warburton}, \citenamefont {Wardell}, \citenamefont {Durkan},\ and\
  \citenamefont {Miller}}]{Albertini:2022rfe}%
  \BibitemOpen
  \bibfield  {author} {\bibinfo {author} {\bibfnamefont {Angelica}\
  \bibnamefont {Albertini}}, \bibinfo {author} {\bibfnamefont {Alessandro}\
  \bibnamefont {Nagar}}, \bibinfo {author} {\bibfnamefont {Adam}\ \bibnamefont
  {Pound}}, \bibinfo {author} {\bibfnamefont {Niels}\ \bibnamefont
  {Warburton}}, \bibinfo {author} {\bibfnamefont {Barry}\ \bibnamefont
  {Wardell}}, \bibinfo {author} {\bibfnamefont {Leanne}\ \bibnamefont
  {Durkan}}, \ and\ \bibinfo {author} {\bibfnamefont {Jeremy}\ \bibnamefont
  {Miller}},\ }\bibfield  {title} {\enquote {\bibinfo {title} {{Comparing
  second-order gravitational self-force, numerical relativity, and effective
  one body waveforms from inspiralling, quasicircular, and nonspinning black
  hole binaries}},}\ }\href {\doibase 10.1103/PhysRevD.106.084061} {\bibfield
  {journal} {\bibinfo  {journal} {Phys. Rev. D}\ }\textbf {\bibinfo {volume}
  {106}},\ \bibinfo {pages} {084061} (\bibinfo {year} {2022})},\ \Eprint
  {http://arxiv.org/abs/2208.01049} {arXiv:2208.01049 [gr-qc]} \BibitemShut
  {NoStop}%
\bibitem [{\citenamefont {van~de Meent}(2016)}]{vanDeMeent2016}%
  \BibitemOpen
  \bibfield  {author} {\bibinfo {author} {\bibfnamefont {Maarten}\ \bibnamefont
  {van~de Meent}},\ }\bibfield  {title} {\enquote {\bibinfo {title}
  {{Gravitational self-force on eccentric equatorial orbits around a Kerr black
  hole}},}\ }\href {\doibase 10.1103/PhysRevD.94.044034} {\bibfield  {journal}
  {\bibinfo  {journal} {Phys. Rev. D}\ }\textbf {\bibinfo {volume} {94}},\
  \bibinfo {pages} {044034} (\bibinfo {year} {2016})},\ \Eprint
  {http://arxiv.org/abs/1606.06297} {arXiv:1606.06297 [gr-qc]} \BibitemShut
  {NoStop}%
\bibitem [{\citenamefont {van De~Meent}(2017)}]{vanDeMeent2017}%
  \BibitemOpen
  \bibfield  {author} {\bibinfo {author} {\bibfnamefont {Maarten}\ \bibnamefont
  {van De~Meent}},\ }\bibfield  {title} {\enquote {\bibinfo {title} {{The mass
  and angular momentum of reconstructed metric perturbations}},}\ }\href
  {\doibase 10.1088/1361-6382/aa71c3} {\bibfield  {journal} {\bibinfo
  {journal} {Class. Quant. Grav.}\ }\textbf {\bibinfo {volume} {34}},\ \bibinfo
  {pages} {124003} (\bibinfo {year} {2017})},\ \Eprint
  {http://arxiv.org/abs/1702.00969} {arXiv:1702.00969 [gr-qc]} \BibitemShut
  {NoStop}%
\bibitem [{\citenamefont {Hinderer}\ and\ \citenamefont
  {Flanagan}(2008)}]{Hinderer2008}%
  \BibitemOpen
  \bibfield  {author} {\bibinfo {author} {\bibfnamefont {Tanja}\ \bibnamefont
  {Hinderer}}\ and\ \bibinfo {author} {\bibfnamefont {Eanna~E.}\ \bibnamefont
  {Flanagan}},\ }\bibfield  {title} {\enquote {\bibinfo {title} {{Two timescale
  analysis of extreme mass ratio inspirals in Kerr. I. Orbital Motion}},}\
  }\href {\doibase 10.1103/PhysRevD.78.064028} {\bibfield  {journal} {\bibinfo
  {journal} {Phys. Rev. D}\ }\textbf {\bibinfo {volume} {78}},\ \bibinfo
  {pages} {064028} (\bibinfo {year} {2008})},\ \Eprint
  {http://arxiv.org/abs/0805.3337} {arXiv:0805.3337 [gr-qc]} \BibitemShut
  {NoStop}%
\bibitem [{\citenamefont {Burke}\ \emph {et~al.}(2023)\citenamefont {Burke},
  \citenamefont {Piovano}, \citenamefont {Warburton}, \citenamefont {Lynch},
  \citenamefont {Speri}, \citenamefont {Kavanagh}, \citenamefont {Wardell},
  \citenamefont {Pound}, \citenamefont {Durkan},\ and\ \citenamefont
  {Miller}}]{Burke:2023lno}%
  \BibitemOpen
  \bibfield  {author} {\bibinfo {author} {\bibfnamefont {Ollie}\ \bibnamefont
  {Burke}}, \bibinfo {author} {\bibfnamefont {Gabriel~Andres}\ \bibnamefont
  {Piovano}}, \bibinfo {author} {\bibfnamefont {Niels}\ \bibnamefont
  {Warburton}}, \bibinfo {author} {\bibfnamefont {Philip}\ \bibnamefont
  {Lynch}}, \bibinfo {author} {\bibfnamefont {Lorenzo}\ \bibnamefont {Speri}},
  \bibinfo {author} {\bibfnamefont {Chris}\ \bibnamefont {Kavanagh}}, \bibinfo
  {author} {\bibfnamefont {Barry}\ \bibnamefont {Wardell}}, \bibinfo {author}
  {\bibfnamefont {Adam}\ \bibnamefont {Pound}}, \bibinfo {author}
  {\bibfnamefont {Leanne}\ \bibnamefont {Durkan}}, \ and\ \bibinfo {author}
  {\bibfnamefont {Jeremy}\ \bibnamefont {Miller}},\ }\bibfield  {title}
  {\enquote {\bibinfo {title} {{Accuracy Requirements: Assessing the Importance
  of First Post-Adiabatic Terms for Small-Mass-Ratio Binaries}},}\ }\href@noop
  {} {\  (\bibinfo {year} {2023})},\ \Eprint {http://arxiv.org/abs/2310.08927}
  {arXiv:2310.08927 [gr-qc]} \BibitemShut {NoStop}%
\bibitem [{\citenamefont {Miller}\ and\ \citenamefont
  {Pound}(2021)}]{Miller2020}%
  \BibitemOpen
  \bibfield  {author} {\bibinfo {author} {\bibfnamefont {Jeremy}\ \bibnamefont
  {Miller}}\ and\ \bibinfo {author} {\bibfnamefont {Adam}\ \bibnamefont
  {Pound}},\ }\bibfield  {title} {\enquote {\bibinfo {title} {{Two-timescale
  evolution of extreme-mass-ratio inspirals: waveform generation scheme for
  quasicircular orbits in Schwarzschild spacetime}},}\ }\href {\doibase
  10.1103/PhysRevD.103.064048} {\bibfield  {journal} {\bibinfo  {journal}
  {Phys. Rev. D}\ }\textbf {\bibinfo {volume} {103}},\ \bibinfo {pages}
  {064048} (\bibinfo {year} {2021})},\ \Eprint
  {http://arxiv.org/abs/2006.11263} {arXiv:2006.11263 [gr-qc]} \BibitemShut
  {NoStop}%
\bibitem [{\citenamefont {Arun}\ \emph {et~al.}(2022)\citenamefont {Arun} \emph
  {et~al.}}]{LISA:2022kgy}%
  \BibitemOpen
  \bibfield  {author} {\bibinfo {author} {\bibfnamefont {K.~G.}\ \bibnamefont
  {Arun}} \emph {et~al.} (\bibinfo {collaboration} {LISA}),\ }\bibfield
  {title} {\enquote {\bibinfo {title} {{New horizons for fundamental physics
  with LISA}},}\ }\href {\doibase 10.1007/s41114-022-00036-9} {\bibfield
  {journal} {\bibinfo  {journal} {Living Rev. Rel.}\ }\textbf {\bibinfo
  {volume} {25}},\ \bibinfo {pages} {4} (\bibinfo {year} {2022})},\ \Eprint
  {http://arxiv.org/abs/2205.01597} {arXiv:2205.01597 [gr-qc]} \BibitemShut
  {NoStop}%
\bibitem [{\citenamefont {Seoane}\ \emph {et~al.}(2023)\citenamefont {Seoane}
  \emph {et~al.}}]{LISA:2022yao}%
  \BibitemOpen
  \bibfield  {author} {\bibinfo {author} {\bibfnamefont {Pau~Amaro}\
  \bibnamefont {Seoane}} \emph {et~al.} (\bibinfo {collaboration} {LISA}),\
  }\bibfield  {title} {\enquote {\bibinfo {title} {{Astrophysics with the Laser
  Interferometer Space Antenna}},}\ }\href {\doibase
  10.1007/s41114-022-00041-y} {\bibfield  {journal} {\bibinfo  {journal}
  {Living Rev. Rel.}\ }\textbf {\bibinfo {volume} {26}},\ \bibinfo {pages} {2}
  (\bibinfo {year} {2023})},\ \Eprint {http://arxiv.org/abs/2203.06016}
  {arXiv:2203.06016 [gr-qc]} \BibitemShut {NoStop}%
\bibitem [{\citenamefont {Afshordi}\ \emph {et~al.}(2023)\citenamefont
  {Afshordi} \emph {et~al.}}]{LISAConsortiumWaveformWorkingGroup:2023arg}%
  \BibitemOpen
  \bibfield  {author} {\bibinfo {author} {\bibfnamefont {Niayesh}\ \bibnamefont
  {Afshordi}} \emph {et~al.} (\bibinfo {collaboration} {LISA Consortium
  Waveform Working Group}),\ }\bibfield  {title} {\enquote {\bibinfo {title}
  {{Waveform Modelling for the Laser Interferometer Space Antenna}},}\
  }\href@noop {} {\  (\bibinfo {year} {2023})},\ \Eprint
  {http://arxiv.org/abs/2311.01300} {arXiv:2311.01300 [gr-qc]} \BibitemShut
  {NoStop}%
\bibitem [{\citenamefont {Colpi}\ \emph {et~al.}(2024)\citenamefont {Colpi}
  \emph {et~al.}}]{Colpi:2024xhw}%
  \BibitemOpen
  \bibfield  {author} {\bibinfo {author} {\bibfnamefont {Monica}\ \bibnamefont
  {Colpi}} \emph {et~al.},\ }\bibfield  {title} {\enquote {\bibinfo {title}
  {{LISA Definition Study Report}},}\ }\href@noop {} {\  (\bibinfo {year}
  {2024})},\ \Eprint {http://arxiv.org/abs/2402.07571} {arXiv:2402.07571
  [astro-ph.CO]} \BibitemShut {NoStop}%
\bibitem [{\citenamefont {Miller}\ \emph {et~al.}(2023)\citenamefont {Miller},
  \citenamefont {Leather}, \citenamefont {Pound},\ and\ \citenamefont
  {Warburton}}]{Miller:2023ers}%
  \BibitemOpen
  \bibfield  {author} {\bibinfo {author} {\bibfnamefont {Jeremy}\ \bibnamefont
  {Miller}}, \bibinfo {author} {\bibfnamefont {Benjamin}\ \bibnamefont
  {Leather}}, \bibinfo {author} {\bibfnamefont {Adam}\ \bibnamefont {Pound}}, \
  and\ \bibinfo {author} {\bibfnamefont {Niels}\ \bibnamefont {Warburton}},\
  }\bibfield  {title} {\enquote {\bibinfo {title} {{Worldtube puncture scheme
  for first- and second-order self-force calculations in the Fourier
  domain}},}\ }\href@noop {} {\  (\bibinfo {year} {2023})},\ \Eprint
  {http://arxiv.org/abs/2401.00455} {arXiv:2401.00455 [gr-qc]} \BibitemShut
  {NoStop}%
\bibitem [{\citenamefont {Katz}\ \emph {et~al.}(2021)\citenamefont {Katz},
  \citenamefont {Chua}, \citenamefont {Speri}, \citenamefont {Warburton},\ and\
  \citenamefont {Hughes}}]{Katz:2021yft}%
  \BibitemOpen
  \bibfield  {author} {\bibinfo {author} {\bibfnamefont {Michael~L.}\
  \bibnamefont {Katz}}, \bibinfo {author} {\bibfnamefont {Alvin J.~K.}\
  \bibnamefont {Chua}}, \bibinfo {author} {\bibfnamefont {Lorenzo}\
  \bibnamefont {Speri}}, \bibinfo {author} {\bibfnamefont {Niels}\ \bibnamefont
  {Warburton}}, \ and\ \bibinfo {author} {\bibfnamefont {Scott~A.}\
  \bibnamefont {Hughes}},\ }\bibfield  {title} {\enquote {\bibinfo {title}
  {{Fast extreme-mass-ratio-inspiral waveforms: New tools for millihertz
  gravitational-wave data analysis}},}\ }\href {\doibase
  10.1103/PhysRevD.104.064047} {\bibfield  {journal} {\bibinfo  {journal}
  {Phys. Rev. D}\ }\textbf {\bibinfo {volume} {104}},\ \bibinfo {pages}
  {064047} (\bibinfo {year} {2021})},\ \Eprint
  {http://arxiv.org/abs/2104.04582} {arXiv:2104.04582 [gr-qc]} \BibitemShut
  {NoStop}%
\bibitem [{\citenamefont {Pound}(2012{\natexlab{a}})}]{Pound:2012nt}%
  \BibitemOpen
  \bibfield  {author} {\bibinfo {author} {\bibfnamefont {Adam}\ \bibnamefont
  {Pound}},\ }\bibfield  {title} {\enquote {\bibinfo {title} {{Second-order
  gravitational self-force}},}\ }\href {\doibase
  10.1103/PhysRevLett.109.051101} {\bibfield  {journal} {\bibinfo  {journal}
  {Phys. Rev. Lett.}\ }\textbf {\bibinfo {volume} {109}},\ \bibinfo {pages}
  {051101} (\bibinfo {year} {2012}{\natexlab{a}})},\ \Eprint
  {http://arxiv.org/abs/1201.5089} {arXiv:1201.5089 [gr-qc]} \BibitemShut
  {NoStop}%
\bibitem [{\citenamefont {Pound}(2012{\natexlab{b}})}]{Pound2012}%
  \BibitemOpen
  \bibfield  {author} {\bibinfo {author} {\bibfnamefont {Adam}\ \bibnamefont
  {Pound}},\ }\bibfield  {title} {\enquote {\bibinfo {title} {Nonlinear
  gravitational self-force: Field outside a small body},}\ }\href {\doibase
  10.1103/physrevd.86.084019} {\bibfield  {journal} {\bibinfo  {journal} {Phys.
  Rev. D}\ }\textbf {\bibinfo {volume} {86}},\ \bibinfo {pages} {084019}
  (\bibinfo {year} {2012}{\natexlab{b}})},\ \Eprint
  {http://arxiv.org/abs/1206.6538} {arXiv:1206.6538 [gr-qc]} \BibitemShut
  {NoStop}%
\bibitem [{\citenamefont {Pound}(2017)}]{Pound:2017psq}%
  \BibitemOpen
  \bibfield  {author} {\bibinfo {author} {\bibfnamefont {Adam}\ \bibnamefont
  {Pound}},\ }\bibfield  {title} {\enquote {\bibinfo {title} {{Nonlinear
  gravitational self-force: second-order equation of motion}},}\ }\href
  {\doibase 10.1103/PhysRevD.95.104056} {\bibfield  {journal} {\bibinfo
  {journal} {Phys. Rev. D}\ }\textbf {\bibinfo {volume} {95}},\ \bibinfo
  {pages} {104056} (\bibinfo {year} {2017})},\ \Eprint
  {http://arxiv.org/abs/1703.02836} {arXiv:1703.02836 [gr-qc]} \BibitemShut
  {NoStop}%
\bibitem [{\citenamefont {Pound}\ \emph {et~al.}(2020)\citenamefont {Pound},
  \citenamefont {Wardell}, \citenamefont {Warburton},\ and\ \citenamefont
  {Miller}}]{Pound2019}%
  \BibitemOpen
  \bibfield  {author} {\bibinfo {author} {\bibfnamefont {Adam}\ \bibnamefont
  {Pound}}, \bibinfo {author} {\bibfnamefont {Barry}\ \bibnamefont {Wardell}},
  \bibinfo {author} {\bibfnamefont {Niels}\ \bibnamefont {Warburton}}, \ and\
  \bibinfo {author} {\bibfnamefont {Jeremy}\ \bibnamefont {Miller}},\
  }\bibfield  {title} {\enquote {\bibinfo {title} {{Second-Order Self-Force
  Calculation of Gravitational Binding Energy in Compact Binaries}},}\ }\href
  {\doibase 10.1103/PhysRevLett.124.021101} {\bibfield  {journal} {\bibinfo
  {journal} {Phys. Rev. Lett.}\ }\textbf {\bibinfo {volume} {124}},\ \bibinfo
  {pages} {021101} (\bibinfo {year} {2020})},\ \Eprint
  {http://arxiv.org/abs/1908.07419} {arXiv:1908.07419 [gr-qc]} \BibitemShut
  {NoStop}%
\bibitem [{\citenamefont {Warburton}\ \emph {et~al.}(2021)\citenamefont
  {Warburton}, \citenamefont {Pound}, \citenamefont {Wardell}, \citenamefont
  {Miller},\ and\ \citenamefont {Durkan}}]{Warburton2021}%
  \BibitemOpen
  \bibfield  {author} {\bibinfo {author} {\bibfnamefont {Niels}\ \bibnamefont
  {Warburton}}, \bibinfo {author} {\bibfnamefont {Adam}\ \bibnamefont {Pound}},
  \bibinfo {author} {\bibfnamefont {Barry}\ \bibnamefont {Wardell}}, \bibinfo
  {author} {\bibfnamefont {Jeremy}\ \bibnamefont {Miller}}, \ and\ \bibinfo
  {author} {\bibfnamefont {Leanne}\ \bibnamefont {Durkan}},\ }\bibfield
  {title} {\enquote {\bibinfo {title} {{Gravitational-Wave Energy Flux for
  Compact Binaries through Second Order in the Mass Ratio}},}\ }\href {\doibase
  10.1103/PhysRevLett.127.151102} {\bibfield  {journal} {\bibinfo  {journal}
  {Phys. Rev. Lett.}\ }\textbf {\bibinfo {volume} {127}},\ \bibinfo {pages}
  {151102} (\bibinfo {year} {2021})},\ \Eprint
  {http://arxiv.org/abs/2107.01298} {arXiv:2107.01298 [gr-qc]} \BibitemShut
  {NoStop}%
\bibitem [{\citenamefont {Osburn}\ and\ \citenamefont
  {Nishimura}(2022)}]{Osburn2022}%
  \BibitemOpen
  \bibfield  {author} {\bibinfo {author} {\bibfnamefont {Thomas}\ \bibnamefont
  {Osburn}}\ and\ \bibinfo {author} {\bibfnamefont {Nami}\ \bibnamefont
  {Nishimura}},\ }\bibfield  {title} {\enquote {\bibinfo {title} {{New
  self-force method via elliptic partial differential equations for Kerr
  inspiral models}},}\ }\href {\doibase 10.1103/PhysRevD.106.044056} {\bibfield
   {journal} {\bibinfo  {journal} {Phys. Rev. D}\ }\textbf {\bibinfo {volume}
  {106}},\ \bibinfo {pages} {044056} (\bibinfo {year} {2022})},\ \Eprint
  {http://arxiv.org/abs/2206.07031} {arXiv:2206.07031 [gr-qc]} \BibitemShut
  {NoStop}%
\bibitem [{\citenamefont {Wald}(1973)}]{Wald1973}%
  \BibitemOpen
  \bibfield  {author} {\bibinfo {author} {\bibfnamefont {Robert~M.}\
  \bibnamefont {Wald}},\ }\bibfield  {title} {\enquote {\bibinfo {title} {{On
  perturbations of a Kerr black hole}},}\ }\href {\doibase 10.1063/1.1666203}
  {\bibfield  {journal} {\bibinfo  {journal} {J. Math. Phys.}\ }\textbf
  {\bibinfo {volume} {14}},\ \bibinfo {pages} {1453--1461} (\bibinfo {year}
  {1973})}\BibitemShut {NoStop}%
\bibitem [{\citenamefont {Chrzanowski}(1975)}]{Chrzanowski1975}%
  \BibitemOpen
  \bibfield  {author} {\bibinfo {author} {\bibfnamefont {P.~L.}\ \bibnamefont
  {Chrzanowski}},\ }\bibfield  {title} {\enquote {\bibinfo {title} {{Vector
  Potential and Metric Perturbations of a Rotating Black Hole}},}\ }\href
  {\doibase 10.1103/PhysRevD.11.2042} {\bibfield  {journal} {\bibinfo
  {journal} {Phys. Rev. D}\ }\textbf {\bibinfo {volume} {11}},\ \bibinfo
  {pages} {2042--2062} (\bibinfo {year} {1975})}\BibitemShut {NoStop}%
\bibitem [{\citenamefont {Kegeles}\ and\ \citenamefont
  {Cohen}(1979)}]{Kegeles1979}%
  \BibitemOpen
  \bibfield  {author} {\bibinfo {author} {\bibfnamefont {L.~S.}\ \bibnamefont
  {Kegeles}}\ and\ \bibinfo {author} {\bibfnamefont {J.~M.}\ \bibnamefont
  {Cohen}},\ }\bibfield  {title} {\enquote {\bibinfo {title} {{Constructive
  procedure for pertrubations of space-times}},}\ }\href {\doibase
  10.1103/PhysRevD.19.1641} {\bibfield  {journal} {\bibinfo  {journal} {Phys.
  Rev. D}\ }\textbf {\bibinfo {volume} {19}},\ \bibinfo {pages} {1641--1664}
  (\bibinfo {year} {1979})}\BibitemShut {NoStop}%
\bibitem [{\citenamefont {Ori}(2003)}]{Ori2002}%
  \BibitemOpen
  \bibfield  {author} {\bibinfo {author} {\bibfnamefont {Amos}\ \bibnamefont
  {Ori}},\ }\bibfield  {title} {\enquote {\bibinfo {title} {{Reconstruction of
  inhomogeneous metric perturbations and electromagnetic four potential in Kerr
  space-time}},}\ }\href {\doibase 10.1103/PhysRevD.67.124010} {\bibfield
  {journal} {\bibinfo  {journal} {Phys. Rev. D}\ }\textbf {\bibinfo {volume}
  {67}},\ \bibinfo {pages} {124010} (\bibinfo {year} {2003})},\ \Eprint
  {http://arxiv.org/abs/gr-qc/0207045} {arXiv:gr-qc/0207045} \BibitemShut
  {NoStop}%
\bibitem [{\citenamefont {Price}\ \emph {et~al.}(2007)\citenamefont {Price},
  \citenamefont {Shankar},\ and\ \citenamefont {Whiting}}]{Price2006}%
  \BibitemOpen
  \bibfield  {author} {\bibinfo {author} {\bibfnamefont {Larry~R.}\
  \bibnamefont {Price}}, \bibinfo {author} {\bibfnamefont {Karthik}\
  \bibnamefont {Shankar}}, \ and\ \bibinfo {author} {\bibfnamefont
  {Bernard~F.}\ \bibnamefont {Whiting}},\ }\bibfield  {title} {\enquote
  {\bibinfo {title} {{On the existence of radiation gauges in Petrov type II
  spacetimes}},}\ }\href {\doibase 10.1088/0264-9381/24/9/014} {\bibfield
  {journal} {\bibinfo  {journal} {Class. Quant. Grav.}\ }\textbf {\bibinfo
  {volume} {24}},\ \bibinfo {pages} {2367--2388} (\bibinfo {year} {2007})},\
  \Eprint {http://arxiv.org/abs/gr-qc/0611070} {arXiv:gr-qc/0611070}
  \BibitemShut {NoStop}%
\bibitem [{\citenamefont {Keidl}\ \emph {et~al.}(2010)\citenamefont {Keidl},
  \citenamefont {Shah}, \citenamefont {Friedman}, \citenamefont {Kim},\ and\
  \citenamefont {Price}}]{Keidl2010}%
  \BibitemOpen
  \bibfield  {author} {\bibinfo {author} {\bibfnamefont {Tobias~S.}\
  \bibnamefont {Keidl}}, \bibinfo {author} {\bibfnamefont {Abhay~G.}\
  \bibnamefont {Shah}}, \bibinfo {author} {\bibfnamefont {John~L.}\
  \bibnamefont {Friedman}}, \bibinfo {author} {\bibfnamefont {Dong-Hoon}\
  \bibnamefont {Kim}}, \ and\ \bibinfo {author} {\bibfnamefont {Larry~R.}\
  \bibnamefont {Price}},\ }\bibfield  {title} {\enquote {\bibinfo {title}
  {{Gravitational Self-force in a Radiation Gauge}},}\ }\href {\doibase
  10.1103/PhysRevD.82.124012} {\bibfield  {journal} {\bibinfo  {journal} {Phys.
  Rev. D}\ }\textbf {\bibinfo {volume} {82}},\ \bibinfo {pages} {124012}
  (\bibinfo {year} {2010})},\ \bibinfo {note} {[Erratum: Phys.Rev.D 90, 109902
  (2014)]},\ \Eprint {http://arxiv.org/abs/1004.2276} {arXiv:1004.2276 [gr-qc]}
  \BibitemShut {NoStop}%
\bibitem [{\citenamefont {Shah}\ \emph {et~al.}(2011)\citenamefont {Shah},
  \citenamefont {Keidl}, \citenamefont {Friedman}, \citenamefont {Kim},\ and\
  \citenamefont {Price}}]{Shah2010}%
  \BibitemOpen
  \bibfield  {author} {\bibinfo {author} {\bibfnamefont {Abhay~G.}\
  \bibnamefont {Shah}}, \bibinfo {author} {\bibfnamefont {Tobias~S.}\
  \bibnamefont {Keidl}}, \bibinfo {author} {\bibfnamefont {John~L.}\
  \bibnamefont {Friedman}}, \bibinfo {author} {\bibfnamefont {Dong-Hoon}\
  \bibnamefont {Kim}}, \ and\ \bibinfo {author} {\bibfnamefont {Larry~R.}\
  \bibnamefont {Price}},\ }\bibfield  {title} {\enquote {\bibinfo {title}
  {{Conservative, gravitational self-force for a particle in circular orbit
  around a Schwarzschild black hole in a Radiation Gauge}},}\ }\href {\doibase
  10.1103/PhysRevD.83.064018} {\bibfield  {journal} {\bibinfo  {journal} {Phys.
  Rev. D}\ }\textbf {\bibinfo {volume} {83}},\ \bibinfo {pages} {064018}
  (\bibinfo {year} {2011})},\ \Eprint {http://arxiv.org/abs/1009.4876}
  {arXiv:1009.4876 [gr-qc]} \BibitemShut {NoStop}%
\bibitem [{\citenamefont {Pound}\ \emph {et~al.}(2014)\citenamefont {Pound},
  \citenamefont {Merlin},\ and\ \citenamefont {Barack}}]{Pound2013}%
  \BibitemOpen
  \bibfield  {author} {\bibinfo {author} {\bibfnamefont {Adam}\ \bibnamefont
  {Pound}}, \bibinfo {author} {\bibfnamefont {Cesar}\ \bibnamefont {Merlin}}, \
  and\ \bibinfo {author} {\bibfnamefont {Leor}\ \bibnamefont {Barack}},\
  }\bibfield  {title} {\enquote {\bibinfo {title} {{Gravitational self-force
  from radiation-gauge metric perturbations}},}\ }\href {\doibase
  10.1103/PhysRevD.89.024009} {\bibfield  {journal} {\bibinfo  {journal} {Phys.
  Rev. D}\ }\textbf {\bibinfo {volume} {89}},\ \bibinfo {pages} {024009}
  (\bibinfo {year} {2014})},\ \Eprint {http://arxiv.org/abs/1310.1513}
  {arXiv:1310.1513 [gr-qc]} \BibitemShut {NoStop}%
\bibitem [{\citenamefont {Merlin}\ \emph {et~al.}(2016)\citenamefont {Merlin},
  \citenamefont {Ori}, \citenamefont {Barack}, \citenamefont {Pound},\ and\
  \citenamefont {van~de Meent}}]{Merlin2016}%
  \BibitemOpen
  \bibfield  {author} {\bibinfo {author} {\bibfnamefont {Cesar}\ \bibnamefont
  {Merlin}}, \bibinfo {author} {\bibfnamefont {Amos}\ \bibnamefont {Ori}},
  \bibinfo {author} {\bibfnamefont {Leor}\ \bibnamefont {Barack}}, \bibinfo
  {author} {\bibfnamefont {Adam}\ \bibnamefont {Pound}}, \ and\ \bibinfo
  {author} {\bibfnamefont {Maarten}\ \bibnamefont {van~de Meent}},\ }\bibfield
  {title} {\enquote {\bibinfo {title} {{Completion of metric reconstruction for
  a particle orbiting a Kerr black hole}},}\ }\href {\doibase
  10.1103/PhysRevD.94.104066} {\bibfield  {journal} {\bibinfo  {journal} {Phys.
  Rev. D}\ }\textbf {\bibinfo {volume} {94}},\ \bibinfo {pages} {104066}
  (\bibinfo {year} {2016})},\ \Eprint {http://arxiv.org/abs/1609.01227}
  {arXiv:1609.01227 [gr-qc]} \BibitemShut {NoStop}%
\bibitem [{\citenamefont {Barack}\ \emph {et~al.}(2008)\citenamefont {Barack},
  \citenamefont {Ori},\ and\ \citenamefont {Sago}}]{Barack2008}%
  \BibitemOpen
  \bibfield  {author} {\bibinfo {author} {\bibfnamefont {Leor}\ \bibnamefont
  {Barack}}, \bibinfo {author} {\bibfnamefont {Amos}\ \bibnamefont {Ori}}, \
  and\ \bibinfo {author} {\bibfnamefont {Norichika}\ \bibnamefont {Sago}},\
  }\bibfield  {title} {\enquote {\bibinfo {title} {{Frequency-domain
  calculation of the self force: The High-frequency problem and its
  resolution}},}\ }\href {\doibase 10.1103/PhysRevD.78.084021} {\bibfield
  {journal} {\bibinfo  {journal} {Phys. Rev. D}\ }\textbf {\bibinfo {volume}
  {78}},\ \bibinfo {pages} {084021} (\bibinfo {year} {2008})},\ \Eprint
  {http://arxiv.org/abs/0808.2315} {arXiv:0808.2315 [gr-qc]} \BibitemShut
  {NoStop}%
\bibitem [{\citenamefont {Hopper}\ and\ \citenamefont
  {Evans}(2010)}]{Hopper2010}%
  \BibitemOpen
  \bibfield  {author} {\bibinfo {author} {\bibfnamefont {Seth}\ \bibnamefont
  {Hopper}}\ and\ \bibinfo {author} {\bibfnamefont {Charles~R.}\ \bibnamefont
  {Evans}},\ }\bibfield  {title} {\enquote {\bibinfo {title} {{Gravitational
  perturbations and metric reconstruction: Method of extended homogeneous
  solutions applied to eccentric orbits on a Schwarzschild black hole}},}\
  }\href {\doibase 10.1103/PhysRevD.82.084010} {\bibfield  {journal} {\bibinfo
  {journal} {Phys. Rev. D}\ }\textbf {\bibinfo {volume} {82}},\ \bibinfo
  {pages} {084010} (\bibinfo {year} {2010})},\ \Eprint
  {http://arxiv.org/abs/1006.4907} {arXiv:1006.4907 [gr-qc]} \BibitemShut
  {NoStop}%
\bibitem [{\citenamefont {Shah}\ and\ \citenamefont {Pound}(2015)}]{Shah2015}%
  \BibitemOpen
  \bibfield  {author} {\bibinfo {author} {\bibfnamefont {Abhay~G.}\
  \bibnamefont {Shah}}\ and\ \bibinfo {author} {\bibfnamefont {Adam}\
  \bibnamefont {Pound}},\ }\bibfield  {title} {\enquote {\bibinfo {title}
  {{Linear-in-mass-ratio contribution to spin precession and tidal invariants
  in Schwarzschild spacetime at very high post-Newtonian order}},}\ }\href
  {\doibase 10.1103/PhysRevD.91.124022} {\bibfield  {journal} {\bibinfo
  {journal} {Phys. Rev. D}\ }\textbf {\bibinfo {volume} {91}},\ \bibinfo
  {pages} {124022} (\bibinfo {year} {2015})},\ \Eprint
  {http://arxiv.org/abs/1503.02414} {arXiv:1503.02414 [gr-qc]} \BibitemShut
  {NoStop}%
\bibitem [{\citenamefont {Berndtson}(2007)}]{Berndtson:2007gsc}%
  \BibitemOpen
  \bibfield  {author} {\bibinfo {author} {\bibfnamefont {Mark~V.}\ \bibnamefont
  {Berndtson}},\ }\emph {\bibinfo {title} {{Harmonic gauge perturbations of the
  Schwarzschild metric}}},\ \href@noop {} {Ph.D. thesis} (\bibinfo {year}
  {2007}),\ \Eprint {http://arxiv.org/abs/0904.0033} {arXiv:0904.0033 [gr-qc]}
  \BibitemShut {NoStop}%
\bibitem [{\citenamefont {Aksteiner}\ \emph {et~al.}(2019)\citenamefont
  {Aksteiner}, \citenamefont {Andersson},\ and\ \citenamefont
  {B\"ackdahl}}]{Aksteiner:2016pjt}%
  \BibitemOpen
  \bibfield  {author} {\bibinfo {author} {\bibfnamefont {Steffen}\ \bibnamefont
  {Aksteiner}}, \bibinfo {author} {\bibfnamefont {Lars}\ \bibnamefont
  {Andersson}}, \ and\ \bibinfo {author} {\bibfnamefont {Thomas}\ \bibnamefont
  {B\"ackdahl}},\ }\bibfield  {title} {\enquote {\bibinfo {title} {{New
  identities for linearized gravity on the Kerr spacetime}},}\ }\href {\doibase
  10.1103/PhysRevD.99.044043} {\bibfield  {journal} {\bibinfo  {journal} {Phys.
  Rev. D}\ }\textbf {\bibinfo {volume} {99}},\ \bibinfo {pages} {044043}
  (\bibinfo {year} {2019})},\ \Eprint {http://arxiv.org/abs/1601.06084}
  {arXiv:1601.06084 [gr-qc]} \BibitemShut {NoStop}%
\bibitem [{\citenamefont {Dolan}\ \emph {et~al.}(2022)\citenamefont {Dolan},
  \citenamefont {Kavanagh},\ and\ \citenamefont {Wardell}}]{Dolan:2021ijg}%
  \BibitemOpen
  \bibfield  {author} {\bibinfo {author} {\bibfnamefont {Sam~R.}\ \bibnamefont
  {Dolan}}, \bibinfo {author} {\bibfnamefont {Chris}\ \bibnamefont {Kavanagh}},
  \ and\ \bibinfo {author} {\bibfnamefont {Barry}\ \bibnamefont {Wardell}},\
  }\bibfield  {title} {\enquote {\bibinfo {title} {{Gravitational Perturbations
  of Rotating Black Holes in Lorenz Gauge}},}\ }\href {\doibase
  10.1103/PhysRevLett.128.151101} {\bibfield  {journal} {\bibinfo  {journal}
  {Phys. Rev. Lett.}\ }\textbf {\bibinfo {volume} {128}},\ \bibinfo {pages}
  {151101} (\bibinfo {year} {2022})},\ \Eprint
  {http://arxiv.org/abs/2108.06344} {arXiv:2108.06344 [gr-qc]} \BibitemShut
  {NoStop}%
\bibitem [{\citenamefont {Dolan}\ \emph {et~al.}(2023)\citenamefont {Dolan},
  \citenamefont {Durkan}, \citenamefont {Kavanagh},\ and\ \citenamefont
  {Wardell}}]{Dolan:2023enf}%
  \BibitemOpen
  \bibfield  {author} {\bibinfo {author} {\bibfnamefont {Sam~R.}\ \bibnamefont
  {Dolan}}, \bibinfo {author} {\bibfnamefont {Leanne}\ \bibnamefont {Durkan}},
  \bibinfo {author} {\bibfnamefont {Chris}\ \bibnamefont {Kavanagh}}, \ and\
  \bibinfo {author} {\bibfnamefont {Barry}\ \bibnamefont {Wardell}},\
  }\bibfield  {title} {\enquote {\bibinfo {title} {{Metric perturbations of
  Kerr spacetime in Lorenz gauge: Circular equatorial orbits}},}\ }\href@noop
  {} {\  (\bibinfo {year} {2023})},\ \Eprint {http://arxiv.org/abs/2306.16459}
  {arXiv:2306.16459 [gr-qc]} \BibitemShut {NoStop}%
\bibitem [{\citenamefont {Wardell}()}]{Wardell:2023-talk}%
  \BibitemOpen
  \bibfield  {author} {\bibinfo {author} {\bibfnamefont {Barry}\ \bibnamefont
  {Wardell}},\ }\href@noop {} {\enquote {\bibinfo {title} {{Metric
  perturbations of Kerr spacetime in Lorenz gauge}},}\ }\bibinfo {note}
  {{presentation at "Infinity on a Gridshell" conference (Niels Bohr Institute,
  Copenhagen, 2023)}}\BibitemShut {NoStop}%
\bibitem [{\citenamefont {Dolan}\ \emph {et~al.}()\citenamefont {Dolan},
  \citenamefont {Kavanagh},\ and\ \citenamefont
  {Wardell}}]{Wardell:2023-inprep}%
  \BibitemOpen
  \bibfield  {author} {\bibinfo {author} {\bibfnamefont {Sam~R.}\ \bibnamefont
  {Dolan}}, \bibinfo {author} {\bibfnamefont {Chris}\ \bibnamefont {Kavanagh}},
  \ and\ \bibinfo {author} {\bibfnamefont {Barry}\ \bibnamefont {Wardell}},\
  }\href@noop {} {\enquote {\bibinfo {title} {{Sourced metric perturbations of
  Kerr-NUT spacetime in Lorenz gauge}},}\ }\bibinfo {note} {In
  preparation}\BibitemShut {NoStop}%
\bibitem [{\citenamefont {Green}\ \emph {et~al.}(2020)\citenamefont {Green},
  \citenamefont {Hollands},\ and\ \citenamefont {Zimmerman}}]{Green2019}%
  \BibitemOpen
  \bibfield  {author} {\bibinfo {author} {\bibfnamefont {Stephen~R.}\
  \bibnamefont {Green}}, \bibinfo {author} {\bibfnamefont {Stefan}\
  \bibnamefont {Hollands}}, \ and\ \bibinfo {author} {\bibfnamefont {Peter}\
  \bibnamefont {Zimmerman}},\ }\bibfield  {title} {\enquote {\bibinfo {title}
  {{Teukolsky formalism for nonlinear Kerr perturbations}},}\ }\href {\doibase
  10.1088/1361-6382/ab7075} {\bibfield  {journal} {\bibinfo  {journal} {Class.
  Quant. Grav.}\ }\textbf {\bibinfo {volume} {37}},\ \bibinfo {pages} {075001}
  (\bibinfo {year} {2020})},\ \Eprint {http://arxiv.org/abs/1908.09095}
  {arXiv:1908.09095 [gr-qc]} \BibitemShut {NoStop}%
\bibitem [{\citenamefont {Toomani}\ \emph {et~al.}(2022)\citenamefont
  {Toomani}, \citenamefont {Zimmerman}, \citenamefont {Spiers}, \citenamefont
  {Hollands}, \citenamefont {Pound},\ and\ \citenamefont
  {Green}}]{Toomani2021}%
  \BibitemOpen
  \bibfield  {author} {\bibinfo {author} {\bibfnamefont {Vahid}\ \bibnamefont
  {Toomani}}, \bibinfo {author} {\bibfnamefont {Peter}\ \bibnamefont
  {Zimmerman}}, \bibinfo {author} {\bibfnamefont {Andrew}\ \bibnamefont
  {Spiers}}, \bibinfo {author} {\bibfnamefont {Stefan}\ \bibnamefont
  {Hollands}}, \bibinfo {author} {\bibfnamefont {Adam}\ \bibnamefont {Pound}},
  \ and\ \bibinfo {author} {\bibfnamefont {Stephen~R.}\ \bibnamefont {Green}},\
  }\bibfield  {title} {\enquote {\bibinfo {title} {{New metric reconstruction
  scheme for gravitational self-force calculations}},}\ }\href {\doibase
  10.1088/1361-6382/ac37a5} {\bibfield  {journal} {\bibinfo  {journal} {Class.
  Quant. Grav.}\ }\textbf {\bibinfo {volume} {39}},\ \bibinfo {pages} {015019}
  (\bibinfo {year} {2022})},\ \Eprint {http://arxiv.org/abs/2108.04273}
  {arXiv:2108.04273 [gr-qc]} \BibitemShut {NoStop}%
\bibitem [{\citenamefont {Wardell}(2015)}]{Wardell2015}%
  \BibitemOpen
  \bibfield  {author} {\bibinfo {author} {\bibfnamefont {Barry}\ \bibnamefont
  {Wardell}},\ }\bibfield  {title} {\enquote {\bibinfo {title} {{Self-force:
  Computational Strategies}},}\ }\href {\doibase 10.1007/978-3-319-18335-0\_14}
  {\bibfield  {journal} {\bibinfo  {journal} {Fund. Theor. Phys.}\ }\textbf
  {\bibinfo {volume} {179}},\ \bibinfo {pages} {487--522} (\bibinfo {year}
  {2015})},\ \Eprint {http://arxiv.org/abs/1501.07322} {arXiv:1501.07322
  [gr-qc]} \BibitemShut {NoStop}%
\bibitem [{\citenamefont {Upton}\ and\ \citenamefont
  {Pound}(2021)}]{Upton2021}%
  \BibitemOpen
  \bibfield  {author} {\bibinfo {author} {\bibfnamefont {Samuel~D.}\
  \bibnamefont {Upton}}\ and\ \bibinfo {author} {\bibfnamefont {Adam}\
  \bibnamefont {Pound}},\ }\bibfield  {title} {\enquote {\bibinfo {title}
  {{Second-order gravitational self-force in a highly regular gauge}},}\ }\href
  {\doibase 10.1103/PhysRevD.103.124016} {\bibfield  {journal} {\bibinfo
  {journal} {Phys. Rev. D}\ }\textbf {\bibinfo {volume} {103}},\ \bibinfo
  {pages} {124016} (\bibinfo {year} {2021})},\ \Eprint
  {http://arxiv.org/abs/2101.11409} {arXiv:2101.11409 [gr-qc]} \BibitemShut
  {NoStop}%
\bibitem [{\citenamefont {Upton}(2023)}]{Upton:2023tcv}%
  \BibitemOpen
  \bibfield  {author} {\bibinfo {author} {\bibfnamefont {Samuel~D.}\
  \bibnamefont {Upton}},\ }\bibfield  {title} {\enquote {\bibinfo {title}
  {{Second-order gravitational self-force in a highly regular gauge: Covariant
  and coordinate punctures}},}\ }\href@noop {} {\  (\bibinfo {year} {2023})},\
  \Eprint {http://arxiv.org/abs/2309.03778} {arXiv:2309.03778 [gr-qc]}
  \BibitemShut {NoStop}%
\bibitem [{\citenamefont {Detweiler}(2001)}]{Detweiler2000}%
  \BibitemOpen
  \bibfield  {author} {\bibinfo {author} {\bibfnamefont {Steven~L.}\
  \bibnamefont {Detweiler}},\ }\bibfield  {title} {\enquote {\bibinfo {title}
  {{Radiation reaction and the self-force for a point mass in general
  relativity}},}\ }\href {\doibase 10.1103/PhysRevLett.86.1931} {\bibfield
  {journal} {\bibinfo  {journal} {Phys. Rev. Lett.}\ }\textbf {\bibinfo
  {volume} {86}},\ \bibinfo {pages} {1931--1934} (\bibinfo {year} {2001})},\
  \Eprint {http://arxiv.org/abs/gr-qc/0011039} {arXiv:gr-qc/0011039}
  \BibitemShut {NoStop}%
\bibitem [{\citenamefont {Detweiler}\ and\ \citenamefont
  {Whiting}(2003)}]{Detweiler:2002mi}%
  \BibitemOpen
  \bibfield  {author} {\bibinfo {author} {\bibfnamefont {Steven~L.}\
  \bibnamefont {Detweiler}}\ and\ \bibinfo {author} {\bibfnamefont
  {Bernard~F.}\ \bibnamefont {Whiting}},\ }\bibfield  {title} {\enquote
  {\bibinfo {title} {{Selfforce via a Green's function decomposition}},}\
  }\href {\doibase 10.1103/PhysRevD.67.024025} {\bibfield  {journal} {\bibinfo
  {journal} {Phys. Rev. D}\ }\textbf {\bibinfo {volume} {67}},\ \bibinfo
  {pages} {024025} (\bibinfo {year} {2003})},\ \Eprint
  {http://arxiv.org/abs/gr-qc/0202086} {arXiv:gr-qc/0202086} \BibitemShut
  {NoStop}%
\bibitem [{\citenamefont {Poisson}\ \emph {et~al.}(2011)\citenamefont
  {Poisson}, \citenamefont {Pound},\ and\ \citenamefont
  {Vega}}]{Poisson:2011nh}%
  \BibitemOpen
  \bibfield  {author} {\bibinfo {author} {\bibfnamefont {Eric}\ \bibnamefont
  {Poisson}}, \bibinfo {author} {\bibfnamefont {Adam}\ \bibnamefont {Pound}}, \
  and\ \bibinfo {author} {\bibfnamefont {Ian}\ \bibnamefont {Vega}},\
  }\bibfield  {title} {\enquote {\bibinfo {title} {{The Motion of point
  particles in curved spacetime}},}\ }\href {\doibase 10.12942/lrr-2011-7}
  {\bibfield  {journal} {\bibinfo  {journal} {Living Rev. Rel.}\ }\textbf
  {\bibinfo {volume} {14}},\ \bibinfo {pages} {7} (\bibinfo {year} {2011})},\
  \Eprint {http://arxiv.org/abs/1102.0529} {arXiv:1102.0529 [gr-qc]}
  \BibitemShut {NoStop}%
\bibitem [{\citenamefont {Heffernan}\ \emph {et~al.}(2012)\citenamefont
  {Heffernan}, \citenamefont {Ottewill},\ and\ \citenamefont
  {Wardell}}]{Heffernan2012}%
  \BibitemOpen
  \bibfield  {author} {\bibinfo {author} {\bibfnamefont {Anna}\ \bibnamefont
  {Heffernan}}, \bibinfo {author} {\bibfnamefont {Adrian}\ \bibnamefont
  {Ottewill}}, \ and\ \bibinfo {author} {\bibfnamefont {Barry}\ \bibnamefont
  {Wardell}},\ }\bibfield  {title} {\enquote {\bibinfo {title} {High-order
  expansions of the {Detweiler}-{Whiting} singular field in {Schwarzschild}
  spacetime},}\ }\href {\doibase 10.1103/physrevd.86.104023} {\bibfield
  {journal} {\bibinfo  {journal} {Phys. Rev. D}\ }\textbf {\bibinfo {volume}
  {86}},\ \bibinfo {pages} {104023} (\bibinfo {year} {2012})},\ \Eprint
  {http://arxiv.org/abs/1204.0794} {arXiv:1204.0794 [gr-qc]} \BibitemShut
  {NoStop}%
\bibitem [{\citenamefont {Pound}\ and\ \citenamefont
  {Miller}(2014)}]{Pound2014}%
  \BibitemOpen
  \bibfield  {author} {\bibinfo {author} {\bibfnamefont {Adam}\ \bibnamefont
  {Pound}}\ and\ \bibinfo {author} {\bibfnamefont {Jeremy}\ \bibnamefont
  {Miller}},\ }\bibfield  {title} {\enquote {\bibinfo {title} {{Practical,
  covariant puncture for second-order self-force calculations}},}\ }\href
  {\doibase 10.1103/PhysRevD.89.104020} {\bibfield  {journal} {\bibinfo
  {journal} {Phys. Rev. D}\ }\textbf {\bibinfo {volume} {89}},\ \bibinfo
  {pages} {104020} (\bibinfo {year} {2014})},\ \Eprint
  {http://arxiv.org/abs/1403.1843} {arXiv:1403.1843 [gr-qc]} \BibitemShut
  {NoStop}%
\bibitem [{\citenamefont {Vega}\ and\ \citenamefont
  {Detweiler}(2008)}]{Vega:2007mc}%
  \BibitemOpen
  \bibfield  {author} {\bibinfo {author} {\bibfnamefont {Ian}\ \bibnamefont
  {Vega}}\ and\ \bibinfo {author} {\bibfnamefont {Steven~L.}\ \bibnamefont
  {Detweiler}},\ }\bibfield  {title} {\enquote {\bibinfo {title}
  {{Regularization of fields for self-force problems in curved spacetime:
  Foundations and a time-domain application}},}\ }\href {\doibase
  10.1103/PhysRevD.77.084008} {\bibfield  {journal} {\bibinfo  {journal} {Phys.
  Rev. D}\ }\textbf {\bibinfo {volume} {77}},\ \bibinfo {pages} {084008}
  (\bibinfo {year} {2008})},\ \Eprint {http://arxiv.org/abs/0712.4405}
  {arXiv:0712.4405 [gr-qc]} \BibitemShut {NoStop}%
\bibitem [{\citenamefont {Barack}\ and\ \citenamefont
  {Golbourn}(2007)}]{Barack:2007jh}%
  \BibitemOpen
  \bibfield  {author} {\bibinfo {author} {\bibfnamefont {Leor}\ \bibnamefont
  {Barack}}\ and\ \bibinfo {author} {\bibfnamefont {Darren~A.}\ \bibnamefont
  {Golbourn}},\ }\bibfield  {title} {\enquote {\bibinfo {title} {{Scalar-field
  perturbations from a particle orbiting a black hole using numerical evolution
  in 2+1 dimensions}},}\ }\href {\doibase 10.1103/PhysRevD.76.044020}
  {\bibfield  {journal} {\bibinfo  {journal} {Phys. Rev. D}\ }\textbf {\bibinfo
  {volume} {76}},\ \bibinfo {pages} {044020} (\bibinfo {year} {2007})},\
  \Eprint {http://arxiv.org/abs/0705.3620} {arXiv:0705.3620 [gr-qc]}
  \BibitemShut {NoStop}%
\bibitem [{\citenamefont {Warburton}\ and\ \citenamefont
  {Wardell}(2014)}]{Warburton:2013lea}%
  \BibitemOpen
  \bibfield  {author} {\bibinfo {author} {\bibfnamefont {Niels}\ \bibnamefont
  {Warburton}}\ and\ \bibinfo {author} {\bibfnamefont {Barry}\ \bibnamefont
  {Wardell}},\ }\bibfield  {title} {\enquote {\bibinfo {title} {{Applying the
  effective-source approach to frequency-domain self-force calculations}},}\
  }\href {\doibase 10.1103/PhysRevD.89.044046} {\bibfield  {journal} {\bibinfo
  {journal} {Phys. Rev. D}\ }\textbf {\bibinfo {volume} {89}},\ \bibinfo
  {pages} {044046} (\bibinfo {year} {2014})},\ \Eprint
  {http://arxiv.org/abs/1311.3104} {arXiv:1311.3104 [gr-qc]} \BibitemShut
  {NoStop}%
\bibitem [{\citenamefont {Barack}(2009)}]{Barack2009}%
  \BibitemOpen
  \bibfield  {author} {\bibinfo {author} {\bibfnamefont {Leor}\ \bibnamefont
  {Barack}},\ }\bibfield  {title} {\enquote {\bibinfo {title} {{Gravitational
  self force in extreme mass-ratio inspirals}},}\ }\href {\doibase
  10.1088/0264-9381/26/21/213001} {\bibfield  {journal} {\bibinfo  {journal}
  {Class. Quant. Grav.}\ }\textbf {\bibinfo {volume} {26}},\ \bibinfo {pages}
  {213001} (\bibinfo {year} {2009})},\ \Eprint {http://arxiv.org/abs/0908.1664}
  {arXiv:0908.1664 [gr-qc]} \BibitemShut {NoStop}%
\bibitem [{\citenamefont {Wald}(1978)}]{Wald1978}%
  \BibitemOpen
  \bibfield  {author} {\bibinfo {author} {\bibfnamefont {Robert~M.}\
  \bibnamefont {Wald}},\ }\bibfield  {title} {\enquote {\bibinfo {title}
  {{Construction of Solutions of Gravitational, Electromagnetic, Or Other
  Perturbation Equations from Solutions of Decoupled Equations}},}\ }\href
  {\doibase 10.1103/PhysRevLett.41.203} {\bibfield  {journal} {\bibinfo
  {journal} {Phys. Rev. Lett.}\ }\textbf {\bibinfo {volume} {41}},\ \bibinfo
  {pages} {203--206} (\bibinfo {year} {1978})}\BibitemShut {NoStop}%
\bibitem [{\citenamefont {Keidl}\ \emph {et~al.}(2007)\citenamefont {Keidl},
  \citenamefont {Friedman},\ and\ \citenamefont {Wiseman}}]{Keidl2006}%
  \BibitemOpen
  \bibfield  {author} {\bibinfo {author} {\bibfnamefont {Tobias~S.}\
  \bibnamefont {Keidl}}, \bibinfo {author} {\bibfnamefont {John~L.}\
  \bibnamefont {Friedman}}, \ and\ \bibinfo {author} {\bibfnamefont {Alan~G.}\
  \bibnamefont {Wiseman}},\ }\bibfield  {title} {\enquote {\bibinfo {title}
  {{On finding fields and self-force in a gauge appropriate to separable wave
  equations}},}\ }\href {\doibase 10.1103/PhysRevD.75.124009} {\bibfield
  {journal} {\bibinfo  {journal} {Phys. Rev. D}\ }\textbf {\bibinfo {volume}
  {75}},\ \bibinfo {pages} {124009} (\bibinfo {year} {2007})},\ \Eprint
  {http://arxiv.org/abs/gr-qc/0611072} {arXiv:gr-qc/0611072} \BibitemShut
  {NoStop}%
\bibitem [{\citenamefont {Miller}\ \emph {et~al.}(2016)\citenamefont {Miller},
  \citenamefont {Wardell},\ and\ \citenamefont {Pound}}]{Miller:2016hjv}%
  \BibitemOpen
  \bibfield  {author} {\bibinfo {author} {\bibfnamefont {Jeremy}\ \bibnamefont
  {Miller}}, \bibinfo {author} {\bibfnamefont {Barry}\ \bibnamefont {Wardell}},
  \ and\ \bibinfo {author} {\bibfnamefont {Adam}\ \bibnamefont {Pound}},\
  }\bibfield  {title} {\enquote {\bibinfo {title} {{Second-order perturbation
  theory: the problem of infinite mode coupling}},}\ }\href {\doibase
  10.1103/PhysRevD.94.104018} {\bibfield  {journal} {\bibinfo  {journal} {Phys.
  Rev. D}\ }\textbf {\bibinfo {volume} {94}},\ \bibinfo {pages} {104018}
  (\bibinfo {year} {2016})},\ \Eprint {http://arxiv.org/abs/1608.06783}
  {arXiv:1608.06783 [gr-qc]} \BibitemShut {NoStop}%
\bibitem [{\citenamefont {Barry~Wardell}\ and\ \citenamefont
  {Barack}()}]{PuncturePaper}%
  \BibitemOpen
  \bibfield  {author} {\bibinfo {author} {\bibfnamefont {Niels Warburton Samuel
  D.~Upton}\ \bibnamefont {Barry~Wardell}, \bibfnamefont {Adam~Pound}}\ and\
  \bibinfo {author} {\bibfnamefont {Leor}\ \bibnamefont {Barack}},\ }\href@noop
  {} {\enquote {\bibinfo {title} {{Effective source for second-order self-force
  calculations: quasicircular orbits in Schwarzschild spacetime}},}\ }\bibinfo
  {note} {In preparation}\BibitemShut {NoStop}%
\bibitem [{BHP()}]{BHPToolkit}%
  \BibitemOpen
  \href@noop {} {\enquote {\bibinfo {title} {{Black Hole Perturbation
  Toolkit}},}\ }\bibinfo {howpublished}
  {(\href{http://bhptoolkit.org/}{bhptoolkit.org})}\BibitemShut {NoStop}%
\bibitem [{\citenamefont {Wardell}\ and\ \citenamefont
  {Warburton}(2015)}]{Wardell:2015ada}%
  \BibitemOpen
  \bibfield  {author} {\bibinfo {author} {\bibfnamefont {Barry}\ \bibnamefont
  {Wardell}}\ and\ \bibinfo {author} {\bibfnamefont {Niels}\ \bibnamefont
  {Warburton}},\ }\bibfield  {title} {\enquote {\bibinfo {title} {{Applying the
  effective-source approach to frequency-domain self-force calculations:
  Lorenz-gauge gravitational perturbations}},}\ }\href {\doibase
  10.1103/PhysRevD.92.084019} {\bibfield  {journal} {\bibinfo  {journal} {Phys.
  Rev. D}\ }\textbf {\bibinfo {volume} {92}},\ \bibinfo {pages} {084019}
  (\bibinfo {year} {2015})},\ \Eprint {http://arxiv.org/abs/1505.07841}
  {arXiv:1505.07841 [gr-qc]} \BibitemShut {NoStop}%
\bibitem [{\citenamefont {Barack}\ and\ \citenamefont
  {Lousto}(2005)}]{Barack:2005nr}%
  \BibitemOpen
  \bibfield  {author} {\bibinfo {author} {\bibfnamefont {L.}~\bibnamefont
  {Barack}}\ and\ \bibinfo {author} {\bibfnamefont {C.O.}\ \bibnamefont
  {Lousto}},\ }\bibfield  {title} {\enquote {\bibinfo {title} {Perturbations of
  {S}chwarzschild black holes in the {L}orenz gauge: Formulation and numerical
  implementation},}\ }\href {\doibase 10.1103/PhysRevD.72.104026} {\bibfield
  {journal} {\bibinfo  {journal} {Phys. Rev.}\ }\textbf {\bibinfo {volume}
  {D72}},\ \bibinfo {pages} {104026} (\bibinfo {year} {2005})},\ \Eprint
  {http://arxiv.org/abs/gr-qc/0510019} {arXiv:gr-qc/0510019 [gr-qc]}
  \BibitemShut {NoStop}%
\bibitem [{\citenamefont {Barack}\ and\ \citenamefont
  {Sago}(2007)}]{Barack:2007tm}%
  \BibitemOpen
  \bibfield  {author} {\bibinfo {author} {\bibfnamefont {L.}~\bibnamefont
  {Barack}}\ and\ \bibinfo {author} {\bibfnamefont {N.}~\bibnamefont {Sago}},\
  }\bibfield  {title} {\enquote {\bibinfo {title} {{Gravitational self force on
  a particle in circular orbit around a Schwarzschild black hole}},}\ }\href
  {\doibase 10.1103/PhysRevD.75.064021} {\bibfield  {journal} {\bibinfo
  {journal} {Phys. Rev.}\ }\textbf {\bibinfo {volume} {D75}},\ \bibinfo {pages}
  {064021} (\bibinfo {year} {2007})},\ \Eprint
  {http://arxiv.org/abs/gr-qc/0701069} {arXiv:gr-qc/0701069 [gr-qc]}
  \BibitemShut {NoStop}%
\bibitem [{\citenamefont {Leather}\ and\ \citenamefont
  {Warburton}(2023)}]{Leather:2023dzj}%
  \BibitemOpen
  \bibfield  {author} {\bibinfo {author} {\bibfnamefont {Benjamin}\
  \bibnamefont {Leather}}\ and\ \bibinfo {author} {\bibfnamefont {Niels}\
  \bibnamefont {Warburton}},\ }\bibfield  {title} {\enquote {\bibinfo {title}
  {{Applying the effective-source approach to frequency-domain self-force
  calculations for eccentric orbits}},}\ }\href@noop {} {\  (\bibinfo {year}
  {2023})},\ \Eprint {http://arxiv.org/abs/2306.17221} {arXiv:2306.17221
  [gr-qc]} \BibitemShut {NoStop}%
\bibitem [{\citenamefont {Bini}\ and\ \citenamefont
  {Geralico}(2019)}]{Bini:2019xwn}%
  \BibitemOpen
  \bibfield  {author} {\bibinfo {author} {\bibfnamefont {Donato}\ \bibnamefont
  {Bini}}\ and\ \bibinfo {author} {\bibfnamefont {Andrea}\ \bibnamefont
  {Geralico}},\ }\bibfield  {title} {\enquote {\bibinfo {title} {{Gauge-fixing
  for the completion problem of reconstructed metric perturbations of a Kerr
  spacetime}},}\ }\href@noop {} {\  (\bibinfo {year} {2019})},\ \Eprint
  {http://arxiv.org/abs/1908.03191} {arXiv:1908.03191 [gr-qc]} \BibitemShut
  {NoStop}%
\bibitem [{\citenamefont {Poisson}(2009)}]{Poisson:2009pwt}%
  \BibitemOpen
  \bibfield  {author} {\bibinfo {author} {\bibfnamefont {Eric}\ \bibnamefont
  {Poisson}},\ }\href {\doibase 10.1017/CBO9780511606601} {\emph {\bibinfo
  {title} {{A Relativist's Toolkit: The Mathematics of Black-Hole
  Mechanics}}}}\ (\bibinfo  {publisher} {Cambridge University Press},\ \bibinfo
  {year} {2009})\BibitemShut {NoStop}%
\bibitem [{\citenamefont {Detweiler}(2008)}]{Detweiler2008}%
  \BibitemOpen
  \bibfield  {author} {\bibinfo {author} {\bibfnamefont {Steven~L.}\
  \bibnamefont {Detweiler}},\ }\bibfield  {title} {\enquote {\bibinfo {title}
  {{A Consequence of the gravitational self-force for circular orbits of the
  Schwarzschild geometry}},}\ }\href {\doibase 10.1103/PhysRevD.77.124026}
  {\bibfield  {journal} {\bibinfo  {journal} {Phys. Rev. D}\ }\textbf {\bibinfo
  {volume} {77}},\ \bibinfo {pages} {124026} (\bibinfo {year} {2008})},\
  \Eprint {http://arxiv.org/abs/0804.3529} {arXiv:0804.3529 [gr-qc]}
  \BibitemShut {NoStop}%
\bibitem [{\citenamefont {Dolan}\ \emph {et~al.}(2015)\citenamefont {Dolan},
  \citenamefont {Nolan}, \citenamefont {Ottewill}, \citenamefont {Warburton},\
  and\ \citenamefont {Wardell}}]{Dolan:2014}%
  \BibitemOpen
  \bibfield  {author} {\bibinfo {author} {\bibfnamefont {Sam~R.}\ \bibnamefont
  {Dolan}}, \bibinfo {author} {\bibfnamefont {Patrick}\ \bibnamefont {Nolan}},
  \bibinfo {author} {\bibfnamefont {Adrian~C.}\ \bibnamefont {Ottewill}},
  \bibinfo {author} {\bibfnamefont {Niels}\ \bibnamefont {Warburton}}, \ and\
  \bibinfo {author} {\bibfnamefont {Barry}\ \bibnamefont {Wardell}},\
  }\bibfield  {title} {\enquote {\bibinfo {title} {{Tidal invariants for
  compact binaries on quasicircular orbits}},}\ }\href {\doibase
  10.1103/PhysRevD.91.023009} {\bibfield  {journal} {\bibinfo  {journal} {Phys.
  Rev. D}\ }\textbf {\bibinfo {volume} {91}},\ \bibinfo {pages} {023009}
  (\bibinfo {year} {2015})},\ \Eprint {http://arxiv.org/abs/1406.4890}
  {arXiv:1406.4890 [gr-qc]} \BibitemShut {NoStop}%
\bibitem [{\citenamefont {Orszag}(1974)}]{Orszag:1974}%
  \BibitemOpen
  \bibfield  {author} {\bibinfo {author} {\bibfnamefont {S.}~\bibnamefont
  {Orszag}},\ }\bibfield  {title} {\enquote {\bibinfo {title} {Fourier series
  on spheres},}\ }\href@noop {} {\bibfield  {journal} {\bibinfo  {journal}
  {Monthly Weather Review}\ }\textbf {\bibinfo {volume} {102}},\ \bibinfo
  {pages} {56} (\bibinfo {year} {1974})}\BibitemShut {NoStop}%
\bibitem [{\citenamefont {Spiers}\ \emph {et~al.}(2023)\citenamefont {Spiers},
  \citenamefont {Pound},\ and\ \citenamefont {Moxon}}]{Spiers:2023cip}%
  \BibitemOpen
  \bibfield  {author} {\bibinfo {author} {\bibfnamefont {Andrew}\ \bibnamefont
  {Spiers}}, \bibinfo {author} {\bibfnamefont {Adam}\ \bibnamefont {Pound}}, \
  and\ \bibinfo {author} {\bibfnamefont {Jordan}\ \bibnamefont {Moxon}},\
  }\bibfield  {title} {\enquote {\bibinfo {title} {{Second-order Teukolsky
  formalism in Kerr spacetime: Formulation and nonlinear source}},}\ }\href
  {\doibase 10.1103/PhysRevD.108.064002} {\bibfield  {journal} {\bibinfo
  {journal} {Phys. Rev. D}\ }\textbf {\bibinfo {volume} {108}},\ \bibinfo
  {pages} {064002} (\bibinfo {year} {2023})},\ \Eprint
  {http://arxiv.org/abs/2305.19332} {arXiv:2305.19332 [gr-qc]} \BibitemShut
  {NoStop}%
\bibitem [{\citenamefont {Osburn}()}]{OsburnCapra2023}%
  \BibitemOpen
  \bibfield  {author} {\bibinfo {author} {\bibfnamefont {Thomas}\ \bibnamefont
  {Osburn}},\ }\href@noop {} {\enquote {\bibinfo {title} {Kerr self-force via
  elliptic pdes},}\ }\bibinfo {note} {Talk given at 26th Capra Meeting (Niels
  Bohr Institute, Copenhagen, July 2023)}\BibitemShut {NoStop}%
\bibitem [{\citenamefont {Bourg}\ \emph {et~al.}(2024)\citenamefont {Bourg},
  \citenamefont {Pound}, \citenamefont {Upton},\ and\ \citenamefont
  {Panosso~Macedo}}]{Bourg2024}%
  \BibitemOpen
  \bibfield  {author} {\bibinfo {author} {\bibfnamefont {Patrick}\ \bibnamefont
  {Bourg}}, \bibinfo {author} {\bibfnamefont {Adam}\ \bibnamefont {Pound}},
  \bibinfo {author} {\bibfnamefont {Samuel~D.}\ \bibnamefont {Upton}}, \ and\
  \bibinfo {author} {\bibfnamefont {Rodrigo}\ \bibnamefont {Panosso~Macedo}},\
  }\bibfield  {title} {\enquote {\bibinfo {title} {{Simple, efficient method of
  calculating the Detweiler-Whiting singular field to very high order}},}\
  }\href@noop {} {\  (\bibinfo {year} {2024})},\ \Eprint
  {http://arxiv.org/abs/2404.10082} {arXiv:2404.10082 [gr-qc]} \BibitemShut
  {NoStop}%
\bibitem [{\citenamefont {Panosso~Macedo}\ \emph {et~al.}(2024)\citenamefont
  {Panosso~Macedo}, \citenamefont {Bourg}, \citenamefont {Pound},\ and\
  \citenamefont {Upton}}]{Macedo2024}%
  \BibitemOpen
  \bibfield  {author} {\bibinfo {author} {\bibfnamefont {Rodrigo}\ \bibnamefont
  {Panosso~Macedo}}, \bibinfo {author} {\bibfnamefont {Patrick}\ \bibnamefont
  {Bourg}}, \bibinfo {author} {\bibfnamefont {Adam}\ \bibnamefont {Pound}}, \
  and\ \bibinfo {author} {\bibfnamefont {Samuel~D.}\ \bibnamefont {Upton}},\
  }\bibfield  {title} {\enquote {\bibinfo {title} {{Multi-domain spectral
  method for self-force calculations}},}\ }\href@noop {} {\  (\bibinfo {year}
  {2024})},\ \Eprint {http://arxiv.org/abs/2404.10083} {arXiv:2404.10083
  [gr-qc]} \BibitemShut {NoStop}%
\bibitem [{\citenamefont {van~de Meent}(2018)}]{vandeMeent:2017bcc}%
  \BibitemOpen
  \bibfield  {author} {\bibinfo {author} {\bibfnamefont {Maarten}\ \bibnamefont
  {van~de Meent}},\ }\bibfield  {title} {\enquote {\bibinfo {title}
  {{Gravitational self-force on generic bound geodesics in Kerr spacetime}},}\
  }\href {\doibase 10.1103/PhysRevD.97.104033} {\bibfield  {journal} {\bibinfo
  {journal} {Phys. Rev. D}\ }\textbf {\bibinfo {volume} {97}},\ \bibinfo
  {pages} {104033} (\bibinfo {year} {2018})},\ \Eprint
  {http://arxiv.org/abs/1711.09607} {arXiv:1711.09607 [gr-qc]} \BibitemShut
  {NoStop}%
\bibitem [{\citenamefont {Pound}(2015)}]{Pound:2015wva}%
  \BibitemOpen
  \bibfield  {author} {\bibinfo {author} {\bibfnamefont {Adam}\ \bibnamefont
  {Pound}},\ }\bibfield  {title} {\enquote {\bibinfo {title} {{Second-order
  perturbation theory: problems on large scales}},}\ }\href {\doibase
  10.1103/PhysRevD.92.104047} {\bibfield  {journal} {\bibinfo  {journal} {Phys.
  Rev. D}\ }\textbf {\bibinfo {volume} {92}},\ \bibinfo {pages} {104047}
  (\bibinfo {year} {2015})},\ \Eprint {http://arxiv.org/abs/1510.05172}
  {arXiv:1510.05172 [gr-qc]} \BibitemShut {NoStop}%
\bibitem [{\citenamefont {Panosso~Macedo}\ \emph {et~al.}(2022)\citenamefont
  {Panosso~Macedo}, \citenamefont {Leather}, \citenamefont {Warburton},
  \citenamefont {Wardell},\ and\ \citenamefont
  {Zengino\u{g}lu}}]{PanossoMacedo:2022fdi}%
  \BibitemOpen
  \bibfield  {author} {\bibinfo {author} {\bibfnamefont {Rodrigo}\ \bibnamefont
  {Panosso~Macedo}}, \bibinfo {author} {\bibfnamefont {Benjamin}\ \bibnamefont
  {Leather}}, \bibinfo {author} {\bibfnamefont {Niels}\ \bibnamefont
  {Warburton}}, \bibinfo {author} {\bibfnamefont {Barry}\ \bibnamefont
  {Wardell}}, \ and\ \bibinfo {author} {\bibfnamefont {An\i{}l}\ \bibnamefont
  {Zengino\u{g}lu}},\ }\bibfield  {title} {\enquote {\bibinfo {title}
  {{Hyperboloidal method for frequency-domain self-force calculations}},}\
  }\href {\doibase 10.1103/PhysRevD.105.104033} {\bibfield  {journal} {\bibinfo
   {journal} {Phys. Rev. D}\ }\textbf {\bibinfo {volume} {105}},\ \bibinfo
  {pages} {104033} (\bibinfo {year} {2022})},\ \Eprint
  {http://arxiv.org/abs/2202.01794} {arXiv:2202.01794 [gr-qc]} \BibitemShut
  {NoStop}%
\bibitem [{\citenamefont {Da~Silva}\ \emph {et~al.}(2023)\citenamefont
  {Da~Silva}, \citenamefont {Panosso~Macedo}, \citenamefont {Thompson},
  \citenamefont {Kroon}, \citenamefont {Durkan},\ and\ \citenamefont
  {Long}}]{DaSilva:2023xif}%
  \BibitemOpen
  \bibfield  {author} {\bibinfo {author} {\bibfnamefont {Lidia J.~Gomes}\
  \bibnamefont {Da~Silva}}, \bibinfo {author} {\bibfnamefont {Rodrigo}\
  \bibnamefont {Panosso~Macedo}}, \bibinfo {author} {\bibfnamefont
  {Jonathan~E.}\ \bibnamefont {Thompson}}, \bibinfo {author} {\bibfnamefont
  {Juan A.~Valiente}\ \bibnamefont {Kroon}}, \bibinfo {author} {\bibfnamefont
  {Leanne}\ \bibnamefont {Durkan}}, \ and\ \bibinfo {author} {\bibfnamefont
  {Oliver}\ \bibnamefont {Long}},\ }\bibfield  {title} {\enquote {\bibinfo
  {title} {{Hyperboloidal discontinuous time-symmetric numerical algorithm with
  higher order jumps for gravitational self-force computations in the time
  domain}},}\ }\href@noop {} {\  (\bibinfo {year} {2023})},\ \Eprint
  {http://arxiv.org/abs/2306.13153} {arXiv:2306.13153 [gr-qc]} \BibitemShut
  {NoStop}%
\bibitem [{\citenamefont {Panosso~Macedo}(2023)}]{PanossoMacedo:2023qzp}%
  \BibitemOpen
  \bibfield  {author} {\bibinfo {author} {\bibfnamefont {Rodrigo}\ \bibnamefont
  {Panosso~Macedo}},\ }\bibfield  {title} {\enquote {\bibinfo {title}
  {{Hyperboloidal approach for static spherically symmetric spacetimes: a
  didactical introduction and applications in black-hole physics}},}\
  }\href@noop {} {\  (\bibinfo {year} {2023})},\ \Eprint
  {http://arxiv.org/abs/2307.15735} {arXiv:2307.15735 [gr-qc]} \BibitemShut
  {NoStop}%
\bibitem [{\citenamefont {van~de Meent}\ and\ \citenamefont
  {Shah}(2015)}]{van2015metric}%
  \BibitemOpen
  \bibfield  {author} {\bibinfo {author} {\bibfnamefont {Maarten}\ \bibnamefont
  {van~de Meent}}\ and\ \bibinfo {author} {\bibfnamefont {Abhay~G.}\
  \bibnamefont {Shah}},\ }\bibfield  {title} {\enquote {\bibinfo {title}
  {{Metric perturbations produced by eccentric equatorial orbits around a Kerr
  black hole}},}\ }\href {\doibase 10.1103/PhysRevD.92.064025} {\bibfield
  {journal} {\bibinfo  {journal} {Phys. Rev. D}\ }\textbf {\bibinfo {volume}
  {92}},\ \bibinfo {pages} {064025} (\bibinfo {year} {2015})},\ \Eprint
  {http://arxiv.org/abs/1506.04755} {arXiv:1506.04755 [gr-qc]} \BibitemShut
  {NoStop}%
\bibitem [{\citenamefont {Starobinskii}\ and\ \citenamefont
  {Churilov}(1974)}]{Starobinskil:1974nkd}%
  \BibitemOpen
  \bibfield  {author} {\bibinfo {author} {\bibfnamefont {Alexei~A.}\
  \bibnamefont {Starobinskii}}\ and\ \bibinfo {author} {\bibfnamefont {S.~M.}\
  \bibnamefont {Churilov}},\ }\bibfield  {title} {\enquote {\bibinfo {title}
  {{Amplification of electromagnetic and gravitational waves scattered by a
  rotating "black hole"}},}\ }\href@noop {} {\bibfield  {journal} {\bibinfo
  {journal} {Sov. Phys. JETP}\ }\textbf {\bibinfo {volume} {38}},\ \bibinfo
  {pages} {1--5} (\bibinfo {year} {1974})}\BibitemShut {NoStop}%
\bibitem [{\citenamefont {Teukolsky}\ and\ \citenamefont
  {Press}(1974)}]{teukolsky1974perturbations}%
  \BibitemOpen
  \bibfield  {author} {\bibinfo {author} {\bibfnamefont {Saul~A}\ \bibnamefont
  {Teukolsky}}\ and\ \bibinfo {author} {\bibfnamefont {WH}~\bibnamefont
  {Press}},\ }\bibfield  {title} {\enquote {\bibinfo {title} {Perturbations of
  a rotating black hole. iii-interaction of the hole with gravitational and
  electromagnetic radiation},}\ }\href@noop {} {\bibfield  {journal} {\bibinfo
  {journal} {Astrophys. J.}\ }\textbf {\bibinfo {volume} {193}},\ \bibinfo
  {pages} {443--461} (\bibinfo {year} {1974})}\BibitemShut {NoStop}%
\bibitem [{\citenamefont {Loutrel}\ \emph {et~al.}(2021)\citenamefont
  {Loutrel}, \citenamefont {Ripley}, \citenamefont {Giorgi},\ and\
  \citenamefont {Pretorius}}]{Loutrel:2020wbw}%
  \BibitemOpen
  \bibfield  {author} {\bibinfo {author} {\bibfnamefont {Nicholas}\
  \bibnamefont {Loutrel}}, \bibinfo {author} {\bibfnamefont {Justin~L.}\
  \bibnamefont {Ripley}}, \bibinfo {author} {\bibfnamefont {Elena}\
  \bibnamefont {Giorgi}}, \ and\ \bibinfo {author} {\bibfnamefont {Frans}\
  \bibnamefont {Pretorius}},\ }\bibfield  {title} {\enquote {\bibinfo {title}
  {{Second Order Perturbations of Kerr Black Holes: Reconstruction of the
  Metric}},}\ }\href {\doibase 10.1103/PhysRevD.103.104017} {\bibfield
  {journal} {\bibinfo  {journal} {Phys. Rev. D}\ }\textbf {\bibinfo {volume}
  {103}},\ \bibinfo {pages} {104017} (\bibinfo {year} {2021})},\ \Eprint
  {http://arxiv.org/abs/2008.11770} {arXiv:2008.11770 [gr-qc]} \BibitemShut
  {NoStop}%
\end{thebibliography}%

\end{document}